\documentclass[%
 preprint,
 amsmath,
 amssymb,
 aps,
 nofootinbib,
 preprintnumbers
]{revtex4-1}

\usepackage{simplewick}
\usepackage{footnote}
\usepackage{graphicx}
\usepackage{dcolumn}
\usepackage{bm}
\usepackage{stmaryrd}
\usepackage{amsmath}
\usepackage{amssymb}
\usepackage{bbm}
\usepackage{amsfonts}
\usepackage{xcolor}
\usepackage{footmisc}
\usepackage{comment}
\usepackage[linktocpage=true,bookmarksnumbered=true,colorlinks=true,plainpages,linkcolor=blue,citecolor=red]{hyperref}

\newcommand{\matrixbb}[4]{\left(\hspace{-5 pt}\begin{tabular}{ c c } ${#1}$ & ${#2}$ \\ ${#3}$ & ${#4}$ \end{tabular}\hspace{-5 pt}\right)}

\newcommand{\vep}{\varepsilon}

\begin{document}
\preprint{MSUHEP-22-021, ADP-22-20/T1191}
\title{Spin-2 Kaluza-Klein Scattering in a \texorpdfstring{\linebreak}{Lg}
Stabilized Warped Background}

\author{R. Sekhar Chivukula$^{a}$}
\author{Dennis Foren$^{a}$}
\author{Kirtimaan A. Mohan$^{b}$}
\author{Dipan Sengupta$^{a,c}$}
\author{Elizabeth H. Simmons$^{a}$}

\affiliation{$^{a}$ Department of Physics and Astronomy, 9500 Gilman Drive,
 University of California, San Diego }
 \affiliation{$^{b}$ Department of Physics and Astronomy, 567 Wilson Road, Michigan State University, East Lansing}
 \affiliation{$^c$ARC Centre of Excellence for Dark Matter Particle Physics, Department of Physics, University of Adelaide, South Australia 5005, Australia}
\email{rschivukula@physics.ucsd.edu}
\email{dennisforen@gmail.com}
\email{kamohan@msu.edu}
\email{disengupta@physics.ucsd.edu}
\email{ehsimmons@ucsd.edu}

\date{\today}

\begin{abstract}
Scattering amplitudes involving massive spin-2 particles typically grow rapidly with energy. In this paper we demonstrate that the anomalous high-energy growth of the scattering amplitudes cancel for the massive spin-2 Kaluza-Klein modes arising from compactified five-dimensional gravity in a stabilized warped geometry. Generalizing previous work, we show that the two sum rules which enforce the cancellations between the contributions to the scattering amplitudes coming from the exchange of the (massive) radion and those from the exchange of the tower of Goldberger-Wise scalar states (admixtures of the original gravitational and scalar fields of the theory) still persist in the case of the warping which would be required to produce the hierarchy between the weak and Planck scales in a Randall-Sundrum model. We provide an analytic proof of one combination of these generalized scalar sum rules, and show how the sum rule depends on the Einstein equations determining the background geometry and the mode-equations and normalization of the tower of physical scalar states.  Finally, we provide a consistent and self-contained derivation of the equations governing the physical scalar modes and we list, in appendices, the full set of sum rules ensuring proper high-energy growth of all $2\to 2$ massive spin-2 scattering amplitudes.

\end{abstract}
\maketitle

\tableofcontents
\vfill\eject

\everypar{\looseness=-1}
\section{Introduction}

Historically, extra-dimensional theories of gravity were introduced soon after Einstein's discovery of the general theory of relativity. In the original form 
extra dimensions were introduced by Kaluza and Klein (KK) to unify electromagnetism with gravity, the only two fundamental forces known at the time \cite{Kaluza:1921tu,Klein:1926tv}. Extra-dimensional models have continued to evolve since the late 1970's, thanks in large part to the development of string theory. Over the last three decades, low-energy realizations of extra-dimensional models gained prevalence as well-motivated scenarios of physics beyond the standard model. One of the most popular and phenomenologically-viable models of extra dimensions is the Randall-Sundrum model \cite{Randall:1999ee,Randall:1999vf}, wherein a compact extra dimension in Anti-de Sitter space is used to generate relative exponential factors; this factor allows particles fixed to a brane to interact at electroweak strength while also ensuring bulk-propagating gravity is weak in the observed extended four-dimensional space, thus providing a geometric solution for the hierarchy problem of the standard model. \looseness=-1

Low-energy four-dimensional effective field theories arising from compactified theories of gravity involve towers of interacting spin-0 and spin-2 fields (and potentially spin-1 fields as well, though these are often eliminated by imposing an orbifold symmetry on the compact extra dimension).  The massive spin-2 resonances -- sometimes called Kaluza-Klein (KK) gravitons -- are particularly interesting. The existence of self interactions between these KK gravitons is problematic because typically scattering amplitudes between massive spin-2 particles grow far too rapidly with energy to keep unitarity constraints satisfied much beyond the mass of lightest massive spin-2 state involved. For example, theories of massive gravity that extend 4D general relativity by adding a Fierz-Pauli mass term \cite{Fierz:1939ix} result in 2-to-2 scattering amplitudes for the helicity-zero channel (the channel whose amplitude has the highest energy-growth) that grow like $s^{5}/(m_{\text{FP}}^{8}M_{\text{Pl}}^{2})$  \cite{ArkaniHamed:2002sp}, where $m_{\text{FP}}$ is the mass of the graviton, $s$ the squared center-of-mass energy, and $M_{\text{Pl}}$ the reduced Planck mass. Adding carefully-chosen potential terms to the model \cite{Hinterbichler:2011tt,deRham:2014zqa} can soften matrix element growth down to ${\cal O}(s^3)$. However, compactified theories of extra-dimensional gravity should -- due to higher-dimensional diffeomorphism invariance, requiring a smooth $m_{\text{KK}}\to 0$ limit -- grow at most as ${\cal O}(s)$. \looseness =-1

As we  have demonstrated in earlier work \cite{Chivukula:2019rij,Chivukula:2019zkt,Chivukula:2020hvi,Foren:2020egq}, while the individual contributions to massive helicity-zero spin-2 scattering amplitudes do each grow like ${\cal O}(s^5)$ in compactified gravity theories, cancellations occur between these contributions ultimately ensure the overall 2-to-2 scattering amplitudes grow no faster than ${\cal O}(s)$.\footnote{See also \cite{Bonifacio:2019ioc,Hang:2021fmp}.} These cancellations require that the couplings and masses of the Kaluza-Klein spin-2 and spin-0 modes be constrained to satisfy various sum-rule relations, which we have shown to be satisfied in both flat and warped compactifications.

In the original formulation of the RS1 (Randall-Sundrum I) model, extra-dimensional gravity generates a massless radion in the effective 4D theory\footnote{Along with the usual massless 4D graviton.}. A massless radion couples to the trace of the stress-energy tensor, yielding a Brans-Dicke like theory, in odds with predictions of the general theory of relativity. Additionally, this massless radion sources a Casimir force \cite{Hofmann:2000cj}, thereby destabilizing the extra dimension and leading to its collapse.  A naive computation of scattering amplitudes of massive spin-2 KK particles in a compactified theory of gravity, with a radion mass introduced by hand, reveals that scattering amplitudes grow like $m^{2}_{\text{radion}}s^{2}/(m_{\text{KK}}^{4}M_{\text{Pl}}^{2})$, i.e. ${\cal O}(s^{2})$ instead of ${\cal O}(s)$.  In order to stabilize the extra dimension in a way that retains validity of the theory up to the Planck scale, the radion must not only be made massive, but its mass must be generated {\it dynamically}. These dynamics must also provide additional contributions guaranteeing that the overall scattering amplitude grows no faster than ${\cal O}(s)$ -- as we have explicitly shown in \cite{Chivukula:2021xod} and explore here in detail.

In the same year that Randall and Sundrum published their model, Goldberger and Wise published a dynamical mechanism for stabilizing the model's extra dimension \cite{Goldberger:1999uk,Goldberger:1999un}. Their mechanism shares conceptual similarities with the standard technique of generating massive gauge bosons by spontaneously breaking the associated underlying gauge symmetry, wherein dynamics producing a non-zero vacuum expectation value (VEV) for a gauge-variant operator induces mixing between the longitudinal components of the gauge bosons with Goldstone bosons. In the Goldberger-Wise (GW) mechanism, a new 5D bulk scalar field $\hat{\Phi}(x,y)$ is appended to the RS1 model and included in new potential terms that also involve (via standard gravitational factors) the usual RS1 metric fields. That bulk scalar field then spontaneously acquires a background profile $\phi_{0}(y)$ with nonconstant dependence on the extra-dimensional coordinate $y$, causing mixing between background fluctuations $\hat{f}$ of the bulk scalar field and scalar fluctuations $\hat{r}$ of the RS1 metric. While this could in principle yield two physically-relevant superpositions of the bulk scalar field and scalar metric fluctuations, one combination is automatically forced to vanish in unitary gauge, leaving just one physical 5D scalar field in the spectrum.

Quantitatively, the new mixed scalar sector decomposes into a tower of spin-0 modes, all described by a single Sturm-Liouville equation with non-trivial Robin boundary conditions involving delta function contributions at the boundaries of the RS1 geometry. In this way, the GW mechanism ultimately generates an infinite tower of physical massive spin-0 states. But what happened to the radion? If the background profile of the bulk scalar field were, instead, made constant in the extra-dimensional coordinate, no extra-dimensional symmetries would be spontaneously broken and all but the lowest state in the spin-0 tower would cease to mix with the gravitational sector of the theory. In this limit, the lowest state would become massless, and -- indeed -- its couplings would exactly match those of the original unstabilized RS1 radion. Hence, in the GW-stabilized RS1 model we are considering, the radion should be associated with the lightest {\it massive} state among an entire tower of massive spin-0 states.

The assessment of the validity of the GW-stabilized RS1 effective field theory proceeds as for the corresponding unstabilized case, by calculating $2\to 2$ massive KK graviton scattering, now with an extended scalar sector, as compared to only the radion in the unstabilized case. An account of this calculation was provided in \cite{Chivukula:2021xod}, where we formulated an extended set of sum-rules  required to ensure that scattering amplitudes were well behaved in a stabilized theory of extra dimension without reference to any explicit GW model. In addition, we proposed a simple model of a stabilized-but-approximately-flat extra dimension, the ``flat stabilized" model, and we demonstrated that the revised sum-rules were satisfied in this model. \looseness=-1

In this work we extend prior results into a new domain by computing the couplings and masses of the scalar and spin-2 states in a Randall-Sundrum model with a Goldberger-Wise stabilization mechanism in the phenomenologically interesting case in which the warping reproduces the hierarchy between the weak and Planck scales. We provide a self-contained derivation of the equations governing the physical scalar modes. We show how one combination of the generalized sum-rules in particular relies explicitly on the equations determining the background bulk scalar field and metric, including the scalar mode wavefunctions and their normalization conditions. We introduce a model in which the Goldberger-Wise dynamics are a small perturbation away from the unstabilized warped RS1 model. We then demonstrate numerically that all of the sum rules needed to ensure that the anomalous growth of the scattering amplitudes cancel are satisfied to leading non-trivial order in perturbation theory.

The computations performed in the pure gravity sector in this paper and the preceding ones \cite{Chivukula:2019rij,Chivukula:2019zkt,Chivukula:2020hvi,Chivukula:2021xod}, demonstrating that amplitudes in compactified extra dimensional theories grow no faster than $\mathcal{O}(s)$, have important phenomenological consequences. The underlying higher-dimensional diffeomorphism invariance that ensures that scattering amplitudes are well behaved also guarantees that scattering amplitudes involving matter particles should also be compliant with the same principle. For example, in calculations for cosmological observables like relic abundances of dark matter with KK graviton portals (both for freeze-in and freeze-out), the velocity averaged cross sections must be properly estimated at large $\sqrt{s}$. An erroneous estimate with anomalously growing cross sections would lead to inaccurate predictions for cosmological observables within the scope of these models. We will illustrate these issues in detail in future work.

The rest of the paper is organized as follows. In Sec. \ref{sec:theLagrangian} we describe the Lagrangian of the Goldberger-Wise Randall-Sundrum model and set notation for the background geometry. In Sec. \ref{sec:KaluzaKleinExpansions} we describe the spin-0 and spin-2 mode expansions. Our analysis of the Kaluza-Klein expansions for this system follows the computations of \cite{Csaki:2000zn,Kofman:2004tk,Boos:2005dc,Boos:2012zz}, generalized to de Donder gauge, and is presented in detail for completeness and clarity in Appendix \ref{app:Lagrangian}. A review of Kaluza-Klein mode scattering and couplings, and description of the version of the sum rules of \cite{Chivukula:2021xod} used here, as well as a description of the analytic proof of one combination of the sum rules involving the scalar sector is given in Sec. \ref{sec:SumRules}. Details of the analytic proof of the sum rule is given in Appendix \ref{ShortSumRules}. We also provide, in the totality of Appendix \ref{DerivingSumRules}, a complete list of the sum-rule relations which must be satisfied for all $2 \to 2$ massive spin-2 scattering amplitudes to grow no faster than ${\cal O}(s)$ -- completing the analyses begun in \cite{Chivukula:2019rij,Chivukula:2019zkt,Chivukula:2020hvi,Foren:2020egq}.  A description and the analysis of the perturbative warped-stabilized model is given in Sec. \ref{sec:DFGK}. In particular, our numerical checks of the sum rules in this model are illustrated in Fig. \ref{fig:fig3} and Fig. \ref{fig:fig4} of Sec. \ref{sec:NumericalVerification}. Our perturbative analysis requires a slight generalization of Rayleigh-Schr\"odinger perturbation theory to account for perturbations in the weight-function of the corresponding Sturm-Liouville problem, and this formalism is described in Appendix \ref{app:PertTheory}. Our conclusions are given in Sec. \ref{sec:conclusion}. Mathematica \cite{Mathematica} files giving the expressions for all the spin-2 and spin-0 perturbative wavefunctions can be found on \href{https://github.com/kirtimaan/Stabilized-Extra-Dimenion}{GitHub}.\footnote{\url{https://github.com/kirtimaan/Stabilized-Extra-Dimenion}}

\section{The Lagrangian}

\label{sec:theLagrangian}

In this section we outline schematically how a canonical 4D effective Lagrangian is derived from a 5D RS1 model stabilized by the Goldberger-Wise mechanism. We provide a self-contained discussion of the details of this derivation in Appendix~\ref{app:Lagrangian} utilizing arguments similar to those found in refs.~\cite{Kofman:2004tk,Boos:2005dc,Boos:2012zz}, generalized to de Donder gauge to enable consistent scattering amplitude computations for processes involving the (massless) graviton. In this section we specify our notation and outline the results needed to present our computations. 

We begin by writing down the Lagrangian which consists of the following terms. 
\begin{align}
    \mathcal{L}_{\text{5D}} \equiv \mathcal{L}_{\text{EH}} +  \mathcal{L}_{\Phi\Phi} + \mathcal{L}_{\text{pot}} +\mathcal{L}_{\text{GHY}} +\Delta \mathcal{L} ~.
    \label{eq:L5D}
\end{align}
Here $\mathcal{L}_{\text{EH}}$ comes from the usual Einstein-Hilbert action, $\mathcal{L}_{\Phi\Phi}$ and $\mathcal{L}_{\text{pot}}$ are the kinetic and potential terms respectively of a bulk scalar field $\hat{\Phi}(x,y)$, $\mathcal{L}_{\text{GHY}}$ is the Gibbons-Hawking-York (GHY) boundary term \cite{York:1972sj,Gibbons:1976ue}, and $\Delta \mathcal{L}$ is a useful total derivative we define in Appendix~\ref{app:Lagrangian}. The combination of $\mathcal{L}_{\text{GHY}}$ and $\Delta \mathcal{L}$  
is required to have a well-posed variational principle 
for the gravitational action \cite{Dyer:2008hb}. 
This Lagrangian is a function of the 5D metric $G$, which we parameterize in terms a 4D metric perturbation $g_{\mu\nu}$ and a scalar metric perturbation $\hat{r}$ as \cite{Charmousis:1999rg}
\begin{align}
    [G_{MN}] = \matrixbb{\,\,\,g_{\mu\nu} \,\, \exp{\left[-2\left(A(y)+\frac{ e^{2A(y)}}{2\sqrt{6}} \kappa\,\hat{r}(x,y)\right)\right]}}{0}{0}{-\left(1 + \frac{ e^{2A(y)}}{\sqrt{6}} \kappa\,\hat{r}(x,y)\right)^{2}\,\,\,} \ .
    \label{eq:metric}
\end{align}
in terms of coordinates $x^{M} \equiv (x^{\mu},y)$, where $y \in (-\pi r_{c},+\pi r_{c}]$ parameterizes an orbifolded extra-dimension\footnote{That is, the extra-dimension is a circle in which the points with coordinates $y$ and $-y$ are identified. As we will see, this view of the extra dimension (as opposed to treating it as a line-segment) allows us to motivate and use the boundary conditions of the Kaluza-Klein mode equations at the orbifold fixed-points at $y=0$ and $y=\pi$ more easily.}; for convenience we define $\varphi \equiv y/r_{c}$, and use $y$ or $\varphi$ interchangeably as the coordinate of the fifth dimension. Note that this form of the metric was used in our previous works \cite{Chivukula:2019rij,Chivukula:2019zkt,Chivukula:2020hvi} for the unstabilized RS1 metric, following \cite{Charmousis:1999rg} to bring the quadratic Lagrangian to a canonical form from the outset. In the case of the stabilized model, the situation is further complicated by a nontrivial mixing between the bulk scalar field and the scalar metric fluctuations. As we will explain below and demonstrate explicitly in Appendix \ref{app:Lagrangian}, the total derivative $\Delta\mathcal{L}$ helps us bring the Lagrangian into a canonical form. 

The warp factor $A(y)$ encodes the warped background geometry. In RS1 \cite{Randall:1999ee}, in which the extra dimension is unstabilized, $A(y) = k|y|$, where $k$ is related to the spacetime curvature. The Goldberger-Wise mechanism \cite{Goldberger:1999uk,Goldberger:1999un} complicates the background geometry such that the specific form of $A(y)$ becomes dependent on the details of the mechanism's bulk scalar interactions. Crucially, the scalar (bulk and boundary) potential terms in $\mathcal{L}_{\text{pot}}$ are chosen such that the scalar field gains a $y$-dependent background field value, and the trade-off between the contributions to the action from bulk kinetic energy terms $\mathcal{L}_{\Phi\Phi}$ and the scalar potential(s) stabilize the size of the extra-dimension.

The Lagrangian thus far is written in terms of the metric perturbations $\{g_{\mu\nu},\hat{r}\}$ and the bulk scalar field $\hat{\Phi}$. We next expand $g_{\mu\nu}$ and $\hat{\Phi}$ about their background values:
\begin{align}
    g_{\mu\nu}(x,y) & \equiv \eta_{\mu\nu} + \kappa\, \hat{h}_{\mu\nu}(x,y) \nonumber \\
    \hat{\Phi} (x,y) &\equiv \frac{1}{\kappa}\hat{\phi} \equiv \frac{1}{\kappa}\left[\phi_{0}(y) + \hat{f}(x,y)\right] ~.\label{eq:gandPhi}
\end{align}
The background metric $\eta_{\mu\nu} = \text{Diag}(+1,-1,-1,-1)$ of $g_{\mu\nu}$ is determined by demanding Lorentz-invariance along the extended dimensions, while the background value $\phi_{0}/\kappa$ of $\hat{\Phi}$ must be found by solving the theory's field equations. We normalize the Lagrangian such that the 5D gravitational coupling $\kappa$ is related to the 5D Planck mass $M_{\text{Pl,5D}}$ according to $\kappa^2=4/M_{\text{Pl,5D}}^{3}$. Note that the factors of $\kappa$ (units: energy$^{-3/2}$) included in Eq. \eqref{eq:gandPhi} are such that $\phi_{0}$ and $\hat{f}$ are unitless in natural units. Following Kaluza-Klein (KK) decomposition, the 5D tensor field $\hat{h}_{\mu\nu}$, the 5D scalar field $\hat{r}$, and 5D scalar fluctuation field $\hat{f}$ give rise to an infinite tower of 4D states.

Perturbatively expanding the Lagrangian Eq.~\eqref{eq:L5D} order-by-order in $\kappa$ yields terms containing various powers of $h_{\mu\nu}$, $\hat{f}$, and $\hat{r}$. In particular, at quadratic order in the fields, we find a complicated expression, {\it c.f.} Eqs.~\eqref{eq:LrrBreakdown}-\eqref{eq:L5D_quad1}. Thankfully, there are residual five-dimensional diffeomorphism transformations which leave the form of Eq. \eqref{eq:metric} invariant - these transformations allow us to reorganize how the physical content is embedded in the fields and thereby attain explicitly canonical quadratic Lagrangians.  This process will also mix the 5D fields $\hat{r}$ and $\hat{f}$ (and their constituent 4D states) together in a process that eventually leaves a single scalar tower of physical states. In particular, in order to ultimately bring the quadratic 5D Lagrangian into a form suitable for generating canonical 4D Lagrangians, we impose the gauge-fixing constraint\footnote{A demonstration that one can always impose this gauge constraint can be found in Refs.~\cite{Kofman:2004tk, Boos:2012zz}.}
\begin{align}
    (\partial_{\varphi}\phi_{0})\,\hat{f}(x,y) \equiv \sqrt{6}\, e^{2A(\varphi)}\,\big(\partial_{\varphi}\hat{r}\big)~, \label{SRS1GaugeCondition}
\end{align}
to eliminate the field $\hat{f}$ in terms of $\hat{r}$.\footnote{Note that in the limit  in which there is no non-trivial scalar background, $\phi'_0\to 0$, this constraint leaves only the constant ($\varphi$-independent) mode of $\hat{r}$ in the theory - a mode corresponding to the massless radion of the unstabilized theory. \label{footnote:unstabilized}}  
In this gauge the 5D theory's independent field degrees of freedom consist only of the 5D scalar field $\hat{r}$ and the 5D tensor field $\hat{h}_{\mu\nu}$. To yield a 4D effective theory, each of these 5D fields is subsequently decomposed into a tower of 4D Kaluza-Klein (KK) modes.
We emphasize here that bringing the Lagrangian to a canonical form is a non-trivial task, and it is of paramount importance to figure out all the interactions of both the gravitational and the scalar sector that will eventually determine the structure of the matrix elements and the couplings.

In order to calculate the desired matrix elements, we require the cubic and quartic self-interactions of the 5D tensor field $\hat{h}_{\mu\nu}$ as well as the $\hat{h} \hat{h} \hat{r}$ cubic interaction. The $\hat{h}$ self-interactions (and their 4D effective equivalents) are changed from our previous works \cite{Chivukula:2019zkt,Chivukula:2019rij,Foren:2020egq} only in the specific choice of $A(y)$. Following integration-by-parts and the elimination of total derivatives, we find that the $\hat{h}\hat{h}\hat{r}$ interaction is similarly identical to the unstabilized case \cite{Chivukula:2020hvi}:
\begin{align}
    \mathcal{L}_{hhr} &= -\dfrac{\kappa}{2r_{c}^{2}}\sqrt{\dfrac{3}{2}} e^{-2A(\varphi)} \Big[(\hat{h}^{\,\prime})^{2} - \hat{h}^{\,\prime}_{\mu\nu}(\hat{h}^{\mu\nu})^{\,\prime} \Big] \hat{r}~. \label{eq:hhr-coupling}
\end{align}
Thus, the primary difference between the stabilized and unstabilized cases as far as $\hat{h}\hat{h}\hat{r}$ is concerned regards the KK decomposition of the 5D field $\hat{r}(x,y)$. In the unstabilized case, $\hat{r}$ generates only a single massless scalar state $\hat{r}(x)$ (see footnote \ref{footnote:unstabilized})--- the usual RS1 radion. In the stabilized case, $\hat{r}$ has nontrivial $y$-dependence and instead generates an infinite tower of massive scalars $\{\hat{r}^{(i)}(x)\}$, wherein the lightest of these scalars (with KK number $i=0$) is identified as the massive radion and the heavier states are called Goldberger-Wise (GW) scalars. 
\footnote{Note that, having chosen to express the physical degrees of freedom in terms of the scalar field $\hat{r}$ in Eq. \eqref{SRS1GaugeCondition}, the {\it form} of the couplings between the massive spin-2 states and the tower of GW states is precisely the same as the form of the radion coupling in RS1. However, as we will see, the mode equation and normalization conditions for the physical GW scalars lead to additional complications.} From here on, we will drop the argument $\varphi$ in the warp factor $A(\varphi)$ for convenience.

\section{Kaluza-Klein Mode Expansions}

\label{sec:KaluzaKleinExpansions}

\label{sec:SLprob}
Upon Kaluza-Klein (KK) decomposition, the scalar and tensor modes come from extra-dimensional wavefunctions which satisfy one-dimensional Sturm-Liouville (SL) problems. For completeness and notational consistency, we provide details of the derivation of the
SL problems in the tensor and scalar sectors in Appendix \ref{app:Lagrangian} - following
the procedures originally described in \cite{Boos:2005dc,Boos:2012zz}, (see also Ref.~\cite{Chivukula:2021xod}). We report the results here for the convenience of the reader and to highlight the differences that arise when solving these problems in the stabilized RS1 model.
\subsection{Spin-2 Sector}
The tensor field $h_{\mu\nu}(x,y)$ is decomposed into a tower of 4D KK states $\hat{h}^{(n)}_{\mu\nu}(x)$ in the usual way as follows: recalling $\varphi \equiv y/r_{c}$,
\begin{align}
    \hat{h}_{\mu\nu}(x,y) = \dfrac{1}{\sqrt{\pi r_{c}}} \sum_{n=0}^{+\infty} \hat{h}^{(n)}_{\mu\nu}(x)\, \psi_{n}(\varphi)~.
    \label{eq:spin2mode-equation}
\end{align}
Here $r_c$ is the radius of the extra dimension and $\psi_n (\varphi)$ is the 5D wavefunction of the $n^{\text{th}}$ mode that satisfies the following SL differential equation:
\begin{align}
    \partial_{\varphi}[e^{-4A}\,\partial_{\varphi}\psi_{n}]=-\mu_{n}^{2} e^{-2A}\psi_{n}
     \label{eq:spin2-SLeqn}
\end{align}
where the wavefunctions satisfy the boundary conditions where $(\partial_{\varphi}\psi_{n}) = 0$ at $\varphi \in \{0,\pi\}$. 
The eigenvalues $\mu_n = m_{n} r_{c}$ are the masses $m_{n}$ of the $n^{\text{th}}$ spin-2 KK mode. The wavefunctions are normalized as follows 
\begin{align}
    \dfrac{1}{\pi}\int_{-\pi}^{+\pi} d\varphi\hspace{10 pt} e^{-2A}\,\psi_{m}\,\psi_{n} &= \delta_{m,n} \ ,
    \label{eq:spin2-mode-norm}
\end{align}
and satisfy the completeness relation
\begin{align}
    \delta(\varphi_{2}-\varphi_{1}) =  \dfrac{1}{\pi}\, e^{-2A}\sum_{j=0}^{+\infty} \psi_{j}(\varphi_{1})\, \psi_{j}(\varphi_{2})~.
    \label{eq:spin2-completeness}
\end{align}
The form of this SL problem is identical to the unstabilized case and differs only in how the new background geometry influences the warp factor $A$. In the unstabilized case, the warp factor is simply $kr_{c}|\varphi|$. In the stabilized case, the bulk scalar potential modifies the background geometry such that $A(\varphi)$ satisfies the (Einstein) equation
\begin{align}
    A^{\prime\prime} = \dfrac{1}{12}\bigg[(\phi_{0}^{\prime})^{2} + 4 \sum_{i=1,2} V_{i}r_{c}\,\delta_{i}\bigg] \label{eq:App}
\end{align}
in terms of the background scalar field $\phi_0$ and the brane-localized potentials $V_{1,2}$ at $\varphi = 0$ and $\varphi = \pi$ respectively (refer to Appendix \ref{app:Lagrangian} for additional details).

\subsection{Spin-0 Sector}
The spin-0 sector of the model arises from two sources. The first is the scalar metric fluctuation (where even the lightest mode will be $y$-dependent in the stabilized model), and the second is the new bulk scalar field. The two sectors mix via the gauge condition noted in Eq. \eqref{SRS1GaugeCondition}. Consequently, we attain a single physically-relevant 5D physical scalar perturbation $\hat{r}(x,y)$.
 Similar to the tensor perturbation, the KK decomposition  of the 5D scalar field $\hat{r}(x,y)$ into a tower of spin-0 KK modes proceeds by introducing extra-dimensional wavefunctions $\gamma_{i}(\varphi)$ and a tower of 4D scalar fields $\hat{r}^{(i)}(x)$ parameterized as follows:
\begin{align}
    \hat{r}(x,y) = \dfrac{1}{\sqrt{\pi r_{c}}} \sum_{i=0}^{+\infty} \hat{r}^{(i)}(x)\, \gamma_{i}(\varphi)~.
    \label{eq:spin0-mode-expansion}
\end{align}
The mode equation that brings the 5D scalar Lagrangian to canonical form, however, is quite different from the tensor case and involves nontrivial boundary terms
\begin{align}
    \partial_{\varphi}\bigg[\dfrac{e^{2A}}{(\phi_{0}^{\prime})^{2}} (\partial_{\varphi}\gamma_{i})\bigg] - \dfrac{e^{2A}}{6}\gamma_{i} = -\mu_{(i)}^{2}\,\dfrac{e^{4A}}{(\phi_{0}^{\prime})^{2}} \, \gamma_{i} \, \Bigg\{ 1 + \frac{2\,\delta(\varphi)}{\Big[ 2 \ddot{V}_{1} r_{c} - \frac{\phi _{0}^{\prime\prime}}{\phi_{0}^{\prime}} \Big]} +  \frac{2\,\delta(\varphi - \pi )}{\Big[ 2 \ddot{V}_{2} r_{c} + \frac{\phi _{0}^{\prime\prime}}{\phi_{0}^{\prime}} \Big]} \Bigg\}~,\label{eq:spin0-SLeqn}
\end{align}
where $\phi^{\prime}_{0} \equiv (\partial_{\varphi}\phi_{0})$ and the eigenvalues $\mu_{(n)} = m_{(n)} r_{c}$ are the masses $m_{(n)}$ of the $n^{\text{th}}$ scalar KK mode.  The Dirac deltas enforce the following (orbifold) boundary conditions:
\begin{align}
    (\partial_{\varphi}\gamma_{i})\bigg|_{\varphi = 0} &= -\bigg[2\ddot{V}_{1}r_{c} - \dfrac{\phi_{0}^{\prime\prime}}{\phi_{0}^{\prime}} \bigg]^{-1}\,\mu_{(i)}^{2}\,e^{2A}\,\gamma_{i} \bigg|_{\varphi = 0}~,\nonumber\\
    (\partial_{\varphi}\gamma_{i})\bigg|_{\varphi = \pi} &= +\bigg[2\ddot{V}_{2}\,r_{c} + \dfrac{\phi_{0}^{\prime\prime}}{\phi_{0}^{\prime}}\bigg]^{-1}\,\mu_{(i)}^{2}\,e^{2A}\,\gamma_{i} \bigg|_{\varphi = \pi}~,
    \label{eq:scalarBCtext}
\end{align}
where $\ddot{V}_{1,2}$ are second functional derivatives of the brane potentials evaluated at the background-field configuration. Note that these boundary conditions reduce to Neumann form in the ``stiff-wall" limit, $\ddot{V}_{1,2} \to +\infty$, a limit which will be useful to us during our numerical work in Sec. \ref{sec:DFGK}. In the form of Eq. \eqref{eq:spin0-SLeqn}, the Sturm-Liouville nature of the problem is manifest, and the orthogonality and completeness of the wavefunctions follow immediately \cite{Boos:2005dc,Kofman:2004tk,fulton1977two,binding1994sturm}.\footnote{The completeness of the solutions to the scalar Sturm-Liouville problem in Eq. \eqref{eq:spin0-SLeqn} and the positivity of the scalar mass-squared eigenvalues $\mu_{(i)}^{2}$ are only assured if the coefficients of the $\delta$-function terms are {\it non-negative} \cite{fulton1977two,binding1994sturm}. We will assume that the brane-potentials and background field are such as to satisfy this condition (as they do in the stiff-wall limit we use later). Physically this constraint is more easily understood in the analogous case of the modes of a string: in that case the $\delta$-functions can be understood as representing point masses which can freely move at the boundary of the string, and the coefficients are proportional to these masses and must therefore be positive for stability.}

Due to the mixing between the gravitational and bulk scalar sectors in Eq. \eqref{SRS1GaugeCondition}, however, an
unconventional normalization of the scalar wavefunctions is required to
bring the scalar kinetic energy terms to canonical form\footnote{Because $\mu^2_{(i)}>0$ in the GW model, this normalization choice (albeit unusual) is consistent. Taking the unstabilized limit $\phi'_0\to 0$ (in which $\mu_{(0)}\to 0$ and all other scalar states decouple; see Eq \eqref{SRS1GaugeCondition}) however requires care. This limit is discussed in the context of the ``flat-stabilized" model in  \cite{Chivukula:2021xod}.} 
\begin{align}
    \delta_{mn} & = \dfrac{6\mu^2_{(n)}}
    {\pi}\int_{-\pi}^{+\pi} d\varphi\hspace{5 pt} \gamma_m\gamma_n \dfrac{e^{4A}}{(\phi_{0}^{\prime})^{2}} \Bigg\{ 1 + \frac{2\,\delta(\varphi)}{\Big[ 2 \ddot{V}_{1} r_{c} - \frac{\phi _{0}^{\prime\prime}}{\phi_{0}^{\prime}} \Big]} +  \frac{2\,\delta(\varphi - \pi )}{\Big[ 2 \ddot{V}_{2} r_{c} + \frac{\phi _{0}^{\prime\prime}}{\phi_{0}^{\prime}} \Big]} \Bigg\} \label{eqScalar:Norm2}\\
    & = \dfrac{6}{\pi} \int_{-\pi}^{+\pi} d\varphi\hspace{5 pt} \bigg[\dfrac{e^{2A}}{(\phi_{0}^{\prime})^{2}} \gamma_{m}^{\,\prime} \, \gamma_{n}^{\,\prime} + \dfrac{e^{2A}}{6} \gamma_{m}\gamma_{n} \bigg]~, \label{eqScalar:Norm1}
\end{align}
where the second line follows by applying the differential equation (\ref{eq:spin0-SLeqn}) and integration-by-parts on the periodic doubled orbifold. The scalar wavefunction completeness relation follows from Eq. \eqref{eqScalar:Norm2}:
\begin{equation}
    \delta(\varphi_2-\varphi_1) = \dfrac{6}{\pi} \dfrac{e^{4A(\varphi_1)}}{(\phi_{0}^{\prime}(\varphi_1))^{2}} \, \Bigg\{ 1 + \frac{2\,\delta(\varphi_1)}{\Big[ 2 \ddot{V}_{1} r_{c} - \frac{\phi _{0}^{\prime\prime}}{\phi_{0}^{\prime}} \Big]} +  \frac{2\,\delta(\varphi_1 - \pi )}{\Big[ 2 \ddot{V}_{2} r_{c} + \frac{\phi _{0}^{\prime\prime}}{\phi_{0}^{\prime}} \Big]}\Bigg\}~
    \sum_{j=0}^{+\infty} \mu_{(j)}^{2} \gamma_{j}(\varphi_{1})\, \gamma_{j}(\varphi_{2})~.
    \label{eq:spin0-completeness}
\end{equation}
The second form of the normalization condition in Eq. \eqref{eqScalar:Norm1} is useful in computational work, since it does not rely on knowledge of the eigenvalues.  Details are given in Appendix~\ref{app:Lagrangian}.

Because of the Neumann boundary conditions $(\partial_{\varphi}\psi_{n}) = 0$ and the simplicity of Eq.~\eqref{eq:spin2-SLeqn}, there is always a massless spin-2 mode (with a wavefunction constant in $\varphi$) in the tensor tower. This is not the case for the scalar tower. Due to the non-constant potential of the background scalar, along with its vacuum expectation value, the lightest spin-0 state (identified as the radion with a wavefunction $\gamma_{0}$) acquires a mass $\mu_{(0)}$. 

\section{Massive spin-2 couplings, scattering amplitudes, and Sum Rules}

\label{sec:SumRules}


\begin{figure}
    \centering
    \includegraphics[width=0.8\textwidth]{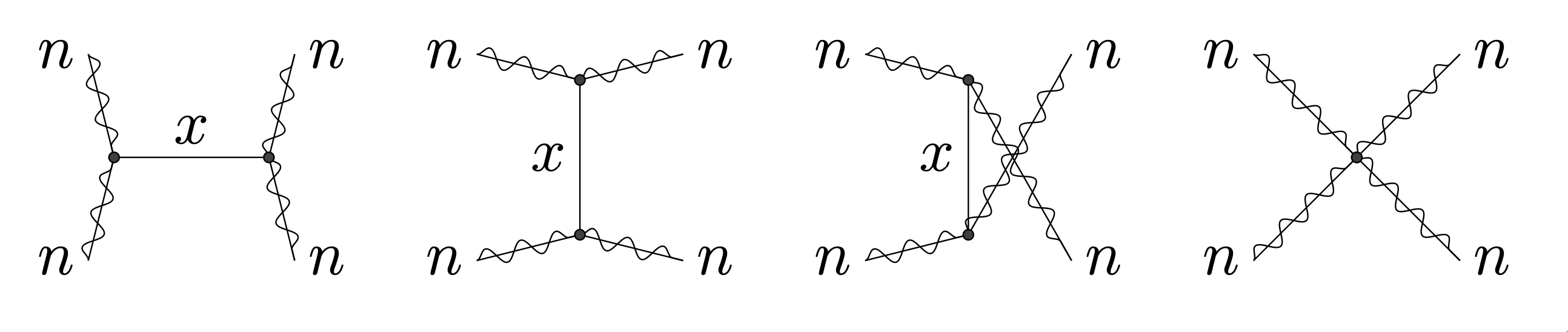}
    \caption{Matrix element diagrams contributing to $n,n \to n,n$ massive spin-2 KK boson scattering. Here $n$ refers to the KK mode number of the external state. The  intermediate states $x$ include a massive radion, the graviton, a tower of massive spin-2 KK bosons, and a tower of GW scalars.}
    \label{fig:kk-scattering}
\end{figure}

As discussed extensively in the literature (see, for example, \cite{Hinterbichler:2011tt,deRham:2014zqa}, and references therein), phenomenological calculations incorporating massive spin-2 states  often generate matrix element diagrams which exhibit anomalous high-energy behavior. 
Extra-dimensional models of gravity possess an underlying 5D diffeomorphism invariance that ensures their amplitudes are well behaved. That is, any overall bad high-energy growth necessarily signals the omission of additional important physics. Such omissions can  produce erroneous phenomenological results.\footnote{Following our previous computations \cite{Chivukula:2019rij,Chivukula:2019zkt,Chivukula:2020hvi} it was shown in \cite{deGiorgi:2021xvm} that in freeze-out computations with KK-graviton portal dark matter scenarios exhibit amplitudes that grow no faster than $\mathcal{O}(s)$ in an unstabilized model. The stabilized and phenomenologically relevant RS1 model is significantly more difficult to calculate, and will be presented in a future work. } \looseness=-1

In our previous work we analyzed the diagrams which contribute to $2\rightarrow 2$ massive spin-2 mode scattering for several variants of the Randall-Sundrum I model; within each of these analyses, we found there exist individual diagrams which diverge as fast as $\mathcal{O}(s^{5})$ at high energies  ($s$ being the usual Mandelstam parameter) and that cancellations between diagrams ensure the total matrix element only grows as fast as $\mathcal{O}(s)$ \cite{Chivukula:2019rij,Chivukula:2019zkt,Chivukula:2020hvi,Foren:2020egq}. This genuine $\mathcal{O}(s)$ growth is important because it insures the 4D effective theory breaks down at an energy scale consistent with the physics of the underlying extra-dimensional theory. The central purpose of this paper is to verify the various coupling relations and sum rules \cite{Chivukula:2021xod} required to ensure these cancellations in a general Goldberger-Wise-stabilized Randall-Sundrum I model, report how most of these rules may be proved in full generality, and (in the next section) numerically demonstrate leading $\mathcal{O}(s)$ growth of the matrix element to second-order in a solvable warped stabilized model.

In this section we review the definitions of the KK mode-couplings relevant to the scattering computations, \ref{subsec:IVA}, describe the sum-rules which apply to these couplings and show how they are related to the physics of the Goldberger-Wise model, \ref{subsec:IVB}, and provide a brief summary of the sum-rules we numerically verify in Sec. \ref{sec:sum-rule-summary}.

\subsection{Scattering Amplitudes and Couplings}
\label{subsec:IVA}
The tree-level diagrams relevant to the aforementioned $2\rightarrow2$ matrix element are shown in Fig.~\ref{fig:kk-scattering}; details can be found in \cite{Chivukula:2019rij,Chivukula:2019zkt,Foren:2020egq, Chivukula:2021xod,Chivukula:2020hvi}. The general strategy employed in this paper follows that of the unstabilized case.  Briefly, the total matrix element $\mathcal{M}$ involves a contact diagram $(\mathcal{M}_{c})$ as well as infinite sums over the diagrams ($\mathcal{M}_{j,X}$, $\mathcal{M}_{(i),X}$) describing $X = s$, $t$, and $u$-channel exchanges of intermediate spin-2 states and spin-0 states:
\begin{align*}
    \mathcal{M} = \mathcal{M}_{c} + \sum_{X \in \{s,t,u\}} \bigg[ \sum_{j=0}^{+\infty} \mathcal{M}_{j,X} + \sum_{i=0}^{+\infty} \mathcal{M}_{(i),X }\bigg]
\end{align*}
where in general KK numbers within parentheses (the $(i)$) refer to those of the spin-0 states, while those without parentheses (the $j$) reference the spin-2 states. We expand the matrix element $\mathcal{M}$ in a
Taylor series in $s$ as \cite{Chivukula:2020hvi}, 
\begin{equation}
    \mathcal{M}(s,\theta) = \sum_{\sigma\in \tfrac{1}{2}\mathbb{Z} } \overline{\mathcal{M}}^{(\sigma)}(\theta)\cdot s^{\sigma}
\end{equation}
and isolate the kinematic factors and the couplings. Generally, 5D diffeomorphism demands that all coefficients of $s^{\sigma}$ with $\sigma > 1$ in this expansion must vanish. At present, because we consider the process involving  helicity-zero external states, half-integer values of $\sigma$ automatically vanish. Demanding $\mathcal{O}(s)$ growth at most in this matrix element necessitates relationships between various masses and coupling structures in the theory.

The couplings present in each of these diagrams come from wavefunction overlap integrals attained following Kaluza-Klein decomposition of the fields in the Lagrangian; this procedure of attaining a 4D effective theory from a Lagrangian such as Eq.~\ref{eq:L5D} via KK decomposition is explained in detail in Refs.~\cite{Chivukula:2020hvi,Foren:2020egq}.
\begin{itemize}
    \item The contact diagram $\mathcal{M}_{c}$ involves the 4-point massive spin-2 vertex and contributes the following wavefunction overlap integrals:
    \begin{align*}
        a_{klmn} &\equiv \dfrac{1}{\pi}\int_{-\pi}^{+\pi} d\varphi\hspace{10 pt}e^{-2A}\,\psi_{k}\,\psi_{l}\,\psi_{m}\,\psi_{n} ~,\\
       b_{k^{\prime}l^{\prime}mn} &\equiv \dfrac{1}{\pi}\int_{-\pi}^{+\pi} d\varphi\hspace{10 pt}e^{-4A}\,(\partial_{\varphi}\psi_{k})(\partial_{\varphi}\psi_{l})\psi_{m}\,\psi_{n} ~.
    \end{align*}
    For helicity-zero, $\mathcal{M}_{c}$ diverges like $\mathcal{O}(s^{5})$.
    \item The spin-2 mediated diagrams $\mathcal{M}_{j,X}$ involve 3-point spin-2 vertices and contribute the following wavefunction overlap integrals:
    \begin{align*}
        a_{lmn} &\equiv \dfrac{1}{\pi}\int_{-\pi}^{+\pi} d\varphi\hspace{10 pt}e^{-2A}\,\psi_{l}\,\psi_{m}\,\psi_{n}~,\\
        b_{l^{\prime}m^{\prime}n} &\equiv \dfrac{1}{\pi}\int_{-\pi}^{+\pi} d\varphi\hspace{10 pt}e^{-4A}\,(\partial_{\varphi}\psi_{l})(\partial_{\varphi}\psi_{m})\psi_{n}~.
    \end{align*}
    For helicity-zero, each $\mathcal{M}_{j,X}$ diverges like $\mathcal{O}(s^{5})$
    \item The spin-0 mediated diagrams $\mathcal{M}_{(i),X}$ involve $3$-point scalar-(spin-2)-(spin-2) couplings and contribute the following wavefunction overlap integrals:
    \begin{align*}
        a_{l^{\prime}m^{\prime}(n)} &\equiv \dfrac{1}{\pi}\int_{-\pi}^{+\pi} d\varphi\hspace{10 pt}e^{-2A}\,(\partial_{\varphi}\psi_{l})(\partial_{\varphi}\psi_{m})\gamma_{n}~.
    \end{align*}
    Due to the structure of the Lagrangian, no corresponding ``$b_{l^{\prime}m^{\prime}(n)}$" or ``$b_{lm^{\prime}(n^{\prime})}$" is generated. For helicity-zero, each $\mathcal{M}_{(i),X}$ diverges like $\mathcal{O}(s^{3})$.
\end{itemize}
In general, (i) couplings labeled with an ``$a$" have an $e^{-2A}$ weight factor whereas those labeled with a ``$b$" involve $e^{-4A}$, (ii) the subscript KK indices indicate the relevant wavefunctions to include in each integral (remembering that parentheses indicate scalar modes), and (iii) a subscript KK index with a prime denotes that the corresponding mode number's wavefunction should be differentiated with respect to the extra-dimensional coordinate $\varphi$.

We can reduce the number of coupling integrals present by using the properties of the KK wavefunctions; namely, the mode equation and completeness relations, Eqs. \eqref{eq:spin2-SLeqn} and \eqref{eq:spin2-completeness}. For example, in prior work \cite{Chivukula:2020hvi,Foren:2020egq} we showed how the 
spin-2 mode equation Eq. \eqref{eq:spin2-SLeqn} and the corresponding completeness relation Eq. \eqref{eq:spin2-completeness} relate some of the $a$ and $b$ couplings:
\begin{align}
    b_{l^{\prime}m^{\prime}n} = \dfrac{1}{2}\Big[\mu_{l}^{2} + \mu_{m}^{2} - \mu_{n}^{2}\Big] a_{lmn}\hspace{35 pt}b_{n^{\prime}n^{\prime}nn} = \dfrac{\mu^2_n}{3}a_{nnnn}~.
    \label{eq:btoatext}
\end{align}
We use these relations and eliminate all $b$-type overlap integrals in favor of $a$-type integrals. Doing so, we may write the sum rules entirely using $a$-type overlap integrals.

\subsection{Sum Rules Insuring Consistency of Scattering Amplitudes}
\label{subsec:IVB}

By requiring the scattering amplitude to grow no faster than $\mathcal{O}(s)$ in the GW model \cite{Chivukula:2021xod}, we previously determined the following general sum rules should be satisfied:
\begin{align}
    \sum_{j=0} a_{nnj}^{2} &= a_{nnnn}~, \label{eq:sr1}\\
    \sum_{j=0} \mu_{j}^{2} a_{nnj}^{2} &= \dfrac{4}{3} \mu_{n}^{2} a_{nnnn}~,\label{eq:sr2}\\
    \sum_{j=0}^{+\infty} \mu_{j}^{4} a^{2}_{nnj} &=  \dfrac{4}{15}\mu_{n}^{4} (4\,a_{nnnn} - 3\,a^{2}_{nn0}) + \dfrac{36}{5} \sum_{i=0}^{+\infty} a^{2}_{n^{\prime}n^{\prime}(i)}~,\label{eq:sr3}\\
    \sum_{j=0}^{+\infty} \mu_{j}^{6} a^{2}_{nnj} &= -4\mu_{n}^{6}a^{2}_{nn0} + 9 \sum_{i=0}^{+\infty}(4\mu_{n}^{2}-\mu_{(i)}^{2}) a^{2}_{n^{\prime}n^{\prime}(i)}~, \label{eq:sr4}
\end{align}
The first two sum rules, Eqs. \eqref{eq:sr1} and \eqref{eq:sr2}, insure that the contributions to the scattering amplitudes growing like ${\cal O}(s^5)$ and ${\cal O}(s^4)$ vanish. They follow directly from the Sturm-Liouville form of the spin-2 KK mode equation, Eq. \eqref{eq:spin2-SLeqn} and the corresponding completeness relation Eq. \eqref{eq:spin2-completeness}- so the proofs given in \cite{Chivukula:2019zkt,Chivukula:2020hvi,Foren:2020egq} apply to any model producing a geometry defined by a warp function $A(y)$. However, the sum rules in Eqs. \eqref{eq:sr3} and \eqref{eq:sr4}, which insure cancellation of the contributions to the amplitude growing like ${\cal O}(s^3)$ and ${\cal O}(s^2)$, involve the scalar tower present in the GW model.




By combining the last two sum rules, Eqs. \eqref{eq:sr3} and \eqref{eq:sr4}, to eliminate the common sum $\sum_{i} a_{n^{\prime}n^{\prime}(i)}^{2}$, we find a mixed rule:
\begin{align}
    \sum_{j=0}^{+\infty} \bigg[5\mu_{n}^{2}-\mu_{j}^{2}\bigg]\mu_{j}^{4} a^{2}_{nnj} &= \dfrac{16}{3}\mu_{n}^{6}a_{nnnn} + 9\sum_{i=0}^{+\infty}\mu_{(i)}^{2} a_{n^{\prime}n^{\prime}(i)}^{2} ~.\label{eq:sr5}
\end{align}
The only scalar tower sum $\sum_{i=0}^{+\infty} \mu_{(i)}^{2} a^{2}_{n^{\prime}n^{\prime}(i)}$ remaining in this particular combination of the $\mathcal{O}(s^{3})$ and $\mathcal{O}(s^{2})$ sum rules can be eliminated using the spin-0 completeness relation Eq. \eqref{eq:spin0-completeness}.  Since the spin-2 wavefunctions satisfy Neumann boundary conditions, $(\partial_\varphi \psi_n)=0$ at $\varphi=0$ and $\pi$, we find
\begin{align}
    \sum_{i=0}^{+\infty} \mu_{(i)}^{2} a_{n^{\prime}n^{\prime}(i)}^{2} &= \dfrac{1}{6}\bigg\{\int d\varphi \hspace{5 pt}  (\partial_{\varphi}\phi_{0})^{2} e^{-8A} (\partial_{\varphi}\psi_{n})^{4} \bigg\}~.
    \label{eq:scalar-sum}
\end{align}
Hence the combination of sum-rules in Eq. \eqref{eq:sr5} does not depend on the explicit form of the scalar wavefunctions $\gamma_i(\varphi)$, but only on the spin-2 wavefunctions $\psi_n(\varphi)$, the exponentiated warp-factor $e^{A(\varphi)}$, and (the derivative of) the background scalar field $\phi_0$.

In Appendix \ref{ShortSumRules} we show, by applying only the {\it spin-2} mode equation \eqref{eq:spin2-SLeqn}, Neumann boundary conditions, and completeness relations \eqref{eq:spin2-completeness}, that
\begin{align}
    \sum_{j=0}^{+\infty} \bigg[5\mu_{n}^{2}-\mu_{j}^{2}\bigg]\mu_{j}^{4} a^{2}_{nnj} &= \dfrac{16}{3}\mu_{n}^{6}a_{nnnn} + 18\bigg\{\dfrac{1}{\pi} \int_{-\pi}^{+\pi} d\varphi\hspace{5 pt} (\partial_{\varphi}^{2} A) e^{-8A}(\partial_{\varphi} \psi_{n})^{4}\bigg\}
    ~.\label{eq:sr-intermediate}
\end{align}
In the GW model we have the Einstein equation (via Eq. \eqref{eq:App}) $(\partial_{\varphi}^{2} A) = (\partial_{\varphi}\phi_{0})^{2}/12 + \sum_{i} V_{i} r_{c}\, \delta_{i}(\varphi)/3$. The Dirac delta terms vanish because $(\partial_{\varphi}\psi_{n}) = 0$ at the boundaries, and
hence
\begin{align}
    \sum_{j=0}^{+\infty} \bigg[5\mu_{n}^{2}-\mu_{j}^{2}\bigg]\mu_{j}^{4} a^{2}_{nnj} &= \dfrac{16}{3}\mu_{n}^{6}a_{nnnn} +
    \dfrac{3}{2}\bigg\{\int d\varphi \hspace{5 pt}  (\partial_{\varphi}\phi_{0})^{2} e^{-8A} (\partial_{\varphi}\psi_{n})^{4} \bigg\}~.\label{eq:sr-intermediatei}
\end{align}
Applying Eq. \eqref{eq:scalar-sum} we immediately obtain
Eq. \eqref{eq:sr5}. Hence Eq. \eqref{eq:sr5} depends non-trivially on the dynamics of the GW model - in particular on the Einstein equations for the warp-factor and on the scalar completeness relation which follows from the mode equation and the scalar mode normalization condition. 

Using the spin-2 completeness relations and the relations between the $a$ and $b$ couplings in Eq. \eqref{eq:btoatext} and the sum-rule in Eq. \eqref{eq:sr1}, we find
\begin{equation}
    \sum_{j=0}^{+\infty} \mu^4_j a^2_{nnj}=4\,c_{n^{\prime}n^{\prime}n^{\prime}n^{\prime}} +\,\dfrac{4}{3}\mu_n^4 a_{nnnn}~.
    \label{eq:completeness-muj4}
\end{equation}
Here we have defined the quantity 
\begin{align}
    c_{n^{\prime}n^{\prime}n^{\prime}n^{\prime}} = \sum_{j=0}^{+\infty} b^2_{n^\prime n^\prime j} \equiv \dfrac{1}{\pi} \int d\varphi\hspace{5 pt}e^{-6A}(\partial_{\varphi}\psi_{n})^4~,
\end{align}
which depends only on the spin-2 wavefunctions.
Plugging this into Eq.~\eqref{eq:sr-intermediatei} we find
\begin{equation}
 \sum_{j=0}^{+\infty}\mu_{j}^{6} a^{2}_{nnj} = \dfrac{4}{3} \mu_{n}^{6}a_{nnnn} + 20 \mu_n^2 c_{n^{\prime}n^{\prime}n^{\prime}n^{\prime}} -\dfrac{3}{2}\bigg\{\int d\varphi \hspace{5 pt}  (\partial_{\varphi}\phi_{0})^{2} e^{-8A} (\partial_{\varphi}\psi_{n})^{4} \bigg\}~,
 \label{eq:sr7}
\end{equation}
which depends only on the spin-2 wavefunctions and the background scalar-field configuration $\phi_0$. \looseness=-1

Finally, having demonstrated that one linear combination of Eqs. \eqref{eq:sr3} and \eqref{eq:sr4} is determined entirely by the spin-2 sector of the GW model, we can also use Eq.~\eqref{eq:completeness-muj4} to isolate the contribution from the GW scalar sector
\begin{align}
    \sum_{i=0}^{+\infty} a^{2}_{n^{\prime}n^{\prime}(i)}=\dfrac{5}{9}c_{n^{\prime}n^{\prime}n^{\prime}n^{\prime}}+\dfrac{1}{9} \mu_n^4 a^2_{nn0}+\dfrac{1}{27} \mu^4_n a_{nnnn}~,
    \label{eq:sr6}
\end{align}
which succinctly summarizes the necessary (but unproven) relationship between the scalar and spin-2 couplings (and hence wavefunctions) which must be satisfied in order for the spin-2 scattering amplitudes to grow no faster than ${\cal O}(s)$.


\subsection{Sum-Rule Summary}
\label{sec:sum-rule-summary}

In summary, the spin-2 coupling relations 
\begin{align}
    \sum_{j=0} a_{nnj}^{2} &= a_{nnnn}~, \tag{\ref{eq:sr1} revisited}\\
    \sum_{j=0} \mu_{j}^{2} a_{nnj}^{2} &= \dfrac{4}{3} \mu_{n}^{2} a_{nnnn}~,\tag{\ref{eq:sr2} revisited}\\
    \sum_{j=0}^{+\infty} \mu^4_j a^2_{nnj} & =4c_{n^{\prime}n^{\prime}n^{\prime}n^{\prime}} +\,\dfrac{4}{3}\mu_n^4 a_{nnnn}~,
    \tag{\ref{eq:completeness-muj4} revisited} \\
     \sum_{j=0}^{+\infty}\mu_{j}^{6} a^{2}_{nnj} &= \dfrac{4}{3} \mu_{n}^{6}a_{nnnn} + 20 \mu_n^2 c_{n^{\prime}n^{\prime}n^{\prime}n^{\prime}} -\dfrac{3}{2}\bigg\{\int d\varphi \hspace{5 pt}  (\partial_{\varphi}\phi_{0})^{2} e^{-8A} (\partial_{\varphi}\psi_{n})^{4} \bigg\}~,
 \tag{\ref{eq:sr7} revisited}
\end{align}
follow from the form of the spin-2 mode equations, spin-2 wavefunction completeness, the Einstein equations for the warp factor, and the scalar completeness relation which follows from the scalar mode equation and the scalar mode normalization condition. Analytic derivations of all of these relations are given above or in Appendix \ref{ShortSumRules}. With respect to guaranteeing that the massive spin-2 scattering amplitudes grow no faster than ${\cal O}(s)$, these relations show that the first-two sum-rules Eqs.~\eqref{eq:sr1}-\eqref{eq:sr2} and one combination of the sum-rules in Eqs.~\eqref{eq:sr3} and \eqref{eq:sr4} are always satisfied.

Separately, the sum-rule
\begin{align}
    \sum_{i=0}^{+\infty} a^{2}_{n^{\prime}n^{\prime}(i)}=\dfrac{5}{9}c_{n^{\prime}n^{\prime}n^{\prime}n^{\prime}}+\dfrac{1}{9} \mu_n^4 a^2_{nn0}+\dfrac{1}{27} \mu^4_n a_{nnnn}~,
    \tag{\ref{eq:sr6} revisited}
\end{align}
which depends on the spin-0 GW scalar couplings must also be satisfied in order for the spin-2 scattering amplitudes to grow no faster than ${\cal O}(s)$. With the methods discussed here, we have been unable to prove analytically that this scalar sum-rule is satisfied in general.\footnote{Generalizations of the proven rules (and their proofs) as well as additional unproven rules necessary for $\mathcal{O}(s)$ growth of the inelastic amplitude $(k,l) \rightarrow (m,n)$ are provided in Appendix \ref{DerivingSumRules}.}

In \cite{Chivukula:2021xod}, we demonstrated that these sum-rules in Eqs.~\eqref{eq:sr1}-\eqref{eq:sr4} were satisfied in the ``flat-stabilized" model - a slight deformation of a flat extra-dimensional model in which the radion is massive and the size of the extra dimension is stable. 
 We now demonstrate numerically that these relations are satisfied in the presence of the warping required to produce the hierarchy between the weak and Planck scales in the Randall-Sundrum model.

\section{Perturbative Analysis of a Warped Stabilized Model}

\label{sec:DFGK}

In the original formulation of the RS1 model~\cite{Randall:1999ee,Randall:1999vf}, Randall and Sundrum constructed a consistent solution to the Einstein field equations by choosing a warp factor $A(\varphi)=k r_c |\varphi|$ sourced by brane and bulk cosmological constants. Once the Goldberger-Wise mechanism is implemented, such a simple functional form becomes unavailable: the background geometry is augmented, the Einstein field equations are changed, and the warp factor $A(\varphi)$ is made more complicated. The background Einstein field equations of the stabilized model (Eqs. \eqref{eq:bgEOM1}, \eqref{eq:jump_cond1a}, \eqref{eq:bgEOM2}, and \eqref{eq:jump_cond1b}) are coupled non-linear equations with respect to $A(\varphi)$ that depend on the derivative of the scalar background $\phi_{0}^{\prime}(\varphi)$ and are generally difficult to solve.
DeWolfe, Freedman, Gubser, and Karch have constructed a specific class of exactly solvable potentials (the DFGK model) \cite{DeWolfe:1999cp} which make calculations in the stabilized RS1 model feasible.

In this section, we review the DFGK class of solutions to set notation, and we detail a specific DFGK model which enables perturbative expansion around the (solved) warped unstabilized RS1 model \cite{Chivukula:2021xod}. Subsequently, for the physical spin-0 and spin-2 towers, we perturbatively compute the wavefunctions and masses using the Sturm-Liouville equations described in section~\ref{sec:SLprob}. Finally, we demonstrate that the sum rules defined in section~\ref{sec:SumRules} are numerically satisfied at second order in the expansion parameter, the lowest order necessary to generate a nonzero radion mass.

\subsection{The DFGK Model}

A key strategy employed in the DFGK model \cite{DeWolfe:1999cp} is the introduction of a superpotential-inspired function $W[\hat{\phi}]$ which is used to simplify the stabilized model's background field equations (Eqs. \eqref{eq:bgEOM1}, \eqref{eq:jump_cond1a}, \eqref{eq:bgEOM2}, and \eqref{eq:jump_cond1b}). In particular, it is assumed that the scalar bulk and brane potentials may be parameterized as:
\begin{align}
    &\hspace{-60 pt}V r_{c}^{2} = \dfrac{1}{8}\bigg(\dfrac{dW}{d\hat{\phi}}\bigg)^{2} - \dfrac{W^2}{24}~,\\
    V_{1}r_{c} = +\dfrac{W}{2} + \beta_{1}^{2}\bigg[\hat{\phi}(\varphi) -\phi_{1}\bigg]^2~,\hspace{20 pt}&\hspace{20 pt}V_{2}r_{c} = -\dfrac{W}{2} + \beta_{2}^{2}\bigg[\hat{\phi}(\varphi)-\phi_{2}\bigg]^2~, \label{eq:branepotentials}
\end{align}
In this case, the background scalar and Einstein equations are solved if
\begin{align}
       (\partial_{\varphi} A) = \frac{W}{12}\bigg|_{\hat{\phi}=\phi_{0}}\text{ sign}(\varphi)~,\hspace{20 pt}&\hspace{20 pt}(\partial_{\varphi} \phi_{0}) = \dfrac{dW}{d\hat{\phi}}\bigg|_{\hat{\phi}=\phi_{0}}\text{ sign}(\varphi)~,\label{AAndPhiInTermsOfW}
\end{align}
where $\phi_{1} \equiv \hat{\phi}(0)$ and $\phi_{2} \equiv \hat{\phi}(\pi)$. Ref. \cite{DeWolfe:1999cp} introduces a convenient $W[\hat{\phi}]$ with the following specific form:\footnote{Here $u$ is a parameter, and not the $\hat{u}$ field of Eq. \ref{eq:uhat}.}
\begin{align}
    W[\hat{\phi}(\varphi)] = 12 kr_{c} - \dfrac{1}{2}\hat{\phi}(\varphi)^{2}\,ur_c~.
\end{align}
Plugging this into Eq. \ref{AAndPhiInTermsOfW} we find solutions for the bulk scalar vacuum and the warp factor:
\begin{align}
    \phi_0(\varphi) &= \phi_{1}e^{-ur_c|\varphi|}~, \label{eq:scalarbackgroundi}\\
    A(\varphi) &= kr_{c}|\varphi| + \dfrac{1}{48}\phi_{1}^{2}\bigg[e^{-2ur_c|\varphi|} - 1\bigg]\ .
    \label{eq:AnPhi}
\end{align}
In the limit that the parameter $u$ vanishes, this reduces to the usual unstabilized RS1, with the bulk field acquiring a constant vacuum expectation value.
The parameters $u$, $\phi_{1}$, and $\phi_{2}$ are related according to
\begin{equation}
    ur_c=\dfrac{1}{\pi}\log\dfrac{\phi_1}{\phi_2}~.
\end{equation}
We next define the small-$u$ limit carefully to facilitate a perturbative analysis.

%
\subsection{The Perturbative DFGK Model}
\label{sec:perturbative}

The forms of $\phi_{0}(\varphi)$ and $A(\varphi)$ in Eqs.~\eqref{eq:scalarbackgroundi}-\eqref{eq:AnPhi} are useful for solving the background equations, but it remains difficult to find general analytic solutions (i.e. the KK wavefunctions and masses) to the differential equations defined in section~\ref{sec:SLprob}. Therefore, we define a limit of the model so that we may perturbatively expand $\phi_{0}$ and $A$ around the unstabilized background (for which analytic solutions are well known), take the stiff wall limit (with $\ddot{V}_{1,2}\to \infty$ so that the scalar boundary
conditions in Eq. \eqref{eq:scalarBCtext} reduce to Neumann conditions at $\varphi=0,\,\pi$), and develop solutions for the Sturm-Liouville problems order-by-order in perturbation theory. Details of the perturbation theory can be found in Appendix~\ref{app:PertTheory}. Here, we now proceed to introduce the effective warp parameter $\tilde{k}$ and perturbation parameter $\epsilon$ \cite{Chivukula:2021xod}.

Suppose we series expand $A(\varphi)$ with respect to the unitless quantity $u r_{c}$. Expanding around $ur_{c} = 0$ yields
\begin{align}
    A(\varphi) &= kr_{c} |\varphi| - \bigg[\dfrac{\phi_{1}^{2}}{24}|\varphi|\bigg](ur_c) + \bigg[\dfrac{\phi_{1}^{2}}{24}|\varphi|^{2}\bigg](ur_c)^{2} + \mathcal{O}\Big((ur_c)^{3}\Big)\\
    &= \bigg[k - \dfrac{\phi_{1}^{2}u}{24}  \bigg]r_{c}|\varphi| + \bigg[\dfrac{\phi_{1}^{2}(ur_c)^{2}}{24} \bigg] |\varphi|^{2} + \mathcal{O}\Big((ur_c)^{3}\Big)~.
\end{align}
The first term in the second line demonstrates that, when $ur_{c}$ is sufficiently small, the stabilized model is a small deformation of an unstabilized Randall-Sundrum I model \cite{Randall:1999ee,Randall:1999vf}. Because we intend to work in the $ur_{c}\rightarrow 0$ limit, we will eliminate the actual warp parameter $k$ in favor of the effective warp parameter,
\begin{equation}\tilde{k}
\equiv k - \phi_{1}^{2}u/24~,
\end{equation}
that applies in that limit.\footnote{In \cite{Chivukula:2021xod} we discussed the properties of the ``flat-stabilized model" with $\tilde{k}=0$.}

In order to simplify various factors that would otherwise be present in multiple equations, we will also replace $ur_{c}$ (and its role as our expansion parameter) with the rescaled dimensionless perturbative parameter $\epsilon\equiv \phi_{1}(ur_c)/\sqrt{24}$. This definition of $\epsilon$ simplifies $A(\varphi)$ at $\mathcal{O}(\epsilon^{2})$
\begin{align}
    A(\varphi) &= \tilde{k}r_{c}|\varphi|+ \dfrac{\phi_{1}^{2}}{48}\bigg[\exp\bigg(-\dfrac{4\sqrt{6}}{\phi_{1}}\epsilon|\varphi|\bigg)-1\bigg] + \dfrac{\phi_{1}}{2\sqrt{6}}\epsilon|\varphi|~,
    \label{eq:AEps}\\
    &= \tilde{k}r_{c}|\varphi| + \epsilon^{2}|\varphi|^{2} + \mathcal{O}(\epsilon^{3})~.
\end{align}
and yields, to all orders in $\epsilon$,
\begin{align}
    W[\hat{\phi}(\varphi)] &= 12 \tilde{k}r_{c} + \dfrac{\sqrt{6}}{\phi_{1}} \bigg[\phi_{1}^{2} -\hat{\phi}(\varphi)^{2}\bigg]\epsilon~,\\
    \phi_{0}(\varphi) &= \phi_{1} \exp\bigg(-\dfrac{2\sqrt{6}}{\phi_{1}}\epsilon|\varphi|\bigg)~= \rm \phi_{1}exp(-\alpha \epsilon |\varphi|). \label{eq:scalarbackground}
\end{align}
where $\alpha=\dfrac{2\sqrt{6}}{\phi_{1}} $. It is the form of the warp-factor shown in Eq. \eqref{eq:AEps} that we use in subsequent perturbative computations.

\subsubsection{Spin-2 wavefunctions and masses in perturbation theory}

\begin{figure}
    \centering
    \includegraphics[width=0.75\textwidth]{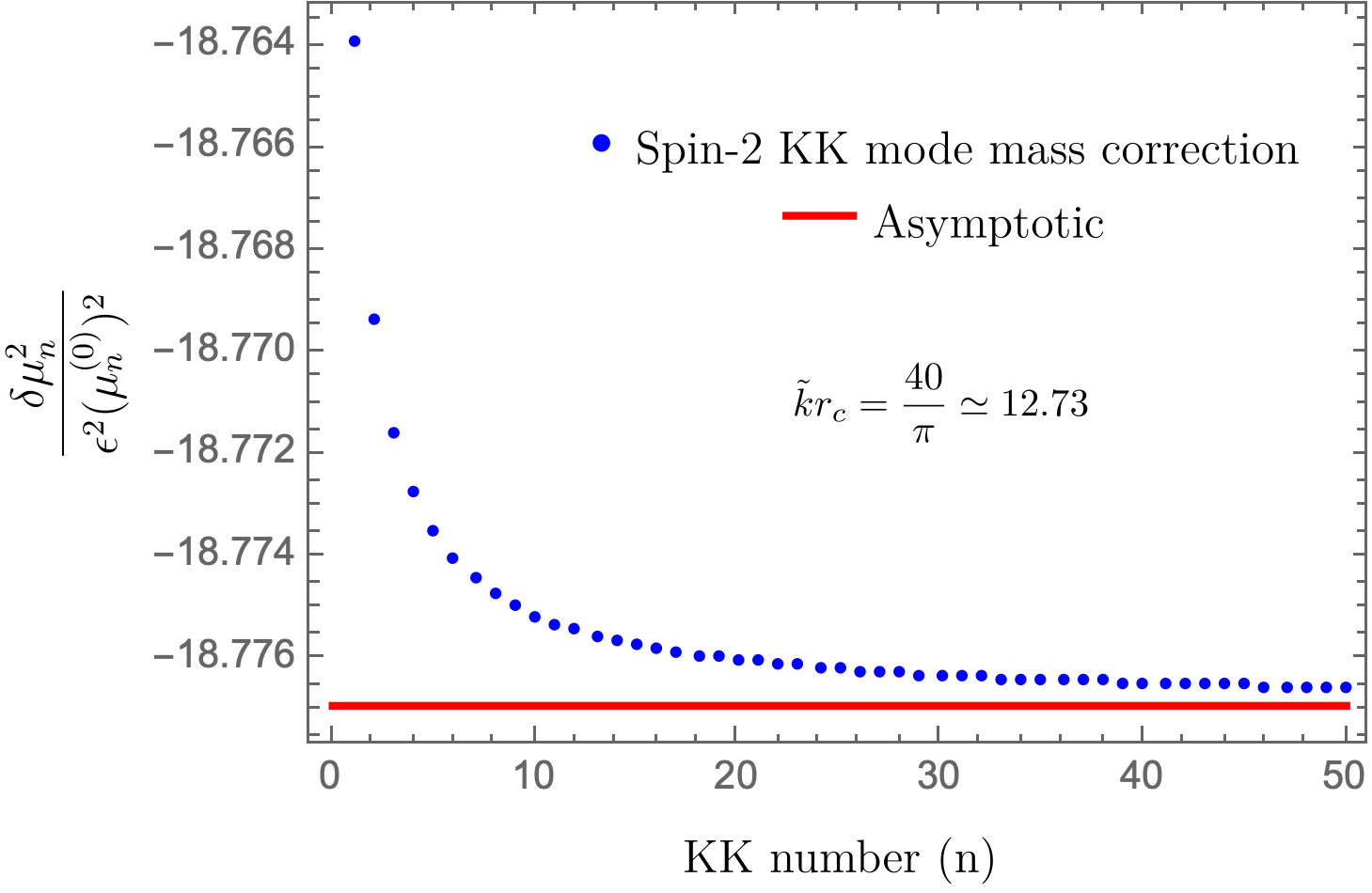}
    \caption{We show for each spin-2 KK mode $n$, the ratio of mass correction to leading order mass times $\epsilon^2$ (dots). The mass correction $\delta \mu_n^2$ is calculated in the large $\tilde{k} r_c$ limit to order $\epsilon^2$, as shown in Eq.~\eqref{eq:delta-mu-squared}. The line represents the asymptotic form of the expression given in Eq.~\eqref{eq:approximate-mass}, which is valid for large spin-2 KK number. }
    \label{fig:plmass}
\end{figure}

To order ${\cal O}(\epsilon^2)$, the spin-2 mode equation Eq.~\eqref{eq:spin2-SLeqn} becomes\footnote{Here, and in subsequent equations related to the perturbative model in this paper, we expand $A(\varphi)$ to ${\cal O}(\epsilon^2)$ using Eq.~\eqref{eq:AEps}, but we do not record the expansion of the exponents in powers of $\epsilon$ to retain succinctness of expressions when writing. However, these exponents are actually expanded during our calculations, e.g. when we define the perturbed differential equations in \eqref{eq:SLi}-\eqref{eq:SLii}.}
\begin{equation}
    \partial_\varphi\left[e^{-4\tilde{k}r_c \varphi-4\epsilon^2\varphi^2}\,\partial_\varphi \psi_n\right]=-\mu^2_n e^{-2\tilde{k}r_c \varphi-2\epsilon^2\varphi^2} \psi_n~.
    \label{eq:spin2-perturbative}
\end{equation}
The form of this equation, along with the Neumann boundary conditions $\partial_\varphi\psi_n=0$ at $\varphi=0,\, \pi$, insures that the zero mode (the graviton) is massless and has a wavefunction which is constant in $\varphi$. Expanding this equation in powers of $\epsilon^2$, we solve Eq.~\eqref{eq:spin2-SLeqn} using perturbation theory as described in Appendix \ref{app:PertTheory}. The perturbative expressions for the spin-2 wavefunctions and masses are quite lengthy and in order to simplify them, we restrict ourselves to the limit when $\tilde{k}r_c$ is large. The expressions for the spin-2 mass and wavefunctions are given to order $\epsilon^2$ in the large-$\tilde{k}r_c$ limit in Appendix~\ref{app:WFandMasses}.

In order to illustrate the effects of the geometry on the spin-2 masses, we calculate the masses for Eq.~\eqref{eq:spin2-SLeqn} using the WKB approximation.
The asymptotic formula for the masses is given by\footnote{There are {\cal O}(1) corrections to $\mu^2_n \ell^2$, that do not grow with $n$, due to the Neumann boundary conditions. These effects can be included, but do not effect the analysis given here.}
\begin{align}
    \mu_n= m_n r_c =\frac{n \pi }{l}~,\hspace{35 pt}
    l=\int_{0}^{\pi} d\varphi\ e^{A} \ .
    \label{eq:spin2-mass-WKB}
\end{align}
The equations above show how the eigenvalues are positive and form an infinite tower of states.
Using the form of $A(\varphi)$ from Eq.~\eqref{eq:AnPhi}, we find the mass of spin-2 KK modes for large mode number to be
\begin{align}
    \mu_n\Big|_{n\gg 1}= \frac{  (n\pi) \tilde{k} r_c}{e^{\pi  \tilde{k} r_c}-1}
    -  (n \pi) \frac{ e^{\pi  \tilde{k} r_c}[\pi  \tilde{k} r_c (\pi  \tilde{k} r_c-2)+2]-2}{\tilde{k} r_c (e^{\pi  \tilde{k}
   r_c}-1)^2}\epsilon ^2+O\left(\epsilon ^3\right)
\end{align}
In the limit when $\tilde{k}r_c \gg 1$, which is the phenomenologically interesting limit, this expression further simplifies to

\begin{equation}
    \mu_n\Big|_{n\gg 1} \simeq  (n \pi) \tilde{k} r_c   e^{-\pi  \tilde{k} r_c} \bigg\{
   1  -\frac{ [\pi  \tilde{k} r_c (\pi  \tilde{k}
   r_c-2)+2]}{\tilde{k}^2 r_c^2}
   \epsilon ^2  \bigg\}\ .
\end{equation}
We can see how the effect of the back reaction on the geometry from the bulk scalar field reduces the masses of the massive spin-2 modes. Further, the square of the ratio of the correction to the mass ($\delta \mu_n$) to the leading order mass $(\mu_n^{(0)})$ is
\begin{equation}
    \left.\frac{\delta \mu_{n}^{2}}{\left(\mu_{n}^{(0)}\right)^{2}}\right|_{n \gg 1 ; \tilde{k} r_{c} \gg 1} \simeq\left[-2 \pi^{2}+\frac{4 \pi}{\tilde{k} r_{c}}-\frac{4}{\tilde{k}^{2} r_{c}^{2}}+\ldots\right] \epsilon^{2} \,.
   \label{eq:approximate-mass}
\end{equation}
We can therefore conclude that the perturbation theory we use is valid when $|\epsilon|\ll \ 1/(\sqrt{2}\  \pi)$. In Fig.~\ref{fig:plmass}, we compare the full expression for the mass corrections (given in Eq. \eqref{eq:delta-mu-squared} of Appendix \ref{app:PertTheory}) to the asymptotic form shown here. Here we see that the full form of the mass, represented by the blue dots approaches the asymptotic value, represented by the bold red line. 
 
While the asymptotic formula is simple and provides a convenient cross-check of our calculations, it is insufficient for our present purposes. In order to demonstrate cancellations and verify sum rules, we need to be able to evaluate the ${\cal O}(\epsilon^2)$ wavefunctions and masses without approximation. We provide exact expressions to order $\epsilon^2$ in the perturbation theory within Appendix~\ref{app:WFandMasses}. These expressions are consistent with the large $\tilde{k} r_c$ limit results. Full expressions (which are valid for arbitrary values of  $\tilde{k} r_c$) are provided as supplementary Mathematica files.

\subsubsection{Spin-0 wavefunctions and masses in perturbation theory}

Eq.~\eqref{eq:scalarbackground} implies that
\begin{equation}
    (\phi'_0)^2=24 \epsilon^2 e^{-2\alpha\epsilon\varphi}~
\end{equation}
in the bulk. To simplify our analysis, we consider the perturbative solution for the scalar tower in the ``stiff-wall" limit, $\ddot{V}_{1,2}\to +\infty$, so that Eq.~\eqref{eq:scalarBCtext} reduces to Neumann conditions 
\begin{equation}
    \partial_\varphi \gamma_i \big|_{\varphi = 0,~\pi}=0~,
\end{equation}
Using this and the expansion of $A(\varphi)$ to order ${\cal O}(\epsilon^2)$ from Eq.~\eqref{eq:AEps}, the spin-0 mode equation Eq.~\eqref{eq:spin0-SLeqn} becomes (after multiplying through by $24\epsilon^2$)
\begin{equation}
    \partial_\varphi\left[e^{(2\tilde{k}r_c\varphi+2\alpha\epsilon\varphi +2\epsilon^2\varphi^2)}\partial_\varphi \gamma_i\right]-4\epsilon^2e^{(2\tilde{k}r_c\varphi+2\epsilon^2\varphi^2)}\gamma_i=-\mu^2_i e^{(4\tilde{k}r_c\varphi+2\alpha\epsilon\varphi+4\epsilon^2\varphi^2)}\gamma_i~,
    \label{eq:spin0-perturbative}
\end{equation}
while the normalization conditions to this order are, 
\begin{equation}
    \delta_{mn}=\dfrac{1}{4\pi \epsilon^2}\int_{-\pi}^{+\pi} d\varphi
    \left[e^{(2\tilde{k}r_c \varphi+2\alpha\epsilon\varphi +2\epsilon^2\varphi^2)} \gamma'_m\gamma'_n+4\epsilon^2 e^{(2\tilde{k}r_c+2\epsilon^2\varphi^2) \varphi}\gamma_m\gamma_n\right]~.
    \label{eq:scalar-pert-norm}
\end{equation}


Expanding the spin-0 mode equation Eq.~\eqref{eq:spin0-perturbative} in powers of $\epsilon$, we solve the  differential Eq.~\eqref{eq:spin0-SLeqn} using perturbation theory described in Appendix~\ref{app:PertTheory}. In particular, using Eq.~\eqref{eq:SL-mass-pert}, we find that the radion (identified as the zero mode of the KK expansion in Eq.~\eqref{eq:spin0-mode-expansion}) acquires a mass-squared at order $\epsilon^2$ in perturbation theory
\begin{equation}
\mu^2_{(0)} = \dfrac{8\epsilon^2}{1+ e^{2 \pi \tilde{k} r_c}} + \mathcal{O}(\epsilon^3) ~,
    \label{eq:RadionMass}
\end{equation}
while the radion wavefunction evaluated to order $\epsilon^2$ is
\begin{equation}
 \gamma_{(0)}= \sqrt{\frac{\tilde{k} r_{c} \pi}{e^{2 \tilde{k} r_{c} \pi } -1}} + C(\tilde{k}r_c) \bigg\{2 \tilde{k} r_{c} \varphi +\text{sech}(\pi  \tilde{k} r_{c}) \sinh[\tilde{k} r_{c} (\pi -2 \varphi )] -\tanh (\pi  \tilde{k} r_{c})\bigg\} \epsilon^{2} + \mathcal{O}(\epsilon^3) \ .
\end{equation}
Here $C(\tilde{k}r_c)$ is  determined through normalization; the normalized radion wave function is provided in the appendix in Eq.~\eqref{eq:RadionWF}. 
In the limit where $\epsilon$ vanishes, the gravitational degrees of freedom and the bulk scalar cease mixing. This results in a massless radion, an unstabilized extra dimension, and an entirely separate tower of scalar states. Note that the leading order radion wavefunction is flat, signalling that it is massless at that order.

The GW scalar wavefunction to order $\epsilon$ is given in Eq.~\eqref{eq:GWscalarWF}. As expected, the wavefunctions are comprised of Bessel functions.
Due to the normalization condition in Eq.~\eqref{eq:scalar-pert-norm}, the massive GW scalar mode wavefunctions have no $\epsilon^0$ terms and start at order $\epsilon$. Since the sum rules in Eq.~\eqref{eq:sr1} through Eq.~\eqref{eq:sr5} have only products of GW scalar wavefunctions, when trying to verify them to order $\epsilon^2$, we only need expressions of the wavefunctions to order $\epsilon$. We therefore, only provide expressions for the GW scalar wavefunction and masses to leading order.\footnote{Note that the limit $\tilde{k} r_{c}=0$, dubbed as the flat stabilized model was studied previously in our work \cite{Chivukula:2021xod}, where we show that the sum rules required for the scattering amplitudes to grow only as $\mathcal{O}(s)$ were satisfied to ${\cal O}(\epsilon^2)$.}

\subsection{Numerical verification of sum rules}

\label{sec:NumericalVerification}

\begin{figure}
    \centering
    \includegraphics[width=0.9\textwidth]{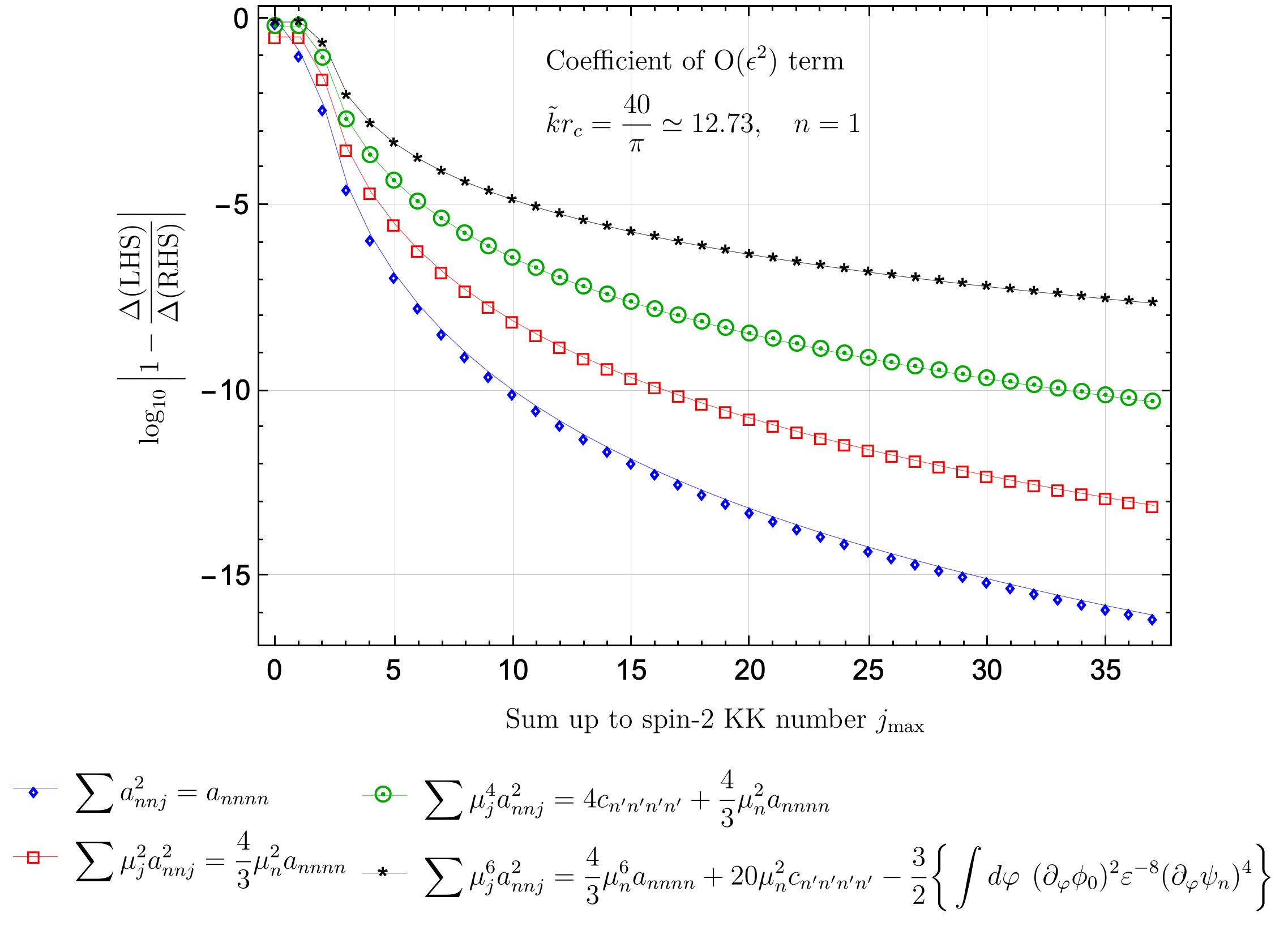}
    \caption{Verification of sum rules (Eqs.~\eqref{eq:sr1},\eqref{eq:sr2},\eqref{eq:completeness-muj4} and \eqref{eq:sr7}, reproduced at the bottom of the diagram) for elastic scattering of KK mode number one ($n=1$). The index $j$, shown on the $x$-axis, is the number of spin-2 modes included in the sum of the left hand side of the sum rules. The $y$-axis indicates the log of relative error $(\log_{10}(1 - \Delta(\text{LHS})/\Delta(\text{RHS})))$ between the ${\cal O}(\epsilon^2)$ {\it corrections} to LHS and RHS of the the relevant equations. This has been evaluated in the large $\tilde{k}r_c$ limit.  }
    \label{fig:fig3}
\end{figure}

We now verify the sum rules summarized in section~\ref{sec:sum-rule-summary} for the warped-stabilized model using our perturbative computations. We can substitute expression for wavefunctions and masses calculated in the DFGK model and provided in Appendix~\ref{app:WFandMasses} and evaluate the overlap integrals numerically. The sum rules in Eq.~\eqref{eq:sr1},\eqref{eq:sr2},\eqref{eq:completeness-muj4} and \eqref{eq:sr7} and Eq. \eqref{eq:sr6} can thereby be evaluated order-by-order in $\epsilon$.  We know that the sum rules are satisfied for the unstabilized RS model \cite{Chivukula:2020hvi}, and therefore these expressions agree to leading-order ($\epsilon^0$). Using our perturbative experessions, we verify here that the sum-rules are satisfied to leading non-trivial order, ${\cal O}(\epsilon^2)$. Equivalently, we show that the ${\cal O}(\epsilon^2)$ contributions on the left- and right-hand sides of
 Eq.~\eqref{eq:sr1},\eqref{eq:sr2},\eqref{eq:completeness-muj4} and \eqref{eq:sr7} and Eq. \eqref{eq:sr6} agree.

Note that the left-hand sides of these expressions are given as infinite sums over different overlap integrals. It is therefore not possible to perform the entire sum. Instead, we perform the sum up a ``cutoff" KK-number and show that the relative error in the ${\cal O}(\epsilon^2)$ contributions to the left hand side (LHS) converge to the ${\cal O}(\epsilon^2)$ contributions to right hand side (RHS) of each expression as the number of KK modes included in the sum increases.

\subsubsection{Spin-2 Sum-Rules and Completeness}

We begin with Eq.~\eqref{eq:sr1}, \eqref{eq:sr2}, \eqref{eq:completeness-muj4}, and \eqref{eq:sr7}. The result of this exercise is shown in Fig.~\ref{fig:fig3}, in the case $n=1$ ({\it e.g.} for elastic scattering of spin-2 modes at KK level 1) and for $\tilde{k}r_c = 40/\pi=12.73$.  We see that each of the series converge nicely with the relative error reducing with the addition of terms to the sum on the LHS of the equation. As described in detail in Section \ref{sec:SumRules}, Eqs.~\eqref{eq:sr1}, \eqref{eq:sr2}, and \eqref{eq:completeness-muj4} can be proven directly using the completeness properties of the solutions of the spin-2 mode equation Eq.~\eqref{eq:spin2-SLeqn}. Their numerical verification demonstrates that accuracy of our perturbative analysis, The first two of these equations demonstrate that the ${\cal O}(s^5)$ and ${\cal O}(s^4)$ contributions to helicity-0 spin-2 elastic scattering vanish to this order in perturbation theory. 

We also see that the sum-rule in Eq.~\eqref{eq:sr7} converges as well. As discussed above, Eq.~\eqref{eq:sr7} depends non-trivially on the dynamics of the Goldberger-Wise model implemented here - in particular on the Einstein equation coupling the scalar-potential to the curvature of the extra dimension Eq.~\eqref{eq:App} and on the completeness conditions of the spin-0 modes Eq.~\eqref{eq:spin0-completeness}. As discussed in Section \ref{sec:sum-rule-summary}, this verifies that one linear-combination of Eqs.~\eqref{eq:sr3} and \eqref{eq:sr4} are also satified.

Finally, in all cases we see that the series only converges rapidly once we have included the $j=2$ term. This is because the overlap integral defining the coupling between two spin-2 level-1 states and a spin-2 state at level $j$ is largest for $j=2$ -- which can be understood as a remnant of the ``discrete" KK momentum-conservation which would be present in a flat extra-dimension.

\subsubsection{The Spin-0 Sum Rule}

\begin{figure}
    \centering
    \includegraphics[width=0.75\textwidth]{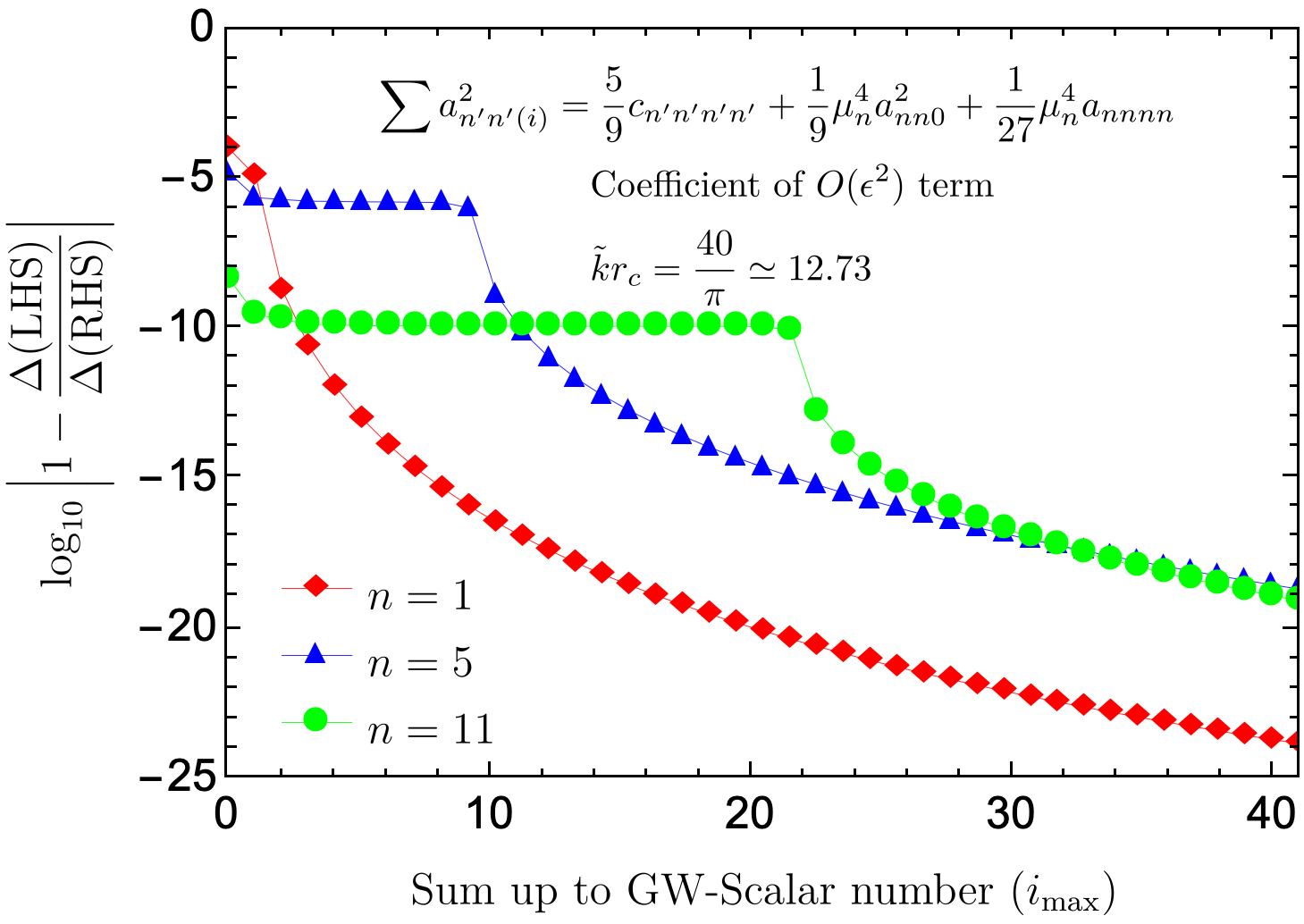}
    \caption{Verification of the scalar sum rule in Eq.~\eqref{eq:sr6} for elastic scattering of spin-2 KK mode number $n=1$ (red), $n=5$ (blue) and $n=11$ (green). Here $n$ represents the KK mode number of the spin-2 particles in the external legs. On the $x$-axis we show the number of GW-scalar modes included in the sum ($i$) of the left hand side of the sum rules in Eq.~\eqref{eq:sr6} versus the log of the relative error $(\log_{10}(1 - \Delta(\text{LHS})/\Delta(\text{RHS})))$ between the ${\cal O}(\epsilon^2)$ contributions to both sides of the relevant equation. This has been evaluated in the large $\tilde{k}r_c$ limit. }
    \label{fig:fig4}
\end{figure}

Finally, we examine  the sum-rule in Eq.~\eqref{eq:sr6} for which we have no analytic proof, and which depends on the couplings of the individual spin-0 states to the massive spin-2 KK modes. The result of this exercise is shown in Fig.~\ref{fig:fig4}, in the case $n=1,\,5,\,11$ ({\it e.g.} for scalar-exchange contributions to elastic scattering of spin-2 modes at KK level 1, 5, and 11) and for $\tilde{k}r_c=12.73$. Again, what is plotted here is the agreement between the ${\cal O}(\epsilon^2)$ contributions to the left- and right-hand sides of Eq.~\eqref{eq:sr6} -- which, at this order, depends explicitly on the forms of the scalar wavefunctions that solve Eq.~\eqref{eq:spin0-SLeqn} and which are subject to the normalization conditions of Eq.~\eqref{eqScalar:Norm1}. 

The non-zero difference between the left- and right-hand sides of the ${\cal O}(\epsilon^2)$ corrections to Eq.~\eqref{eq:sr6} at $i=1$ demonstrate the need to include the tower of Goldberger-Wise scalar states for the spin-2 scattering amplitudes to have the proper high-energy behavior. Again we see that the largest single contribution to the ${\cal O}(\epsilon^2)$ corrections come from the exchange of the GW scalar state whose mode-number is twice that of the incoming particles -- $i=2,10$ and 20, respectively, for incoming mode 1, 5, and 10 spin-2 states. However, the continued convergence when adding additional states is also clear and, formally, the entire tower is needed for the sum-rule to be satisfied.

\section{Conclusion}

\label{sec:conclusion}

In this paper we have presented a thorough analysis of the scattering of massive spin-2 Kaluza-Klein excitations in phenomenologically realistic models based on a warped geometry \cite{Randall:1999ee,Randall:1999vf} stabilized via the Goldberger-Wise \cite{Goldberger:1999uk,Goldberger:1999un} mechanism. These results significantly extend the work presented in \cite{Chivukula:2019rij,Chivukula:2019zkt,Chivukula:2020hvi} on the unstabilized RS1 model and the results in \cite{Chivukula:2021xod} on the ``flat-stabilized" model ($\tilde{k}r_c=0$).  We briefly recap our findings here:

\begin{itemize}
    \item  We provided a complete and self-contained derivation of the mode expansions for the spin-2 and spin-0 states and their interactions.
     Generalizing the presentations in \cite{Kofman:2004tk,Boos:2005dc,Boos:2012zz}, our computations are given in de-Donder gauge for massless gravitons -- allowing us to consistently compute scattering amplitudes involving intermediate off-shell states. 

\item In previous work \cite{Chivukula:2021xod} we had demonstrated the extended sum-rule relationships between spin-2 and spin-0 modes, and their masses and couplings, which must be satisfied in order for elastic massive spin-2 KK scattering to grow no faster than ${\cal O}(s)$. Here, we have provided an analytic proof for one combination of these sum rules, and showed its relation both to the Einstein and scalar background field equations implementing the Goldberger-Wise dynamics and also to the properties of the mode-equations for the physical scalar fields (fields which are admixtures of bulk scalar and gravitational modes in the original theory). 

\item We have provided, in an appendix to this work, a complete list of the sum-rule relations which must be satisfied if all $2 \to 2$ massive spin-2 scattering amplitudes, elastic or inelastic, are to grow no faster than ${\cal O}(s)$ -- completing the analyses begun in \cite{Chivukula:2019rij,Chivukula:2019zkt,Chivukula:2020hvi,Foren:2020egq}. 

\item Finally, using a version of the DFGK model \cite{DeWolfe:1999cp} in which the Goldberger-Wise dynamics can be treated perturbatively \cite{Chivukula:2021xod}, we have checked numerically that the sum rules which enforce the proper high-energy behavior of massive spin-2 scattering continue to be satisfied in the case of the large warping that would be required to produce the hierarchy between the weak and Planck scales. These numerical computations demonstrate that, in models with a massive radion, proper cancellation is achieved only after including the contributions from the tower of scalar states present in the Goldberger-Wise model.

\end{itemize}

 In subsequent work we will take up the issues of providing an analytic proof for the one remaining elastic sum-rule relationship for which we presently have only numerical confirmation (Eq. \eqref{eq:sr6}) \cite{Xing}. In future work we will also explore the phenomenological consequences of the fact that all spin-2 scattering amplitudes in models of compactified gravity can grow no faster than ${\cal O}(s)$; specifically, we will study the implications for the computation of relic abundances of dark matter particles in KK graviton-portal theories and related theories.

\medskip

\noindent {\bf Acknowledgements:} This material is based upon work supported by the National Science Foundation under Grant No. PHY- 1915147 .The work of DS for this paper was supported in part by the University of Adelaide and the Australian Research Council through the Centre of Excellence for Dark Matter Particle Physics (CE200100008). 

\newpage
\appendix

\section{The Derivation of the Scalar Kaluza-Klein Modes}
\label{app:Lagrangian}

The canonical quadratic Lagrangian for the Kaluza-Klein modes in a Goldberger-Wise model has been derived in Refs.~\cite{Kofman:2004tk,Boos:2005dc,Boos:2012zz}. Here, for completeness and as a guide to the interested reader, we present our derivation of the canonical quadratic Lagrangian. In addition, since our computation includes diagrams with off-shell gravitons, we are careful to derive our results in de Donder gauge.\footnote{If one is only concerned with external gravitons, and not doing scattering computations, one can impose the transverse-traceless conditions on all of the fields to simplify the computations of the interactions.} We include all the necessary details needed in order to reproduce our results: deriving the background equations of motion for a Goldberger-Wise-stabilized Randall-Sundrum I model, showing how the 5D scalar field $\hat{r}$ and 5D tensor field $\hat{h}$ decouple, motivating the gauge condition relating the fluctuation field $\hat{f}$ and the 5D scalar field $\hat{r}$, describing how the linear equations of motion inspire the Kaluza-Klein decomposition of the 5D scalar field $\hat{r}$, and demonstrating that the 5D scalar field $\hat{r}(x,y)$ generates a tower of canonical spin-$0$ fields $\hat{r}^{(i)}(x)$ with masses $m_{(i)} \equiv \mu_{(i)}/r_{c}$.

\subsection{The Lagrangian}
The Goldberger-Wise-stabilized Randall-Sundrum I Lagrangian is constructed from several elements, including the spacetime metric. Focusing our attention on the spin-2 ($\hat{h}_{\mu\nu}$) and scalar ($\hat{r}$) fluctuations about a geometry determined by the warp-factor $A(y)$, we use the following parameterization of the metric $G_{MN}$ and its 4D projection $\overline{G}_{MN}$:
\begin{align}
    [G_{MN}] = \matrixbb{w\,g_{\mu\nu}}{0}{0}{-v^{2}}\hspace{35 pt} [\overline{G}_{\mu \nu}] = w\,g_{\mu\nu}
    \label{eq:Metric}
\end{align}
where (taking our parameterization from \cite{Charmousis:1999rg}),
\begin{align}
     w = e^{-2[A(y)+\hat{u}(x,y)]}~,\hspace{35 pt}&v = 1 + 2\hat{u}(x,y)~,\\  g_{\mu\nu} = \eta_{\mu\nu} + \kappa\, \hat{h}_{\mu\nu}(x,y)~,\hspace{35 pt}&\hat{u} = \dfrac{ e^{2A(y)}}{2\sqrt{6}} \kappa\,\hat{r}(x,y)~. \label{eq:uhat}
\end{align}
The compact extra dimension is parameterized on a circle of radius $r_{c}$ such that $y \in (-\pi r_{c}, +\pi r_{c}]$ and the 5D coordinates $X^{M} = (x^{\mu},y)$ define a 4D Minkowski spacetime ``slice" at each fixed value of $y$. We impose an orbifold invariance $y \leftrightarrow -y$ on the infinitesimal spacetime interval $G_{MN} dX^{M}dX^{N}$, and identify $y = 0$ and $y = \pi r_{c}$ as orbifold fixed points. For our spacetime signatures, we use the mostly-minus convention, i.e. $\eta_{\mu\nu}\equiv \text{Diag}(+1,-1,-1,-1)$. Lowercase Greek letters (e.g. $\mu$, $\nu$, $\rho$, $\sigma$) denote 4D indices, uppercase Latin letters (e.g. $M$, $N$, $R$, $S$) denote 5D indices, and lowercase Latin letters (e.g. $m$, $n$, $i$, $j$) denote Kaluza-Klein mode numbers. 

We then define the Lagrangian as
\begin{align}
    \mathcal{L}_{\text{5D}} \equiv \mathcal{L}_{\text{EH}} +  \mathcal{L}_{\Phi\Phi} + \mathcal{L}_{\text{pot}} +\mathcal{L}_{\text{GHY}} + \Delta \mathcal{L}~, \label{L5D_Definition}
\end{align}
where the Einstein-Hilbert (EH), Gibbons-Hawking-York (GHY), scalar kinetic terms ($\Phi \Phi$), and scalar potential terms (pot) are defined as
\begin{align}
    \mathcal{L}_{\text{EH}} &\equiv -\dfrac{2}{\kappa^{2}}\, \sqrt{G} \, R\ ,\\
    \mathcal{L}_{\text{GHY}} &\equiv -\dfrac{4}{\kappa^{2}}\, \partial_{y}\Big[\sqrt{\overline{G}}\, K\Big] = \dfrac{2}{\kappa^2} \partial_y\left[
    \dfrac{w^2}{v} \sqrt{-\text{det}g}\left(
    \tilde{g}^{\mu\nu}\partial_y(g_{\mu\nu}) + 4\frac{\partial_y(w)}{w}
    \right)
    \right] \ ,\\
    \mathcal{L}_{\Phi\Phi} &\equiv \sqrt{G}\bigg[\dfrac{1}{2}\tilde{G}^{MN} (\partial_{M} \hat{\Phi})(\partial_{N}\hat{\Phi}) \bigg]\ ,\\ 
    \mathcal{L}_{\text{pot}} &\equiv -\dfrac{4}{\kappa^{2}}\bigg[\sqrt{G}\, V[\hat{\Phi}] + \sum_{i=1,2} \sqrt{\overline{G}} \,V_{i}[\hat{\Phi}]\,\delta_{i}(\varphi) \bigg]\ .
\end{align}
respectively. Here $R$ is the Ricci scalar and $K$ is the extrinsic curvature at the boundaries. $G$ and $\bar{G}$ are the determinants of metric and the induced metric respectively. The Dirac deltas on the branes are defined in the limit as we approach the branes from within the $y\in[0,\pi r_{c}]$ half of the bulk: $\delta_{1}(\varphi) \equiv \delta(\varphi - 0^{+})$ and $\delta_{2}(\varphi)\equiv \delta(\varphi-\pi^{-})$, where $\varphi=y/r_c$. 

Note that no dynamics are given to describe the physics of the branes themselves, which are assumed to arise through unspecified dynamics at some higher scale ({\it e.g.} string physics) -- and, in particular, no modes arising from that physics (fluctuations of the branes themselves) are included. More on this below, when we discuss the effect of taking the so-called ``stiff-wall" limit on the scalar mode-expansions.

Meanwhile, the contribution $\Delta \mathcal{L}$ in $\mathcal{L}_{\text{5D}}$ is a total derivative we add for convenience which generalizes a total derivative from our unstabilized analysis \cite{Chivukula:2020hvi,Foren:2020egq}. It cancels terms in the Lagrangian at linear order, eliminates mixing between tensor and scalar 5D fields at quadratic order, and simplifies the vertices relevant to this paper.\footnote{If such a term were not introduced here, we would recover its effects (at least to quadratic order) as additional total derivative terms needed to make the Lagrangian explicitly canonical. This $\Delta\mathcal{L}$ naively generalized the $\Delta\mathcal{L}$ introduced in refs.~\cite{Chivukula:2020hvi,Foren:2020egq}, and (unlike the latter) does not eliminate Dirac deltas or twice-differentiated quantities to all orders.} Explicitly, we define it as
\begin{align}
    \Delta \mathcal{L} \equiv \dfrac{2}{\kappa^2} \partial_y\left[
    \dfrac{w^2}{v} \sqrt{-\text{det}g}\left( -3\frac{\partial_y(w)}{w}
    \right)
    \right] ~. \label{eq:DeltaLDefinition}
\end{align}

The perturbative expansion of the gravitational contributions as series in $\kappa$ proceeds as usual. For convenience, we rewrite the bulk scalar field $\hat{\Phi}(x,y)$ such that it is perturbed about the background $\phi_{0}(\varphi)/\kappa$ by an amount $\hat{f}/\kappa$, i.e. $\hat{\Phi} \equiv \hat{\phi}/\kappa \equiv (\phi_{0} + \hat{f})/\kappa$. This rewrite allows us to expand the bulk and brane potentials in $\mathcal{L}_{\text{pot}}$ about $\hat{\phi} = \phi_{0}$ with respect to the dimensionless scalar fluctuation field $\hat{f}$ like so:
\begin{align}
    V[\hat{\Phi}] &= V + \dot{V}\, \hat{f} + \dfrac{1}{2} \ddot{V}\, (\hat{f})^2 + \mathcal{O}\Big((\hat{f})^{3}\Big)~,\\
    V_{i}[\hat{\Phi}] &= V_{i} + \dot{V}_{i}\, \hat{f} + \dfrac{1}{2} \ddot{V}_{i}\, (\hat{f})^2 + \mathcal{O}\Big((\hat{f})^{3}\Big)~,
\end{align}
where dots denote $\hat{\phi}$ functional derivatives, and $V$ and $V_{i}$ (and their $\hat{\phi}$ derivatives) are set to $\hat{\Phi} = \phi_{0}(y)/\kappa$ when their functional arguments are unspecified.

The path forward is, in principle, clear: find the appropriate solutions for the background fields $A(y)$ and $\phi_0(y)$, compute the Lagrangian that describes the dynamics of fluctuations about these background fields, and then diagonalize the quadratic terms in this Lagrangian to establish the (canonically normalized) physical modes and their interactions. In practice this is difficult because of the complicated algebraic structures involved in the mixing between the scalar components of the metric and Goldberger-Wise scalar field in the presence of a non-trivial scalar background.
Some simplification results from the fact that diffeomorphism invariance implies that only one linear combination of the fields $\hat{r}$ and $\hat{f}$ are physical, and we can set the gauge of the calculation such that these fields are related according to:
\begin{align}
    \phi^{\prime}_{0}\,\hat{f} = \sqrt{6}\, e^{2A}\, \hat{r}^{\,\prime}~. \label{eq:GaugeCondition}
\end{align}
We begin our analysis in general, without imposing this gauge condition and only use the gauge constraint Eq. \eqref{eq:GaugeCondition} to identify the physical scalar modes after deriving the background equations, which we turn to now.

\subsection{The Lagrangian to Quadratic Order}
After weak field expanding the Lagrangian Eq. \eqref{L5D_Definition} without applying the gauge condition Eq. \eqref{eq:GaugeCondition} or any background field equations, we obtain
\begin{align}
    \mathcal{L}_{\text{5D}} &\equiv \mathcal{L}_{\text{5D},\text{bkgd}} + \mathcal{L}_{\text{5D},h} + \mathcal{L}^{*}_{\text{5D},r} + \mathcal{L}^{*}_{\text{5D},f} + \mathcal{L}^{*}_{\text{5D},hr} + \mathcal{L}^{*}_{\text{5D},hf} + \mathcal{L}^{*}_{\text{5D},rf}\nonumber\\
    &\hspace{10 pt}+ \mathcal{L}_{\text{5D},hh} + \mathcal{L}^{*}_{\text{5D},rr} + \mathcal{L}^{*}_{\text{5D},ff} + \mathcal{O}(\kappa)~,
\end{align}
to all orders in the background fields and up to quadratic order in the fluctuations. 
The background-only terms in the Lagrangian are
\begin{align}
    \mathcal{L}_{\text{5D},\text{bkgd}} &\equiv  \dfrac{e^{-4A}}{2\,r_{c}^{2}\kappa^{2}} \bigg[24 A^{\prime\prime} - 48 (A^{\prime})^{2} - (\phi_{0}^{\prime})^{2}-8\bigg(V r_{c}^{2} + \sum_{i=1,2}V_{i} r_{c}\, \delta_{i}\bigg)\bigg]~. \label{L5D:bkgd:NoEOM}
\end{align}
The terms linear in the fluctuations are
\begin{align}
    \mathcal{L}_{\text{5D},h} &\equiv \dfrac{e^{-4A}}{4\,r_{c}^{2}\kappa} \bigg[24 A^{\prime\prime} - 48 (A^{\prime})^{2} - (\phi_{0}^{\prime})^{2}-8\bigg(V r_{c}^{2} + \sum_{i=1,2}V_{i} r_{c}\, \delta_{i}\bigg)\bigg]\hat{h} ~, \label{L5D:h:NoEOM}\\
    \mathcal{L}_{\text{5D},r}^{*} &\equiv \dfrac{\sqrt{6}\, e^{-2A}}{r_{c}^{2}\kappa} \bigg[\hat{r}^{\,\prime\prime} - 2 A^{\prime}\,\hat{r}^{\,\prime}\bigg] \nonumber\\ &\hspace{10 pt}+\dfrac{e^{-2A}}{2\sqrt{6}\,r_{c}^{2}\kappa} \bigg[-48 A^{\prime\prime} + 48 (A^{\prime})^{2} + 3 (\phi_{0}^{\prime})^{2}+8\bigg(V r_{c}^{2} + 2\sum_{i=1,2}V_{i} r_{c}\, \delta_{i}\bigg)\bigg]\hat{r} ~, \label{L5D:r:NoEOM}\\
    \mathcal{L}_{\text{5D},f}^{*} &\equiv -\dfrac{e^{-4A}}{r_{c}^{2}\kappa} \phi_{0}^{\prime}\,\hat{f}^{\,\prime} -\dfrac{4\,e^{-4A}}{r_{c}^{2}\kappa} \bigg[\dot{V} r_{c}^{2} + \sum_{i=1,2}\dot{V}_{i} r_{c}\, \delta_{i}\bigg]\hat{f} ~, \label{L5D:f:NoEOM}
\end{align}
where $\hat{h} = \eta^{\mu\nu}\hat{h}_{\mu\nu}$. 
At quadratic order in the fluctuations, 
``off-diagonal" (mode-mixing) quadratic terms are
\begin{align}
    \mathcal{L}_{\text{5D},hr}^{*} &\equiv \dfrac{\kappa}{2}\, \mathcal{L}_{\text{5D},r}^{*} \, \hat{h} ~, \label{L5D:hr:NoEOM}\\ 
    \mathcal{L}_{\text{5D},hf}^{*} &\equiv \dfrac{\kappa}{2}\, \mathcal{L}_{\text{5D},f}^{*} \, \hat{h} ~, \label{L5D:hf:NoEOM}\\
    \mathcal{L}_{\text{5D},rf}^{*} &\equiv \sqrt{\dfrac{3}{2}}\dfrac{e^{-2A}}{r_{c}^{2}} \phi_{0}^{\prime}\,\hat{r}\,\hat{f}^{\,\prime} +\sqrt{\dfrac{8}{3}}\dfrac{e^{-2A}}{r_{c}^{2}} \bigg[\dot{V} r_{c}^{2} + 2 \sum_{i=1,2}\dot{V}_{i} r_{c}\, \delta_{i}\bigg]\hat{r}\,\hat{f} \label{L5D:rf:NoEOM} ~,
\end{align}
and the ``on-diagonal" quadratic terms are given by
\begin{align}
    \mathcal{L}_{\text{5D},hh} &\equiv e^{-2A} \bigg[ (\partial^{\nu} \hat{h}_{\mu\nu})\, (\partial^{\mu} \hat{h})  - (\partial^{\nu}\hat{h}_{\mu\nu})^{2} + \dfrac{1}{2} (\partial_{\mu}\hat{h}_{\nu\rho})^{2}- \dfrac{1}{2} (\partial_{\mu}\hat{h})^{2}\bigg] + \dfrac{e^{-4A}}{2\, r_{c}^{2}}\bigg[ ( \hat{h}^{\prime})^2 - ( \hat{h}^{\prime}_{\mu\nu})^{2} \bigg] \nonumber\\
    &\hspace{10 pt}+\dfrac{e^{-4A}}{16\, r_{c}^{2}} \bigg[24 A^{\prime\prime} - 48 (A^{\prime})^{2} - (\phi_{0}^{\prime})^{2}-8\bigg(V r_{c}^{2} + \sum_{i=1,2}V_{i} r_{c}\, \delta_{i}\bigg)\bigg] \bigg[(\hat{h})^{2} - 2 (\hat{h}_{\mu\nu})^{2} \bigg]~, \label{L5D:hh:NoEOM}\\
    \mathcal{L}_{\text{5D},rr}^{*} &\equiv \dfrac{1}{2}\,e^{+2A}\,(\partial_{\mu}\hat{r})^{2} - \dfrac{1}{r_{c}^{2}}\bigg[2(\hat{r}^{\,\prime})^{2} + 3\,\hat{r}\,\hat{r}^{\,\prime\prime} \bigg] \nonumber\\
    &\hspace{10 pt}- \dfrac{1}{12\, r_{c}^{2}} \bigg[-48 A^{\prime\prime} + 5 (\phi_{0}^{\prime})^{2}+ 16 \sum_{i=1,2}V_{i} r_{c}\, \delta_{i}\bigg](\hat{r})^{2}  ~, \label{L5D:rr:NoEOM}\\ 
    \mathcal{L}_{\text{5D},ff}^{*} &\equiv \dfrac{1}{2}\,e^{-2A}\,(\partial_{\mu}\hat{f})^{2} - \dfrac{e^{-4A}}{2\, r_{c}^{2}}(\hat{f}^{\,\prime})^{2} - \dfrac{2 e^{-4A}}{r_{c}^{2}} \bigg[\ddot{V} r_{c}^{2} + \sum_{i=1,2}\ddot{V}_{i} r_{c}\, \delta_{i}\bigg](\hat{f})^{2}  ~. \label{L5D:ff:NoEOM}
\end{align}
Here we use an asterisk to denote that we have not yet applied a gauge condition relating $\hat{r}$ and $\hat{f}$. 

The first line of terms in Eq. \eqref{L5D:hh:NoEOM} will yield the usual canonical spin-2 Lagrangians after Kaluza-Klein decomposition. As we will soon demonstrate, the other terms in Eq. \eqref{L5D:hh:NoEOM} will be cancelled when the background fields satisfy their equations of motion. However, as close as $\mathcal{L}_{\text{5D},hh}$ is to the desired spin-2 result, the quadratic analysis overall is complicated by the presence of the mixing terms $\mathcal{L}^{*}_{\text{5D},hr}$ and $\mathcal{L}^{*}_{\text{5D},hf}$, which seemingly imply kinetic mixing between the tensor field $\hat{h}$ and the scalar fields $\hat{f}$ and $\hat{r}$. In order to eliminate these mixing terms, we must derive the equations of motion for the background fields and for the fluctuations, which we discuss next.

\subsection{Equations of Motion}

The Einstein field equations derived from $\mathcal{L}_{\text{5D}}$ equal
\begin{align}
    \mathcal{G}_{MN} - V[\hat{\Phi}] G_{MN} - \bigg[V_{1}[\hat{\Phi}]\dfrac{\delta_{1}(\varphi)}{r_{c}}  + V_{2}[\hat{\Phi}]\dfrac{\delta_{2}(\varphi)}{r_{c}}  \bigg] \dfrac{\sqrt{\overline{G}}}{\sqrt{G}}\, \overline{G}_{MN} = \dfrac{\kappa^{2}}{4} T_{MN}~,
\end{align}
where $\mathcal{G}_{MN} =  R_{MN} - \tfrac{1}{2} G_{MN}\,R$ is the Einstein tensor, $R_{MN}$ and $R = \tilde{G}^{AB} R_{AB}$ are the Ricci tensor and Ricci scalar respectively, and the stress-energy tensor equals
\begin{align}
    T_{MN} &= 2 \dfrac{\delta \mathcal{L}_{\Phi\Phi}}{\delta \tilde{G}^{MN}} - G_{MN} \mathcal{L}_{\Phi\Phi}~,\nonumber\\
    &= (\partial_{M}\hat{\Phi})(\partial_{N}\hat{\Phi}) - G_{MN}\bigg[ \dfrac{1}{2}\tilde{G}^{AB}(\partial_{A}\hat{\Phi})\,(\partial_{B}\hat{\Phi})\bigg] ~.
\end{align}
Recall that $\hat{\Phi} \equiv \hat{\phi}/\kappa \equiv (\phi_{0} + \hat{f})/\kappa$. We will discuss the Einstein field equations in terms of their decomposition as $(M,N) = \left\{(\mu,\nu), (\mu,5), (5,5)\right\}$, to the first two orders in $\kappa$.

\subsubsection{Background Equations of Motion}

To lowest order in $\kappa$, in which no fluctuation fields are present, only the $(\mu,\nu)$ and $(5,5)$ Einstein field equations are nontrivial (because of the Lorentz-invariance of
the constant-$y$ subspaces, the $(\mu,\nu)$ components of the curvature are proportional to $\eta_{\mu\nu}$), and they imply
\begin{align}
    A^{\prime\prime} &= 2 (A^{\prime})^{2} + \dfrac{1}{24}(\phi_{0}^{\prime})^{2} + \dfrac{1}{3}\bigg[ V r_{c}^{2}  + \sum_{i=1,2} V_{i} r_{c}\,\delta_{i} \bigg]~, \label{eq:bgEOM1} \\
    V r_{c}^{2} &=  -6 (A^{\prime})^{2} + \dfrac{1}{8} (\phi_{0}^{\prime})^{2}~,
    \nonumber
\end{align}
respectively for the background fields. The first of these equations implies the boundary conditions (integrating over the endpoints using an $S^1/Z_2$ orbifold construction where $A(y)$ is assumed to be ``even" under orbifold reflection):
\begin{align}
    V_{1}r_{c}\, \delta_{1} = +6 A^{\prime}\, \delta_{1}\,~,\hspace{35 pt}V_{2}r_{c} \, \delta_{2} = -6 A^{\prime}\, \delta_{2} ~.
    \label{eq:jump_cond1a}
\end{align}
By combining the equations of \eqref{eq:bgEOM1}, we may also write
\begin{align}
    A^{\prime\prime} = \dfrac{1}{12}\bigg[(\phi_{0}^{\prime})^{2} + 4 \sum_{i=1,2} V_{i}r_{c}\,\delta_{i}\bigg] \tag{\eqref{eq:App} revisited}
\end{align}

Note that Eq. \eqref{eq:bgEOM1} ensures $\mathcal{L}_{\text{5D},\text{bkgd}}$ and $\mathcal{L}_{\text{5D},h}$ from Eqs. \eqref{L5D:bkgd:NoEOM}-\eqref{L5D:h:NoEOM} vanish, and ensures $\mathcal{L}_{\text{5D},hh}$ yields canonical spin-2 Lagrangians after Kaluza-Klein decomposition. Eq. \eqref{eq:bgEOM1} also simplifies the various pieces of the Lagrangian, including the linear $\hat{r}$ terms:
\begin{align}
    \mathcal{L}_{\text{5D},r}^{*} &= \dfrac{\sqrt{6}}{r_{c}^{2}\kappa} \partial_{\varphi}\bigg[ e^{-2A} \hat{r}^{\,\prime}\bigg]~. \label{L5D:r:EOM1} 
\end{align}
While the mixing terms $\mathcal{L}^{*}_{\text{5D},hr}$ and $\mathcal{L}^{*}_{\text{5D},hf}$ remain at this point, these will vanish once we have analyzed the scalar sector, which we discuss now. 

We obtain another background equation by considering the Euler-Lagrange equation of the scalar field. The terms independent of the fluctuations yield
\begin{align}
    \phi_{0}^{\prime\prime} &= 4 A^{\prime} \phi_{0}^{\prime} + 4\dot{V}r_{c}^{2} + 4\sum_{i=1,2} \dot{V}_{i} r_{c} \, \delta_{i} ~,
    \label{eq:bgEOM2}
\end{align}
which implies its own boundary conditions (again, assuming the background scalar field configuration is even under the orbifold projection)
\begin{align}
    \dot{V}_{1}r_{c}\, \delta_{1} = +\dfrac{1}{2} \phi_{0}^{\prime}\, \delta_{1}~,\hspace{35 pt}  \dot{V}_{2}r_{c}\, \delta_{2} = -\dfrac{1}{2} \phi_{0}^{\prime}\, \delta_{2}~.
    \label{eq:jump_cond1b}
\end{align}
This simplifies $\mathcal{L}_{\text{5D},f}^{*}$, such that
\begin{align}
    \mathcal{L}_{\text{5D},f}^{*} = -\dfrac{1}{r_{c}^{2}\kappa} \partial_{\varphi} \bigg[e^{-4A}\,\phi_{0}^{\prime}\,\hat{f}\bigg]~.
    \label{L5D:f:EOM1}
\end{align}
This completes our derivation of background equations.

Recall that whenever we write a quantity multiplying $\delta_{1}(\varphi)$ or $\delta_{2}(\varphi)$, it is understood that the quantity is evaluated {\it in the limit} as $\varphi$ approaches the appropriate orbifold fixed point from inside the $[0,\pi r_{c}]$ half of the bulk. This implies, for example, via Eq. \eqref{eq:bgEOM2},
\begin{align}
    \phi_{0}^{\prime\prime}\,\delta_{i} = (\phi_{0}^{\prime\prime})_{\text{bulk}}\,\delta_{i} \equiv \bigg[4 A^{\prime} \phi_{0}^{\prime} + 4\,\dot{V}r_{c}^{2} \bigg]\delta_{i}~. \label{eq:phi0ppTimesDiracDelta}
\end{align}
This also ensures quantities like $A'(\varphi)\, \delta_{i}(\varphi)$ in Eq. \eqref{eq:jump_cond1a} and Eq. \eqref{eq:phi0ppTimesDiracDelta} are written unambiguously, despite $A^{\prime}(\varphi)$ being orbifold odd across the orbifold fixed points.

\subsubsection{Lagrangian at Quadratic Order: Mode Equations}

Next, we examine the equations of motion derived from considering terms in the Lagrangian that are
quadratic or lower in the fluctuations. These will give us the equations which will define the mode
expansions  -- the Kaluza-Klein decomposition -- for the fluctuating fields.
As mentioned in the previous subsubsection, we will be ignoring the spin-2 fields -- they will ultimately decouple from the scalar fields after having performed the correct scalar-field mode expansions.

We begin with the scalar fields in the metric.
Simplifying the expressions using the background equations Eqs. \eqref{eq:bgEOM1}-\eqref{eq:bgEOM2}, the $(\mu,\nu)$, $(\mu,5)$, and $(5,5)$ Einstein field equations at $\mathcal{O}(\kappa)$ are satisfied only if, respectively,
\begin{align}
    0 &= \bigg[\partial_{\varphi} - 4 A^{\prime}\bigg] \bigg[ \sqrt{6}\,e^{2A}\,\hat{r}^{\,\prime} - \phi_{0}^{\prime} \, \hat{f}\bigg]~, \label{eq:linEOM1}\\
    0 &= \partial_{\mu} \bigg[ \sqrt{6}\, e^{2A} \, \hat{r}^{\,\prime} - \phi_{0}^{\prime}\, \hat{f}\bigg]~, \label{eq:linEOM2}\\
    \partial_{\varphi}\bigg[\dfrac{e^{-2A}}{\sqrt{6}}\,\phi_{0}^{\prime}\, \hat{f} \bigg] &= e^{2A} \, r_{c}^{2} \, (\square \hat{r}) + 2A^{\prime}\bigg\{2\, \hat{r}^{\,\prime} + \bigg[\dfrac{e^{-2A}}{\sqrt{6}}\,\phi_{0}^{\prime}\, \hat{f} \bigg] \bigg\}\nonumber\\
    &\hspace{15 pt}  + \dfrac{8\dot{V}r_{c}^{2}}{\phi_{0}^{\prime}}\bigg[\dfrac{e^{-2A}}{\sqrt{6}}\,\phi_{0}^{\prime}\, \hat{f}\bigg] + \dfrac{1}{6} (\phi_{0}^{\prime})^{2}\,\hat{r} + 2\,(\delta_{1} - \delta_{2})\bigg[\dfrac{e^{-2A}}{\sqrt{6}}\,\phi_{0}^{\prime}\, \hat{f} \bigg]~, \label{eq:linEOM3}
\end{align}
where the final equation has also utilized the jump conditions of Eq. \eqref{eq:jump_cond1b}.
As noted in Ref.~\cite{Csaki:2000zn}, integrating Eq.~\eqref{eq:linEOM3}, we end up with a tautology and end up with boundary terms that provide no additional physical information. By moving the Dirac deltas of Eq. \eqref{eq:linEOM3} to the LHS and evaluating the derivatives, we derive an alternative form of the equation which lacks Dirac deltas (implicit or explicit):
\begin{align}
    \dfrac{e^{-2A}}{\sqrt{6}}\,\phi_{0}^{\prime}\,\hat{f}^{\,\prime} = e^{2A} \, r_{c}^{2} \, (\square \hat{r}) + 4 A^{\prime}\,\hat{r}^{\,\prime} + \dfrac{1}{6} (\phi_{0}^{\prime})^{2}\,\hat{r} + \sqrt{\dfrac{8}{3}}\,e^{-2A}\,\dot{V}r_{c}^{2}\,\hat{f}~. \label{eq:linEOM3Alt}
\end{align}


We may also consider the Euler-Lagrangian equation of the fluctuation field $\hat{f}$ at this order, which yields
\begin{align}
    \hat{f}^{\,\prime\prime} &= e^{2A}\,r_{c}^{2}\,(\square \hat{f}) +4 A^{\prime}\,\hat{f}^{\prime} + 4\ddot{V}r_{c}^{2}\,\hat{f} + \sqrt{\dfrac{3}{2}}\,e^{2A}\, \phi_{0}^{\prime}\, \hat{r}^{\prime} + \sqrt{\dfrac{2}{3}}\,e^{2A}\,\bigg[4 \dot{V} r_{c}^{2} + 3 A^{\prime}\phi_{0}^{\prime} \bigg] \hat{r}\nonumber\\
    &\hspace{15 pt} + \bigg[ 4 \ddot{V}_{1} r_{c}\, \hat{f} + \sqrt{\dfrac{2}{3}}\,e^{2A}\,\phi_{0}^{\prime}\,\hat{r}\bigg] \delta_{1} + \bigg[ 4 \ddot{V}_{2} r_{c}\, \hat{f} - \sqrt{\dfrac{2}{3}}\,e^{2A}\,\phi_{0}^{\prime}\,\hat{r}\bigg] \delta_{2}~.  \label{eq:SLEqFluctuation}
\end{align}
This equation implies, via the orbifold construction, the boundary conditions:
\begin{align}
    \hat{f}^{\,\prime}\delta_{1} = \bigg[2\ddot{V}_{1}r_{c}\,\hat{f} + \dfrac{1}{\sqrt{6}}\, e^{2A}\,\phi_{0}^{\prime}\,\hat{r}\bigg]\delta_{1}~, \hspace{35 pt} \hat{f}^{\,\prime}\delta_{2} = -\bigg[2\ddot{V}_{2}r_{c}\,\hat{f} - \dfrac{1}{\sqrt{6}}\, e^{2A}\,\phi_{0}^{\prime}\,\hat{r}\bigg]\delta_{2}~.
    \label{eq:SBCs}
\end{align}
Multiply these jump conditions by $e^{-2A}\phi_{0}^{\prime}/\sqrt{6}$ and use Eq. \eqref{eq:linEOM3Alt} to get
\begin{align}
    \bigg\{e^{2A} \, r_{c}^{2} \, (\square \hat{r}) + 4 A^{\prime}\,\hat{r}^{\,\prime} +
    \bigg[\dfrac{4\dot{V}r_{c}^{2}}{\phi_{0}^{\prime}} - 2 \ddot{V}_{1}r_{c}\bigg]\bigg[\dfrac{e^{-2A}}{\sqrt{6}}\,\phi_{0}^{\prime}\,\hat{f}\bigg]\bigg\}\delta_{1} &= 0~,\\ \bigg\{e^{2A} \, r_{c}^{2} \, (\square \hat{r}) + 4 A^{\prime}\,\hat{r}^{\,\prime} +
    \bigg[\dfrac{4\dot{V}r_{c}^{2}}{\phi_{0}^{\prime}} + 2 \ddot{V}_{2}r_{c}\bigg]\bigg[\dfrac{e^{-2A}}{\sqrt{6}}\,\phi_{0}^{\prime}\,\hat{f}\bigg]\bigg\}\delta_{2} &= 0~.
\end{align}
Using Eq. \eqref{eq:phi0ppTimesDiracDelta}, we may instead write the jump conditions Eq. \eqref{eq:SBCs} as
\begin{align}
    \bigg\{e^{2A} \, r_{c}^{2} \, (\square \hat{r}) + 4 A^{\prime}\,\hat{r}^{\,\prime} +
    \bigg[\dfrac{\phi_{0}^{\prime\prime}}{\phi_{0}^{\prime}} - 4A^{\prime} - 2 \ddot{V}_{1}r_{c}\bigg]\bigg[\dfrac{e^{-2A}}{\sqrt{6}}\,\phi_{0}^{\prime}\,\hat{f}\bigg]\bigg\}\delta_{1} &= 0~,\nonumber\\ \bigg\{e^{2A} \, r_{c}^{2} \, (\square \hat{r}) + 4 A^{\prime}\,\hat{r}^{\,\prime} +
    \bigg[\dfrac{\phi_{0}^{\prime\prime}}{\phi_{0}^{\prime}} - 4A^{\prime} + 2 \ddot{V}_{2}r_{c}\bigg]\bigg[\dfrac{e^{-2A}}{\sqrt{6}}\,\phi_{0}^{\prime}\,\hat{f}\bigg]\bigg\}\delta_{2} &= 0~.\label{eq:linEOM3BC}
\end{align}
This form is more common in the literature.

The linear field equations Eqs. \eqref{eq:linEOM1}-\eqref{eq:linEOM3} and Eq. \eqref{eq:SLEqFluctuation} describe the scalar modes of the theory.  Note, in particular, the recurring quantity $\sqrt{6}\, e^{2A}\, \hat{r}^{\,\prime} - \phi_{0}^{\prime}\,\hat{f}$. This will vanish once we impose the gauge condition
\eqref{eq:GaugeCondition}, which is our next focus.

\subsubsection{The Gauge Condition}

The form of the metric specified by \eqref{eq:Metric} - \eqref{eq:uhat} does not completely fix the ``gauge" for this calculation: we have access to various five-dimensional diffeomorphism transformations which maintain the form of the metric and with which we can choose to simplify our computations. In particular, as shown in \cite{Boos:2005dc}, we can always perform a change of coordinate to impose
the gauge condition introduced previously
\begin{align}
    \sqrt{6}\, e^{2A}\,\hat{r}^{\,\prime} = \phi_{0}^{\prime} \, \hat{f}~.
\tag{\ref{eq:GaugeCondition}}
\end{align}
One immediate consequence of this gauge choice is that the {\it sum} of the mixing terms $\mathcal{L}^{*}_{\text{5D},hr}$ $\mathcal{L}^{*}_{\text{5D},hf}$ vanishes, eliminating (as promised) any problematic mixing between the scalar and spin-2 mode sectors.

The physical implication of the gauge condition \eqref{eq:GaugeCondition} is that one combination of the scalar fields is a gauge-artifact, and does not correspond to a propagating degree of freedom.\footnote{The precise combination of Lagrangian fields which is physical and the corresponding form of its interactions depend on the gauge choice -- although all physical amplitudes are gauge-invariant.} 
Note that the ``mixing" of the scalar degree of freedom in the five-dimensional metric $\hat{r}$ with the bulk scalar field $\hat{f}$ only occurs in the presence of a $y$-dependent scalar background field configuration ($\phi'_0\neq 0$). It is precisely this mixing between the two sectors that enables the dynamics which stabilize the size of the extra-dimension in the Goldberger-Wise mechanism \cite{Goldberger:1999uk,Goldberger:1999un} and which simultaneously give rise to a ``radion" mass.
One advantage of working in this ``unitary" gauge and eliminating the fluctuations of the scalar field $\hat{f}$ in favor of scalar fluctuations of the metric $\hat{r}$ is that all couplings linear in the physical scalar fields have the same algebraic form as couplings linear in the (massless) radion within the unstabilized model -- simplifying the required coupling computations.

Having imposed this gauge condition, the $(\mu,\nu)$ and $(\mu,5)$ linear Einstein field equations Eqs. \eqref{eq:linEOM1}-\eqref{eq:linEOM2} vanish, the $(5,5)$ linear Einstein field equation Eq. \eqref{eq:linEOM3} simplifies to
\begin{align}
    \hat{r}^{\,\prime\prime} &= e^{2A} \, r_{c}^{2} \, (\square \hat{r}) + \bigg[ 6A^{\prime} + \dfrac{8\dot{V}r_{c}^{2}}{\phi_{0}^{\prime}}\bigg]\hat{r}^{\,\prime} + \dfrac{1}{6} (\phi_{0}^{\prime})^{2}\, \hat{r} + 2\,(\delta_{1} - \delta_{2})\,\hat{r}^{\,\prime}~, \label{eq:g55Eqn}
\end{align}
and the jump conditions Eq. \eqref{eq:linEOM3BC} reduce to
\begin{align}
    \bigg\{e^{2A} \, r_{c}^{2} \, (\square \hat{r}) -
    \bigg[2 \ddot{V}_{1}r_{c} - \dfrac{\phi_{0}^{\prime\prime}}{\phi_{0}^{\prime}} \bigg]\hat{r}^{\,\prime}\bigg\}\delta_{1} = \bigg\{e^{2A} \, r_{c}^{2} \, (\square \hat{r}) +
    \bigg[2 \ddot{V}_{2}r_{c}  + \dfrac{\phi_{0}^{\prime\prime}}{\phi_{0}^{\prime}} \bigg]\hat{r}^{\,\prime}\bigg\}\delta_{2} = 0~.\label{eq:ScalarBC}
\end{align}
These equations of motion for the field $\hat{r}$ at quadratic order will define the Kaluza-Klein decomposition of the $\hat{r}$ field. Note that, being careful about Dirac deltas,\footnote{Refer to the discussion after Eq. \eqref{eq:AugmentedL5D} for more details. In short, the quantity $1/(\phi_{0}^{\prime})^{2}$ cannot generate Dirac deltas upon differentiation.}
\begin{align}
    \partial_{\varphi}\bigg[\dfrac{e^{2A}}{(\phi_{0}^{\prime})^{2}} \hat{r}^{\,\prime}\bigg] &= \dfrac{2\,e^{2A}}{(\phi_{0}^{\prime})^{2}}A^{\prime}\, \hat{r}^{\,\prime} - \dfrac{2\, e^{2A}}{(\phi_{0}^{\prime})^{3}}(\phi_{0}^{\prime\prime})_{\text{bulk}}\,\hat{r}^{\,\prime} + \dfrac{e^{2A}}{(\phi_{0}^{\prime})^{2}} \hat{r}^{\,\prime\prime}~,\nonumber\\
    &= \dfrac{e^{2A}}{(\phi_{0}^{\prime})^{2}}\bigg\{ -\Big[6 A^{\prime} + \dfrac{8 \dot{V} r_{c}^{2}}{\phi_{0}^{\prime}}  \Big] \hat{r}^{\,\prime} + \hat{r}^{\,\prime\prime}\bigg\}~,
\end{align}
such that Eq. \eqref{eq:g55Eqn} may also be written in a more conventional form:
\begin{align}
    \partial_{\varphi}\bigg[\dfrac{e^{2A}}{(\phi_{0}^{\prime})^{2}} \hat{r}^{\,\prime}\bigg] - \dfrac{e^{2A}}{6}\hat{r} + 2\Big[\delta_{2}(\varphi) - \delta_{1}(\varphi)\Big]
    \dfrac{e^{2A}}{(\phi_{0}^{\prime})^{2}}\hat{r}^{\,\prime} = \dfrac{e^{4A}}{(\phi_{0}^{\prime})^{2}} r_{c}^{2} (\square \hat{r}) \label{eq:ScalarEOM}
\end{align}
In the next subsection, we use Eqs. \eqref{eq:ScalarEOM} and \eqref{eq:ScalarBC} to define the Kaluza-Klein decomposition of the 5D scalar field. (Note again that the jump conditions of the field $\hat{r}$ are trivial in this form of the equation, and boundary conditions in Eq. \eqref{eq:ScalarBC} are required.)

\subsection{Kaluza-Klein Decomposition of the Scalar Field}
Next, we assume we can decompose the 5D scalar field $\hat{r}(x,y)$ into a tower of 4D fields $\hat{r}_{i}(x)$ and extra-dimensional wavefunctions $\gamma_{i}(\varphi)$:
\begin{align}
    \hat{r}(x,y) &= \dfrac{1}{\sqrt{\pi r_{c}}} \sum_{i = 0}^{+\infty} \hat{r}^{(i)}(x)\,\gamma_{i}(\varphi) ~, \label{eq:ScalarKKDecomp}
\end{align}
where the states are arranged in order of increasing mass and $\varphi = y/r_{c}$. We will show that if the $\gamma_{i}$ satisfies the Sturm-Liouville-like equation (compare to Eq. \eqref{eq:ScalarEOM}):
\begin{align}
    \partial_{\varphi}\bigg[\dfrac{e^{2A}}{(\phi_{0}^{\prime})^{2}} (\partial_{\varphi}\gamma_{i})\bigg] - \dfrac{e^{2A}}{6}\gamma_{i} + 2\Big[\delta_{2}(\varphi) - \delta_{1}(\varphi)\Big]
    \dfrac{e^{2A}}{(\phi_{0}^{\prime})^{2}}(\partial_{\varphi}\gamma_{i}) = -\mu_{(i)}^{2}\,\dfrac{e^{4A}}{(\phi_{0}^{\prime})^{2}} \, \gamma_{i}~, \label{eq:ScalarEOMWfxn}
\end{align}
with boundary conditions (compare to Eq. \eqref{eq:ScalarBC}):
\begin{align}
    (\partial_{\varphi}\gamma_{i})\bigg|_{\varphi = 0+} &= -\bigg[2\ddot{V}_{1}r_{c} - \dfrac{\phi_{0}^{\prime\prime}}{\phi_{0}^{\prime}} \bigg]^{-1}\,\mu_{(i)}^{2}\,e^{2A}\,\gamma_{i} \bigg|_{\varphi = 0+}~,\nonumber\\
    (\partial_{\varphi}\gamma_{i})\bigg|_{\varphi = \pi-} &= +\bigg[2\ddot{V}_{2}\,r_{c} + \dfrac{\phi_{0}^{\prime\prime}}{\phi_{0}^{\prime}}\bigg]^{-1}\,\mu_{(i)}^{2}\,e^{2A}\,\gamma_{i} \bigg|_{\varphi = \pi-}~, \label{eq:ScalarBCWfxn}
\end{align}
the $\hat{r}^{(i)}(x)$ are the properly normalized scalar Kaluza-Klein fields. 

These scalar boundary conditions can alternatively be enforced (recalling that $\hat{r}$ and hence $\gamma_i$ are orbifold-even) using the equation introduced in the body of the paper:
\begin{align}
    \partial_{\varphi}\bigg[\dfrac{e^{2A}}{(\phi_{0}^{\prime})^{2}} (\partial_{\varphi}\gamma_{i})\bigg] - \dfrac{e^{2A}}{6}\gamma_{i} = -\mu_{(i)}^{2}\,\dfrac{e^{4A}}{(\phi_{0}^{\prime})^{2}} \, \gamma_{i} \, \Bigg\{ 1 + \frac{2\,\delta(\varphi)}{\Big[ 2 \ddot{V}_{1} r_{c} - \frac{\phi _{0}^{\prime\prime}}{\phi_{0}^{\prime}} \Big]} +  \frac{2\,\delta(\varphi - \pi )}{\Big[ 2 \ddot{V}_{2} r_{c} + \frac{\phi _{0}^{\prime\prime}}{\phi_{0}^{\prime}} \Big]} \Bigg\}~.\tag{\ref{eq:spin0-SLeqn}}
\end{align}
In this form, the Sturm-Liouville nature of the problem is manifest \cite{Boos:2005dc,Kofman:2004tk,fulton1977two,binding1994sturm}. 
We will choose to normalize the wavefunctions such that
\begin{align}
    \delta_{m,n} & = \dfrac{6\mu^2_n}{\pi}\int_{-\pi}^{+\pi} d\varphi\, \gamma_m\gamma_n \dfrac{e^{4A}}{(\phi_{0}^{\prime})^{2}} \Bigg\{ 1 + \frac{2\,\delta(\varphi)}{\Big[ 2 \ddot{V}_{1} r_{c} - \frac{\phi _{0}^{\prime\prime}}{\phi_{0}^{\prime}} \Big]} +  \frac{2\,\delta(\varphi - \pi )}{\Big[ 2 \ddot{V}_{2} r_{c} + \frac{\phi _{0}^{\prime\prime}}{\phi_{0}^{\prime}} \Big]} \Bigg\}\\
    & = \dfrac{6}{\pi} \int_{-\pi}^{+\pi} d\varphi\hspace{5 pt} \bigg[\dfrac{e^{2A}}{(\phi_{0}^{\prime})^{2}} \gamma_{m}^{\,\prime} \, \gamma_{n}^{\,\prime} + \dfrac{e^{2A}}{6} \gamma_{m}\gamma_{n} \bigg]~, \label{Scalar:Norm1}
\end{align}
where the second line follows by applying the differential equation (\ref{eq:spin0-SLeqn}) and integration by parts on the periodic doubled ``orbifold". We will show that this normalization will yield properly normalized scalar Kaluza-Klein modes.\footnote{Note that this choice is consistent since we have not massless physical scalar modes in this model.}
For our numerical investigations, we consider the ``stiff-wall" limit $\ddot{V}_{1,2}\to \infty$, in which case the eigenmodes $\gamma_i$ satisfy Neumann boundary conditions.  While the stiff-wall limit is (ultimately) unphysical, it is consistent with the simplification we made in ignoring the dynamics of the brane itself - and we can expect the results of our analysis correctly describe low-energy properties of the system. Outside of numerical investigations, we do not take the stiff-wall limit.



To facilitate manipulations at the 5D level, define the following useful auxillary field:
\begin{align}
    \hat{z} &\equiv \dfrac{1}{\sqrt{\pi r_{c}}} \sum_{i = 0}^{+\infty} \mu_{(i)}^{2}\, \hat{r}^{(i)}(x)\,\gamma_{i}(\varphi) ~. \label{eq:zKKDecomp}
\end{align}
Using $\hat{z}$ and the decomposition in \eqref{eq:ScalarKKDecomp}, the wavefunction differential equation becomes
\begin{align}
    \hat{r}^{\,\prime\prime} &=  - e^{2A} \, \hat{z} + \bigg[ 6A^{\prime} + \dfrac{8\dot{V}r_{c}^{2}}{\phi_{0}^{\prime}}\bigg]\hat{r}^{\,\prime} + \dfrac{1}{6} (\phi_{0}^{\prime})^{2}\, \hat{r} + 2\,(\delta_{1} - \delta_{2})\,\hat{r}^{\,\prime}~, \label{eq:rEOM}
\end{align}
such that
\begin{align}
    (\hat{r}^{\prime\prime})_{\text{bulk}} &\equiv - e^{2A} \, \hat{z} + \bigg[ 6A^{\prime} + \dfrac{8\dot{V}r_{c}^{2}}{\phi_{0}^{\prime}}\bigg]\hat{r}^{\,\prime} + \dfrac{1}{6} (\phi_{0}^{\prime})^{2}\, \hat{r}~,
\end{align}
and the wavefunction boundary conditions imply
\begin{align}
    \bigg\{ -e^{2A} \, \hat{z} + \bigg[ 4 A^{\prime} + \dfrac{4 \dot{V} r_{c}^{2}}{\phi_{0}^{\prime}} \bigg]\hat{r}^{\,\prime}\bigg\}\delta_{1} &= \bigg\{ +2 \ddot{V}_{1}r_{c}\, \hat{r}^{\,\prime} \bigg\} \delta_{1}~,\nonumber\\
    \bigg\{- e^{2A} \, \hat{z} + \bigg[ 4 A^{\prime} + \dfrac{4 \dot{V} r_{c}^{2}}{\phi_{0}^{\prime}} \bigg]\hat{r}^{\,\prime}\bigg\}\delta_{2} &= \bigg\{ -2 \ddot{V}_{2}r_{c}\, \hat{r}^{\,\prime}  \bigg\} \delta_{2}~.\label{eq:rBC}
\end{align}
These boundary conditions are written in such a way to most easily replace away $\ddot{V}_{i}$ for future convenience. Let us now return to the Lagrangian.

\subsection{The Canonical Scalar Mode Expansion}
After applying the background equations of motion Eqs. \eqref{eq:bgEOM1}, \eqref{eq:jump_cond1a}, \eqref{eq:bgEOM2}, and \eqref{eq:jump_cond1b}, as well as the gauge condition \eqref{eq:GaugeCondition}, we find the collections of Lagrangian terms Eqs. \eqref{L5D:bkgd:NoEOM}-\eqref{L5D:ff:NoEOM} are simplified. Most contributions now explicitly vanish:
\begin{align}
    \mathcal{L}_{\text{5D},\text{bkgd}} \hspace{8 pt} = \hspace{8 pt} \mathcal{L}_{\text{5D},h} \hspace{8 pt} = \hspace{8 pt} \mathcal{L}_{\text{5D},r}^{*} + \mathcal{L}_{\text{5D},f}^{*} \hspace{8 pt} = \hspace{8 pt} \mathcal{L}_{\text{5D},hr}^{*} + \mathcal{L}_{\text{5D},hf}^{*} \hspace{8 pt} = \hspace{8 pt} 0~.
\end{align}
The tensor quadratic Lagrangian is now of the desired form to yield a tower of canonical spin-2 states after Kaluza-Klein decomposition:
\begin{align}
    \mathcal{L}_{\text{5D},hh} &\equiv e^{-2A} \bigg[ (\partial^{\nu} \hat{h}_{\mu\nu})\, (\partial^{\mu} \hat{h})  - (\partial^{\nu}\hat{h}_{\mu\nu})^{2} + \dfrac{1}{2} (\partial_{\mu}\hat{h}_{\nu\rho})^{2}- \dfrac{1}{2} (\partial_{\mu}\hat{h})^{2}\bigg] + \dfrac{e^{-4A}}{2\, r_{c}^{2}}\bigg[ ( \hat{h}^{\prime})^2 - ( \hat{h}^{\prime}_{\mu\nu})^{2} \bigg]
\end{align}
The scalar quadratic Lagrangian, however, remains quite complicated. We organize the terms from each part of the quadratic scalar Lagrangian as follows:
\begin{align}
    \mathcal{L}_{\text{5D},rr} &= \mathcal{L}_{\text{5D},rr}^{*} + \mathcal{L}_{\text{5D},rf}^{*} + \mathcal{L}_{\text{5D},ff}^{*}\\
    &= \mathcal{L}_{\text{EH},rr} + \mathcal{L}_{\text{GHY},rr} + \mathcal{L}_{\Phi\Phi,rr} + \mathcal{L}_{\text{pot},rr} + \Delta \mathcal{L}_{rr} \label{eq:LrrBreakdown}
\end{align}
where
\begin{align}
    \mathcal{L}_{\text{EH},rr} &= -\dfrac{1}{6}\,e^{2A}\, (\partial_{\mu} \hat{r})^{2} -\dfrac{2}{3}\,e^{2A}\, \hat{r}\, (\square \hat{r}) - \dfrac{3}{r_{c}^{2}}\, (\hat{r}^{\,\prime})^{2}- \dfrac{4}{r_{c}^{2}}\, \hat{r} \,\hat{r}^{\,\prime\prime} + \dfrac{8}{3\,r_{c}^{2}} A^{\prime}\, \hat{r}\, \hat{r}^{\,\prime} + \dfrac{16}{3\,r_{c}^{2}}\, A^{\prime\prime} \, \hat{r}^{\,2}\\
    \mathcal{L}_{\text{GHY},rr} &= \dfrac{4}{r_{c}^{2}}\, (\hat{r}^{\,\prime})^{2} + \dfrac{4}{r_{c}^{2}}\, \hat{r}\,\hat{r}^{\,\prime\prime} - \dfrac{32}{3\,r_{c}^{2}} A^{\prime}\, \hat{r}\, \hat{r}^{\,\prime} - \dfrac{16}{3\,r_{c}^{2}}\, A^{\prime\prime}\, \hat{r}^{\,2}\\
    \mathcal{L}_{\Phi\Phi,rr}  &= \dfrac{3}{(\phi_{0}^{\prime})^{2}}\, e^{2A}\, (\partial_{\mu}\hat{r}^{\,\prime})^{2}  -  \dfrac{3}{r_{c}^{2} (\phi_{0}^{\prime})^{4}} \Big[\phi_{0}^{\prime}\,\hat{r}^{\,\prime\prime} - \phi_{0}^{\prime\prime}\,\hat{r}^{\,\prime} + 2\, A^{\prime}\,(\phi_{0}^{\prime})\, \hat{r}^{\,\prime} - \dfrac{1}{2} (\phi_{0}^{\prime})^{3} \, \hat{r}\Big]^{2} + \dfrac{(\phi_{0}^{\prime})^{2}}{3\, r_{c}^{2}}\,  \hat{r}^{\,2}\\
    \mathcal{L}_{\text{pot},rr} &= \dfrac{4\,\dot{V}}{\phi_{0}^{\prime}} \,\hat{r} \, \hat{r}^{\,\prime} - \dfrac{12\,\ddot{V}}{(\phi_{0}^{\prime})^{2}}  \, (\hat{r}^{\,\prime})^{2} - \sum_{i=1,2} \bigg[ -\dfrac{4 \,V_{i}}{3\,r_{c}} \, \hat{r}^{\,2} + \dfrac{8\,\dot{V}_{i}}{r_{c}\,\phi_{0}^{\prime}} \, \hat{r}\,\hat{r}^{\,\prime} -\dfrac{12\,\ddot{V}_{i}}{r_{c}\,(\phi_{0}^{\prime})^{2}} \, (\hat{r}^{\,\prime})^{2}\bigg] \delta_{i}\\
    \Delta\mathcal{L}_{rr} &= -\dfrac{3}{r_{c}^{2}} (\hat{r}^{\,\prime})^{2} - \dfrac{3}{r_{c}^{2}} \hat{r}\,\hat{r}^{\,\prime\prime} + \dfrac{8}{r_{c}^{2}} A^{\prime} \, \hat{r} \, \hat{r}^{\,\prime} + \dfrac{4}{r_{c}^{2}} A^{\prime\prime}\,\hat{r}^{\,2}
    \label{eq:L5D_quad1}
\end{align}
For ease of comparison, we present these results without yet applying the background equations of motion or integration-by-parts. Note that the squared quantity in $\mathcal{L}_{\Phi\Phi,rr}$ is not singular because the delta-functions in $\phi^{\prime}_{0}\, \hat{r}^{\,\prime\prime} - \phi^{\prime\prime}_{0}\, \hat{r}^{\,\prime}$ cancel.

Rather than consider this quadratic scalar Lagrangian directly, we first add a convenient total derivative (which we determined through trial-and-error). Generally, adding a total derivative to a Lagrangian reorganizes how information is stored in the fields, but ultimately does not change the physics described by the Lagrangian; a classic example of this is the Gibbons-Hawking-York total derivative, which is used to make the Einstein-Hilbert Lagrangian of 4D gravity into a Lagrangian which only depends on fields and their first derivatives \cite{Gibbons:1976ue}. The total derivative we add to $\mathcal{L}_{\text{5D},rr}$ (in addition to the total derivative $\Delta \mathcal{L}$ defined in Eq. \eqref{eq:DeltaLDefinition}, which is already folded into $\mathcal{L}_{\text{5D},rr}$) is
\begin{align}
    \overline{\Delta} \mathcal{L}_{rr} &= \dfrac{1}{r_{c}^{2}} \partial_{\varphi} \bigg\{ \hat{r}\,\hat{r}^{\,\prime} - \dfrac{3\, e^{2A}}{(\phi_{0}^{\prime})^{2}} \hat{z}\,\hat{r}^{\,\prime} + \dfrac{12}{(\phi_{0}^{\prime})^{4}} \Big[ A^{\prime} (\phi_{0}^{\prime})^{2} + V^{\prime} r_{c}^{2}\Big] (\hat{r}^{\,\prime})^{2} \bigg\}
\end{align}
such that we consider the combination
\begin{align}
    \mathcal{L}_{\text{5D},rr} + \overline{\Delta} \mathcal{L}_{rr}~. \label{eq:AugmentedL5D}
\end{align}
The $\varphi$-derivative in $\overline{\Delta} \mathcal{L}_{rr}$ must be evaluated with care, lest we generate spurious singularities. In particular, Dirac delta-function singularities will be generated whenever $\varphi$-differentiating a discontinuity in a function's slope. In the present calculation, such a discontinuity only ever happens at the orbifold fixed points. The standard example of this from Randall-Sundrum models is the twice-differentiated quantity $|\varphi|^{\prime\prime} = \partial_{\varphi}(|\varphi|^{\prime}) = \partial_{\varphi}(\text{sign }\varphi)$, which equals $2(\delta_{1} - \delta_{2})$ in our scheme. If we are not careful when taking $\varphi$-derivatives more generally, we can accidentally generate spurious Dirac deltas which contradict our scheme. Consider $\varphi$-differentiating a quantity which is a square of a $\varphi$-differentiated quantity, such as $1/(\phi_{0}^{\prime})^{2}$. Naively, we attain $-2\,\phi_{0}^{\prime\prime}/(\phi_{0}^{\prime})^{3}$, which generates nonzero Dirac deltas through the $\phi_{0}^{\prime\prime}$. These Dirac deltas are spurious. First, note that $\phi_{0}$ is a function of $|\varphi|$, which means $\phi_{0}^{\prime}$ is proportional to $|\varphi|^{\prime} = \text{sign}(\varphi)$. Thus $1/(\phi_{0}^{\prime})^{2} \propto 1/(\text{sign }\varphi)^{2} = 1$ and $1/(\phi_{0})^{2}$ lacks the overall factor of $|\varphi|^{\prime}$ necessary to generate Dirac deltas upon $\varphi$-differentiation. While naive differentiation yields $-2\,\phi_{0}^{\prime\prime}/(\phi_{0}^{\prime})^{3}$, careful analysis reveals the $\varphi$-derivative of $1/(\phi_{0}^{\prime})^{2}$ is actually the Dirac delta-free quantity $-2\,(\phi_{0}^{\prime\prime})_{\text{bulk}}/(\phi_{0}^{\prime})^{3}$.

For these reasons, evaluation of the $\varphi$-derivative present in the total derivative $\overline{\Delta} \mathcal{L}_{rr}$ yields fewer Dirac deltas than naively expected. Namely, they are only generated upon differentiating $\hat{r}^{\prime}$ in the first two terms and $A^{\prime}$ \& $V^{\prime}$ in the third term. Explicitly, we thus calculate\footnote{Technically these same considerations are important when calculating, for example, $\Delta \mathcal{L}$; however, $\Delta\mathcal{L}_{rr} = \partial_{\varphi}[(4A^{\prime}\hat{r} - 3\hat{r}^{\,\prime})\hat{r}]/r_{c}^{2}$, and naive differentiation yields the correct result.}
\begin{align}
    \overline{\Delta} \mathcal{L}_{rr} &= \dfrac{1}{r_{c}^{2}} \bigg\{ (\hat{r}^{\,\prime})^{2} + \hat{r}\,\hat{r}^{\,\prime\prime} + \dfrac{6\, e^{2A}}{(\phi_{0}^{\prime})^{3}}\,(\phi_{0}^{\prime\prime})_{\text{bulk}}\, \hat{z}\,\hat{r}^{\,\prime} - \dfrac{6\, e^{2A}}{(\phi_{0}^{\prime})^{2}}\, A^{\prime}\, \hat{z}^{\,\prime}\,\hat{r}^{\,\prime} - \dfrac{3\, e^{2A}}{(\phi_{0}^{\prime})^{2}} \hat{z}\,\hat{r}^{\,\prime\prime}\nonumber\\
    &\hspace{35 pt} -48 \dfrac{(\phi_{0}^{\prime\prime})_{\text{bulk}}}{(\phi_{0}^{\prime})^{5}} \Big[ A^{\prime} (\phi_{0}^{\prime})^{2} + V^{\prime} r_{c}^{2}\Big] (\hat{r}^{\,\prime})^{2} + \dfrac{24}{(\phi_{0}^{\prime})^{4}} \Big[ A^{\prime} (\phi_{0}^{\prime})^{2} + V^{\prime} r_{c}^{2}\Big] \hat{r}^{\,\prime}\, (\hat{r}^{\,\prime\prime})_{\text{bulk}} \nonumber\\
    &\hspace{35 pt} + \dfrac{12}{(\phi_{0}^{\prime})^{4}} \Big[ A^{\prime\prime} (\phi_{0}^{\prime})^{2} + 2 A^{\prime} \phi_{0}^{\prime}\, (\phi_{0}^{\prime\prime})_{\text{bulk}} + V^{\prime\prime} r_{c}^{2}\Big] (\hat{r}^{\,\prime})^{2} \bigg\}
\end{align}
Having evaluated $\overline{\Delta} \mathcal{L}_{rr}$ as above, we next consider the quadratic scalar terms $\mathcal{L}_{\text{5D},rr} + \overline{\Delta} \mathcal{L}_{rr}$ after performing the following sequence of manipulations:
\begin{enumerate}
    \item Use 4D integration-by-parts to eliminate any 4D d'Alembertian operators $\square = \partial_{t}^{2} - \vec{\nabla}^{\,2}$, e.g. taking $\hat{r} (\square \hat{r})$ to $-(\partial_{\mu} \hat{r})^{2}$.
    \item Eliminate all instances of $\hat{r}^{\,\prime\prime}$, $A^{\prime\prime}$, $V$, and $\phi_{0}^{\prime\prime}$ (and their Dirac delta-free bulk forms) via Eqs. \eqref{eq:rEOM}, \eqref{eq:bgEOM1}, and \eqref{eq:bgEOM2} respectively. Having done so, all Dirac deltas in the original weak field expanded Lagrangian have been made explicit.
    \item Eliminate $V_{i}$ and $\dot{V}_{i}$ via the background equations Eqs. \eqref{eq:jump_cond1a} and \eqref{eq:jump_cond1b} respectively.
    \item Eliminate $\ddot{V}_{i}$ (which always multiplies an $\hat{r}^{\,\prime}$) via the boundary conditions, Eq. \eqref{eq:rBC}.
    \item Eliminate all instances of $\ddot{V}$, $V^{\prime}$, and $V^{\prime\prime}$, and in favor of $\dot{V}$ and $\dot{V}^{\prime}$ via chain rule relations, i.e.
    \begin{align}
        V^{\prime\prime} = \dot{V}^{\prime}\, \phi_{0}^{\prime} + \dot{V}\, \phi_{0}^{\prime\prime}\hspace{55 pt}
        V^{\prime} = \dot{V}\,\phi_{0}^{\prime}\hspace{55 pt}
        \ddot{V} = \dfrac{\dot{V}^{\prime}}{\phi_{0}^{\prime}}
    \end{align}
    where $\phi_{0}^{\prime\prime}$ is then replaced by Eq. \eqref{eq:bgEOM2}, as done earlier. With this, all Dirac deltas in $\overline{\Delta} \mathcal{L}$ are also explicit.
\end{enumerate}
After performing these replacements, we find all $\dot{V}$ and $\dot{V}^{\prime}$ terms cancel, all Dirac deltas cancel, and we are left with very few terms\footnote{An alternate way of deriving the canonical quadratic Lagrangian is to start with the expression on the right hand side of Eq.~\eqref{eq:CanonQuadA}, and substituting $\hat{z}$ from the (5,5) Einstein equation~\eqref{eq:linEOM3Alt}, as well as a similar expression for $\hat{z}^\prime$ derived from the Euler-Lagrange equation~\eqref{eq:SLEqFluctuation}. The resulting expression can be shown to be equal to the combination $\mathcal{L}_{\text{5D},rr} + \overline{\Delta} \mathcal{L}_{rr}$. It is useful, when performing these manipulations, to remove explicit Dirac delta terms by using the background equations of motion given in Eq.~\eqref{eq:bgEOM1} and Eq.~\eqref{eq:bgEOM2}. }
\begin{align}
    \mathcal{L}_{\text{5D},rr} + \Delta \mathcal{L}_{rr} &= \dfrac{e^{2A}}{2} \bigg[ (\partial_{\mu}\hat{r})^{2} - \dfrac{\hat{z}\,\hat{r}}{r_{c}^{2}}\bigg] + \dfrac{3\,e^{2A}}{(\phi_{0}^{\prime})^{2}} \bigg[(\partial_{\mu}\hat{r}^{\,\prime})^{2} - \dfrac{\hat{z}^{\,\prime}\,\hat{r}^{\,\prime}}{r_{c}^{2}}\bigg]~.
    \label{eq:CanonQuadA}
\end{align}

Upon Kaluza-Klein decomposition via Eqs. \eqref{eq:ScalarKKDecomp} and \eqref{eq:zKKDecomp}, $\mathcal{L}_{\text{5D},rr} + \Delta \mathcal{L}_{rr}$ immediately yields:
\begin{align}
    \sum_{m,n=0}^{+\infty} \dfrac{1}{2}\bigg[(\partial_{\mu}\hat{r}^{(m)})(\partial^{\mu}\hat{r}^{(n)}) - \mu_{(m)}^{2}\,\hat{r}^{(m)}\,\hat{r}^{(n)} \bigg] \cdot \dfrac{6}{\pi} \int_{-\pi}^{+\pi} d\varphi\hspace{5 pt} \bigg[\dfrac{e^{2A}}{(\phi_{0}^{\prime})^{2}} \gamma_{m}^{\,\prime} \, \gamma_{n}^{\,\prime} + \dfrac{e^{2A}}{6} \gamma_{m}\gamma_{n} \bigg]~.
\end{align}
Recall that the scalar state wavefunctions are normalized according to Eq. \eqref{Scalar:Norm1}, such that the integral on the right above (including the $6/\pi$) equals $\delta_{m,n}$. Consequently, we finally achieve our desired result:
\begin{align}
    \mathcal{L}_{\text{5D},rr} + \Delta \mathcal{L}_{rr} &= \sum_{n=0}^{+\infty} \bigg\{ \dfrac{1}{2}(\partial^{\mu}\hat{r}^{(n)})^2 - \dfrac{1}{2}\mu_{(n)}^{2}\,(\hat{r}^{(n)})^{2} \bigg\} ~.
\end{align}
That is, the 5D field $\hat{r}(x,y)$ generates a scalar tower of canonical 4D scalar states $\{\hat{r}^{(n)}(x)\}$, each having mass $m_{(n)}\equiv \mu_{(n)}/r_{c}$, where $\mu_{(n)}$ is determined by solving the differential equation problem for the wavefunctions $\{\gamma_{n}\}$ laid out between Eqs. \eqref{eq:ScalarEOMWfxn} and \eqref{eq:ScalarBCWfxn}.

\newpage
\section[The Inelastic Sum Rules Relating Couplings and Masses]{The Inelastic Sum Rules Relating Couplings and Masses
} \label{DerivingSumRules}
This section derives \& summarizes relationships between couplings and mass spectra that are relevant to ensuring at-most $\mathcal{O}(s)$ growth of tree-level inelastic 2-to-2 helicity-zero massive spin-2 KK mode scattering amplitudes (i.e. the process $(k,l)\rightarrow (m,n)$) in the Goldberger-Wise-stabilized Randall-Sundrum I model. We briefly consider the implications of completeness before deriving a means of expressing all cubic and quartic (spin-2 exclusive) B-type couplings in terms of A-type couplings and special objects $B_{(kl)(mn)}$. These B-to-A formulas reduce the problem of finding amplitude-relevant formulas to the problem of simplifying sums of the form $\sum_{j} \mu_{j}^{2p}\, a_{klj}\, a_{mnj}$. The relevant (inelastic and elastic) sum rules are derived and then summarized in their own subsections. The final subsection describes the remaining set of (unproven) sum rules necessary for at-most $\mathcal{O}(s)$ growth in the fully inelastic process. 

This appendix is written as a stand-alone report of the sum-rule relationships needed to insure that all {\it inelastic} scattering amplitudes (all $2\to 2$ scattering amplitudes with massive spin-2 fields of arbitrary mode number in the external states) grow no faster than ${\cal O}(s)$ and report which we have succeeded in proving -- completing the program begin in \cite{Chivukula:2019rij,Chivukula:2019zkt,Chivukula:2020hvi,Foren:2020egq}. Section \ref{ShortSumRules} derives relationships used in Sec. \ref{subsec:IVB} of the main body of this paper and can be read independently.

\subsection{Definitions}\label{sup-subsection-definitions}
It is convenient to define generalized ``couplings" to be overlap integrals of spin-2 and spin-0 wavefunctions of the form
\begin{align}
    x^{(p)}_{(k^{\prime}\cdots l) \cdots m^{\prime} \cdots n} &\equiv \dfrac{1}{\pi} \int_{-\pi}^{+\pi} d\varphi\hspace{10 pt}\vep^{p}\, (\partial_{\varphi}\gamma_{k})\cdots \gamma_{l}\cdots (\partial_{\varphi}\psi_{m})\cdots \psi_{n} \label{generalizedx}
\end{align}
where $A(\varphi)$ is the warp-factor, $\varepsilon\equiv\exp(-A)$, and  we add an additional factor of $(\partial_{\varphi}A)/kr_{c}$ to the integrand if only an odd number of differentiated wavefunctions are present in the integrand otherwise. The most common integrands appearing in the 4D effective Lagrangian carry powers $\varepsilon^{-2}$ and $\varepsilon^{-4}$, and are given special symbols:
\begin{align}
    a_{(k^{\prime}\cdots l) \cdots m^{\prime} \cdots n} \equiv x^{(-2)}_{(k^{\prime}\cdots l) \cdots m^{\prime} \cdots n} \hspace{35 pt} b_{(k^{\prime}\cdots l) \cdots m^{\prime} \cdots n} = x^{(-4)}_{(k^{\prime}\cdots l) \cdots m^{\prime} \cdots n}
\end{align}
We will also encounter the label ``$c$", which is associated with $p=-6$. In particular, we encounter this integral often:
\begin{align}
    c_{k^{\prime}l^{\prime}m^{\prime}n^{\prime}} & \equiv x^{(-6)}_{k^{\prime}l^{\prime}m^{\prime}n^{\prime}} = \dfrac{1}{\pi} \int d\varphi\hspace{5 pt} \vep^{-6}(\partial_{\varphi}\psi_{k})(\partial_{\varphi}\psi_{l})(\partial_{\varphi}\psi_{m})(\partial_{\varphi}\psi_{n}) \label{sup-cklmnDEF}
\end{align}
Another object that will be useful throughout the rest of this document is the symbol $\mathcal{D} \equiv \vep^{-4}\partial_{\varphi}$, which is a combination of quantities that is often present as a result of the spin-2 Sturm-Liouville equation. When desperate for space, we will nest the notation even further, utilizing $\mathcal{D}_{n} \equiv \vep^{-4}(\partial_{\varphi}\psi_{n})$.

We will ultimately derive sum rules that allow us to rewrite certain useful sums of intermediate masses and couplings in terms of just the quartic A-type coupling $a_{klmn}$, three $B_{(kl)(mn)}$ objects (of which any two fix the value of the third), and integrals $c_{k^{\prime}l^{\prime}m^{\prime}n^{\prime}}$ and $x^{(-8)}_{\phi^{\prime}_{0}\phi^{\prime}_{0}n^{\prime}n^{\prime}n^{\prime}n^{\prime}}$.

\subsection{Applications of Completeness}
The spin-2 mode completeness relation is
\begin{align}
    \delta(\varphi_{2}-\varphi_{1}) =  \dfrac{1}{\pi}\, \varepsilon(\varphi_{1})^{-2}\sum_{j=0}^{+\infty} \psi_{j}(\varphi_{1})\, \psi_{j}(\varphi_{2}) \label{eq:sup-completeness}
\end{align}
where $\varepsilon(\varphi) \equiv e^{A(\varphi)}$ and the wavefunctions $\psi_{n}$ satisfy $\partial_{\varphi}\mathcal{D}\psi_{n}=\partial_{\varphi}[\varepsilon^{-4}(\partial_{\varphi}\psi_{n})]=-\mu_{n}^{2}\varepsilon^{-2}\psi_{n}$. Spin-2 mode completeness allows us to collapse certain sums of cubic coupling products into a single quartic coupling, e.g.
\begin{align}
    a_{klmn} &= \sum_{j} a_{klj}\, a_{mnj} = \sum_{j} a_{kmj}\,a_{lnj} = \sum_{j} a_{knj}\,a_{lmj} \label{sup-acompleteness}\\
    b_{k^{\prime}l^{\prime} mn} &= \sum_{j} b_{k^{\prime} l^{\prime} j}\, a_{mnj}
\end{align}
Furthermore, by combining cubic B-type couplings in this same way, we arrive at
\begin{align}
    c_{k^{\prime}l^{\prime}m^{\prime}n^{\prime}} &= \sum_{j} b_{k^{\prime}l^{\prime}j}\,b_{m^{\prime}n^{\prime}j} = \sum_{j} b_{k^{\prime}m^{\prime}j}\,b_{l^{\prime}n^{\prime}j} = \sum_{j} b_{k^{\prime}n^{\prime}j}\,b_{l^{\prime}m^{\prime}j}\nonumber
\end{align}
This is as far as direct applications of completeness can get us for now.

\subsection{B-to-A Formulas}
This subsection details how to eliminate all B-type couplings (e.g. $b_{l^{\prime}m^{\prime}n}$ and $b_{k^{\prime}l^{\prime}mn}$) in favor of A-type couplings (e.g. $a_{lmn}$ and $a_{klmn}$) and new structures $B_{(kl)(mn)}$. To begin, we note we can absorb a factor of $\mu^{2}$ into A-type couplings with help from the Sturm-Liouville equation. A standard application of this technique proceeds as follows:
\begin{align}
    \mu_{n}^{2}\, a_{lmn} &= \dfrac{1}{\pi}\int d\varphi\hspace{5 pt} \vep^{-2} \psi_{l} \psi_{m} \left[\mu_{n}^{2}\psi_{n}\right]\\
    &= \dfrac{1}{\pi} \int d\varphi\hspace{5 pt}\vep^{-2}  \psi_{l}\psi_{m}\bigg[-\vep^{+2}\partial_{\varphi}(\mathcal{D}\psi_{n})\bigg]\label{sup-286}\\
    &= \dfrac{1}{\pi} \int d\varphi\hspace{5 pt}\partial_{\varphi}\left[\psi_{l}\psi_{m}\right] (\mathcal{D}\psi_{n}) \label{sup-287}\\
    &= \dfrac{1}{\pi} \int d\varphi\hspace{5 pt} \vep^{-4} (\partial_{\varphi}\psi_{l})\psi_{m} (\partial_{\varphi} \psi_{n}) + \dfrac{1}{\pi} \int d\varphi\hspace{5 pt} \vep^{-4} \psi_{l}(\partial_{\varphi}\psi_{m}) (\partial_{\varphi} \psi_{n})\\
    &= b_{l^{\prime}mn^{\prime}} + b_{lm^{\prime}n^{\prime}}
\end{align}
where integration by parts was utilized between Eqs. \eqref{sup-286} and \eqref{sup-287}; because $(\mathcal{D}\psi_{n})$ vanishes on the boundaries, there is no surface term. This and the equivalent calculation with the quartic A-type coupling yield
\begin{align}
    \mu_{n}^{2}\, a_{lmn} &= b_{l^{\prime}mn^{\prime}} + b_{lm^{\prime}n^{\prime}} \label{sup-AIMS3}\\
    \mu_{n}^{2}\, a_{klmn} &= b_{k^{\prime}lmn^{\prime}} + b_{kl^{\prime}mn^{\prime}} + b_{klm^{\prime}n^{\prime}} \label{sup-AIMS4}
\end{align}
By considering different permutations of KK indices, each of these equations corresponds to three and four unique constraints respectively. Because there are only three unique cubic B-type couplings with KK indices $l$, $m$, and $n$ (specifically, $b_{l^{\prime}m^{\prime}n}$, $b_{l^{\prime}mn^{\prime}}$, and $b_{lm^{\prime}n^{\prime}}$), Eq. \eqref{sup-AIMS3} can be inverted to yield
\begin{align}
    \boxed{b_{l^{\prime} m^{\prime} n} = \dfrac{1}{2}\left[\mu_{l}^{2} + \mu_{m}^{2} - \mu_{n}^{2} \right] a_{lmn}} \label{sup-blmnRED}
\end{align}
with which we can eliminate all cubic B-type couplings in favor of the cubic A-type coupling.

There are {\it six} unique quartic B-type couplings with KK indices $k$, $l$, $m$, and $n$. We first halve this set by rewriting each quartic B-type coupling $b_{k^{\prime}l^{\prime}mn}$ in terms of new objects $B_{(kl)(mn)}$. These new objects are motivated as follows: note that Eq. \eqref{sup-AIMS4} implies
\begin{align}
    \dfrac{1}{2} \bigg[ \mu_{k}^{2} + \mu_{l}^{2} - \mu_{m}^{2} - \mu_{n}^{2} \bigg]\, a_{klmn} &= b_{k^{\prime}l^{\prime}mn} - b_{klm^{\prime}n^{\prime}}
\end{align}
Equivalently, we may write this as
\begin{align}
    b_{k^{\prime}l^{\prime}mn} + \dfrac{1}{2} \bigg[\mu_{m}^{2} + \mu_{n}^{2}\bigg]\, a_{klmn} &= b_{klm^{\prime}n^{\prime}} + \dfrac{1}{2}\bigg[\mu_{k}^{2} + \mu_{l}^{2} \bigg]\, a_{klmn} \label{sup-BMotivation}
\end{align}
In other words, the quantity on the LHS possesses a symmetry under the pair swap $(k,l)\leftrightarrow (m,n)$. Furthermore, this symmetry is maintained under the addition of any quantity $\tilde{B}_{(kl)(mn)}$ which is also symmetric under this pair swap. Inspired by Eq. \eqref{sup-BMotivation}, we define
\begin{align}
    B_{(kl)(mn)} \equiv b_{k^{\prime}l^{\prime}mn} + \dfrac{1}{2} \bigg[\mu_{m}^{2} + \mu_{n}^{2}\bigg]\, a_{klmn} + \tilde{B}_{(kl)(mn)}
\end{align}
We will choose the quantity $\tilde{B}_{(kl)(mn)}$ momentarily. Because the B-type couplings satisfy Eq. \eqref{sup-AIMS4}, the sum of all unique $B$ objects satisfies
\begin{align}
    B_{(kl)(mn)} + B_{(km)(ln)} + B_{(kn)(lm)} = \vec{\mu}^{\,2} \, a_{klmn} + \tilde{B}_{(kl)(mn)} + \tilde{B}_{(km)(ln)} + \tilde{B}_{(kn)(lm)} 
\end{align}
where $\vec{\mu}^{\,2} \equiv \mu_{k}^2+\mu_{l}^2+\mu_{m}^{2}+\mu_{n}^{2}$. That is, we can ensure the convenient property
\begin{align}
    B_{(kl)(mn)} + B_{(km)(ln)} + B_{(kn)(lm)} \dot{=} 0
\end{align}
as long as we choose $\tilde{B}_{(kl)(mn)}$ such that
\begin{align}
     \tilde{B}_{(kl)(mn)} + \tilde{B}_{(km)(ln)} + \tilde{B}_{(kn)(lm)} = -\vec{\mu}^{\,2} \, a_{klmn}
\end{align}
One immediate choice (and the choice we take now) is to set each $\tilde{B}$ equal to one third of $-\vec{\mu}^{\,2}\, a_{klmn}$
\begin{align}
    \tilde{B}_{(kl)(mn)} \dot{=} -\dfrac{1}{3}\, a_{klmn}
\end{align}
This yields (as a replacement rule for $b_{k^{\prime}l^{\prime}mn}$ and definition of $B_{(kl)(mn)}$)
\begin{align}
    \boxed{b_{k^{\prime}l^{\prime}mn} = B_{(kl)(mn)}+\dfrac{1}{6}\bigg[2(\mu_{k}^{2}+\mu_{l}^{2})-(\mu_{m}^{2}+\mu_{n}^{2})\bigg] a_{klmn}} \label{sup-bklmnRED}
\end{align}
where $B$ is symmetric within each pair and between pairs:
\begin{align}
    B_{(kl)(mn)} = B_{(mn)(kl)} = B_{(mn)(lk)}
\end{align}
and satisfies the additional constraint
\begin{align}
    \boxed{B_{(kl)(mn)} + B_{(km)(ln)} + B_{(kn)(lm)} = 0}\label{sup-SumOfB}
\end{align}
such that only two among $\{B_{(kl)(mn)},B_{(km)(ln)},B_{(kn)(lm)}\}$ are linearly independent. Note that $B_{(kl)(mn)}$ has the same symmetry properties as $\sum_{j} \mu_{j}^{2p}\,a_{klj}\, a_{mnj}$. Because Eq. \eqref{sup-bklmnRED} reduces B-type couplings to A-type couplings as much is as possible, we refer to it as the quartic B-to-A rule. This and Eq. \eqref{sup-blmnRED} comprise the desired B-to-A formulas.

The above rules are sufficient as-is for reducing the sum $\sum_{j} \mu_{j}^{2}\, a_{klj}\, a_{mnj}$ and yielding the first non-trivial sum rule. Using the cubic coupling equation Eq. \eqref{sup-AIMS3} with completeness yields,
\begin{align}
    b_{k^{\prime}l^{\prime}mn} &= \dfrac{1}{2}\left[\mu_{k}^{2} + \mu_{l}^{2}\right]a_{klmn} - \dfrac{1}{2} \sum_{j=0} \mu_{j}^{2}\, a_{klj}\, a_{mnj}\label{sup-ElimB4Intermediate}
\end{align}
Meanwhile, the LHS can be simplified via  Eq. \eqref{sup-bklmnRED}. Solving for the undetermined sum then gives us,
\begin{align}
    \boxed{\sum_{j=0} \mu_{j}^{2} \, a_{klj}\, a_{mnj} = -2\, B_{(kl)(mn)} + \dfrac{1}{3}\,\vec{\mu}^{\, 2}\, a_{klmn}} \label{sup-A2B4}
\end{align}
where $\vec{\mu}^{\,2} \equiv \mu_{k}^2+\mu_{l}^2+\mu_{m}^{2}+\mu_{n}^{2}$. We  next turn our attention to $\sum_{j} \mu_{j}^{4}\, a_{klj}\, a_{mnj}$ and then $\sum_{j} \mu_{j}^{6}\, a_{klj} \,a_{mnj}$.

\subsection{The \texorpdfstring{$\mu_{j}^{4}$}{Lg} Sum Rule}
The $\sum_{j} \mu_{j}^{4}\,a_{klj}\, a_{mnj}$ relation is relatively straightforward. As defined in Eq. \eqref{sup-cklmnDEF}, we can rewrite $c_{k^{\prime}l^{\prime}m^{\prime}n^{\prime}}$ in terms of B-type cubic couplings, to which we can then apply the B-to-A formulas:
\begin{align}
    c_{k^{\prime}l^{\prime}m^{\prime}n^{\prime}} &= \sum_{j=0} b_{k^{\prime}l^{\prime}j}b_{m^{\prime}n^{\prime}j}\\
    &\hspace{-35 pt}= \dfrac{1}{4}\sum_{j}\left[\mu_{k}^{2} + \mu_{l}^{2} -\mu_{j}^{2}\right]\left[\mu_{m}^{2} + \mu_{n}^{2} -\mu_{j}^{2} \right] a_{klj}a_{mnj}\\
    &\hspace{-35 pt}= \dfrac{1}{4}(\mu_{k}^{2} + \mu_{l}^{2})(\mu_{m}^{2} + \mu_{n}^{2}) a_{klmn} - \dfrac{1}{4}(\vec{\mu}^{\,2}) \sum_{j} \mu_{j}^{2} a_{klj} a_{mnj} + \dfrac{1}{4}\sum_{j} \mu_{j}^{4} a_{klj} a_{mnj}
\end{align}
such that, using Eq. \eqref{sup-A2B4} and solving for the undetermined sum $\sum_{j}\mu_{j}^{4}\,a_{klj}\,a_{mnj}$,
\begin{align}
    \boxed{\sum_{j} \mu_{j}^{4} a_{klj} a_{mnj} = 4\, c_{k^{\prime}l^{\prime}m^{\prime}n^{\prime}} - 2\, (\vec{\mu}^{\,2})\, B_{(kl)(mn)} + \left[\dfrac{1}{3} (\vec{\mu}^{\,2})^{2} - (\mu_{k}^{2} + \mu_{l}^{2})(\mu_{m}^{2} + \mu_{n}^{2}) \right] a_{klmn}} \label{sup-muj4sumruleX}
\end{align}
as desired. Deriving the $\sum_{j} \mu_{j}^{6}\,a_{klj}\, a_{mnj}$ relation requires significantly more work. 

\subsection{The \texorpdfstring{$\mu_{j}^{6}$}{Lg} Sum Rule}

\subsubsection{Elastic}
\label{ShortSumRules}
As a warm-up to the inelastic case, let us first derive the $\mu_{j}^{6}$ sum rule (and review the other sum rules) as they appear in the elastic case, i.e. when $k=l=m=n$. This will provide the general flow of the argument which is made more complicated in the inelastic case. Definitions for $x$, $a$, $b$, etc. are included in subsection  \ref{sup-subsection-definitions}.

Using the spin-2 completeness relation and differential equation alone, we have previously derived many elastic coupling relations \cite{Chivukula:2019rij,Chivukula:2019zkt,Chivukula:2020hvi,Foren:2020egq}. For example, there are the elastic B-to-A formulas
\begin{align}
    b_{n^{\prime}n^{\prime}j} = \dfrac{1}{2}\big[2\mu_{n}^{2} - \mu_{j}^{2}\big] a_{nnj} \hspace{35 pt} b_{j^{\prime} n^{\prime}n} = \dfrac{1}{2} \mu_{j}^{2} a_{nnj} \hspace{35 pt} b_{n^{\prime}n^{\prime}nn} = \dfrac{1}{3} \mu_{n}^{2} a_{nnnn} \label{eq:bnnnnBtoA}
\end{align}
which allow us to rewrite any spin-2 exclusive B-type couplings in terms of A-type couplings. Using these in combination with completeness, we find
\begin{align*}
    a_{nnnn} &= \sum_{j} a_{nnj}^{2}\\
    b_{n^{\prime}n^{\prime}nn} &= \sum_{j} b_{n^{\prime}n^{\prime}j} a_{nnj} = \mu_{n}^{2} \sum_{j} a_{nnj}^{2} - \dfrac{1}{2}\sum_{j} \mu_{j}^{2} a_{nnj}^{2} \\
     c_{n^{\prime}n^{\prime}n^{\prime}n^{\prime}} &\equiv \sum_{j} b_{n^{\prime}n^{\prime}j}^{2} = \mu_{n}^{4} \sum_{j} a_{nnj}^{2} - \mu_{n}^{2}\sum_{j} \mu_{j}^{2} a_{nnj}^{2} + \dfrac{1}{4} \sum_{j} \mu_{j}^{4} a_{nnj}^{2}
\end{align*}
which imply various expressions for sums of the form $\sum_{j} \mu_{j}^{2p} a_{nnj}^{2}$:
\begin{align}
    \sum_{j=0}^{+\infty} a_{nnj}^{2} &= a_{nnnn} \label{eq:mu0annj2}\\
    \sum_{j=0}^{+\infty} \mu_{j}^{2} a_{nnj}^{2} &= \dfrac{4}{3} \mu_{n}^{2} a_{nnnn}\label{eq:mu2annj2}\\
    \sum_{j=0}^{+\infty} \mu_{j}^{4} a_{nnj}^{2} &= 4 c_{n^{\prime}n^{\prime}n^{\prime}n^{\prime}} + \dfrac{4}{3} \mu_{n}^{2} a_{nnnn}\label{eq:mu4annj2}
\end{align}

These relations allow us to quickly rewrite various sums between B-type couplings, including
\begin{align}
    \sum_{j=0}^{+\infty} b_{n^{\prime}n^{\prime}j} b_{j^{\prime}n^{\prime}n} &= \dfrac{1}{3} \mu_{n}^{4} a_{nnnn} - c_{n^{\prime}n^{\prime}n^{\prime}n^{\prime}}~. \label{eq:bnnjbjnn}
\end{align}

As discussed in the main text, a combination of the GW model sum-rules insuring cancellation of the ${\cal O}(s^3)$ and ${\cal O}(s^2)$ contributions to amplitude
may be written
\begin{align}
    \text{Stabilized RS1:}\hspace{15 pt}\sum_{j=0}^{+\infty} \big[5\mu_{n}^{2} - \mu_{j}^{2} \big] \mu_{j}^{4} a_{nnj}^{2} = \dfrac{16}{3} \mu_{n}^{6} a_{nnnn} + 9 \sum_{i=0}^{+\infty} \mu_{(i)}^{2} a_{n^{\prime}n^{\prime}(i)}^{2}~, \tag{\ref{eq:sr5}}
\end{align}
We can use our existing relations to rewrite this expression as:
\begin{align}
    \sum_{j=0}^{+\infty} \mu_{j}^{6} a_{nnj}^{2} &= 20 \mu_{n}^{2} c_{n^{\prime}n^{\prime}n^{\prime}n^{\prime}} + \dfrac{4}{3} \mu_{n}^{6} a_{nnnn} + 9 \sum_{i=0}^{+\infty} \mu_{(i)}^{2} a_{n^{\prime}n^{\prime}(i)}^{2}
\end{align}
It is this variant we seek now to prove.

To begin, note that the B-to-A formulas relate the sum $\sum_{j} \mu_{j}^{6}a_{nnj}^{2}$ to the sum $\sum_{j} \mu_{j}^{2} b_{n^{\prime}n^{\prime}j}^{2}$ like so:
\begin{align}
    \sum_{j=0}^{+\infty} \mu_{j}^{6} a_{nnj}^{2} = 16 \mu_{j}^{2} c_{n^{\prime}n^{\prime}n^{\prime}n^{\prime}} + 4 \sum_{j=0}^{+\infty} \mu_{j}^{2} b_{n^{\prime}n^{\prime}j}^{2} \label{eq:muj6annj2intermediate}
\end{align}
Thus, if we determine a means of rewriting $\sum_{j} \mu_{j}^{2} b_{n^{\prime}n^{\prime}j}^{2}$ in terms of $c_{n^{\prime}n^{\prime}n^{\prime}n^{\prime}}$ and $a_{nnnn}$, then we will have a means of doing the same for the desired sum.

We will arrive at the desired form by considering two integrals of total derivatives, each of which vanishes because $(\partial_{\varphi}\psi_{n})$ vanishes at the orbifold fixed points $\varphi \in \{0,\pi\}$.

\underline{\bf Integral 1:} To begin, consider the following trivial integral:
\begin{align}
    \dfrac{1}{\pi} \int_{-\pi}^{+\pi} d\varphi\hspace{5 pt} \partial_{\varphi}\bigg\{ \bigg[ (\partial_{\varphi} \psi_{j}) - 6 (\partial_{\varphi} A)\psi_{j} \bigg] \varepsilon^{-6}(\partial_{\varphi}\psi_{n})^{2} \bigg\} = 0
\end{align}
By evaluating the net derivative and using the spin-2 mode differential equation to simplify second derivatives of spin-2 wavefunctions,\footnote{It is useful to repackage each $(\partial_{\varphi}\psi)$ instead as $\varepsilon^{+4}(\mathcal{D}\psi)$ where $\mathcal{D} \equiv \varepsilon^{-4}\partial_{\varphi}$ because then the spin-2 mode equation may be utilized more directly in the form $\partial_{\varphi}\mathcal{D}\psi_{n} = - \mu_{n}^{2} \varepsilon^{-2} \psi_{n}$.}  we attain
\begin{align}
    0 &= -12 \bigg\{\dfrac{1}{\pi} \int_{-\pi}^{+\pi} d\varphi\hspace{5 pt} (\partial_{\varphi} A)^{2} \varepsilon^{-6}(\partial_{\varphi} \psi_{n})^{2} \psi_{j}\bigg\} + 12(kr_{c})\mu_{n}^{2} x^{(-4)}_{n^{\prime}nj} - 2\mu_{n}^{2} b_{j^{\prime}n^{\prime}n}\nonumber\\
    &\hspace{35 pt} - 6 \bigg\{\dfrac{1}{\pi} \int_{-\pi}^{+\pi} d\varphi\hspace{5 pt} (\partial_{\varphi}^{2} A) \varepsilon^{-6}(\partial_{\varphi} \psi_{n})^{2} \psi_{j}\bigg\} - \mu_{j}^{2}b_{n^{\prime}n^{\prime}j} \label{eq:mu6nnnnCUBIC1}
\end{align}
We can then construct an instance of $\sum_{j} \mu_{j}^{2} b_{n^{\prime}n^{\prime}j}^{2}$ within this by multiplying it by $b_{n^{\prime}n^{\prime}j}$ and summing over $j$. This yields
\begin{align}
    0 &= -12 \bigg\{\dfrac{1}{\pi} \int_{-\pi}^{+\pi} d\varphi\hspace{5 pt} (\partial_{\varphi} A)^{2} \varepsilon^{-8}(\partial_{\varphi} \psi_{n})^{4}\bigg\} + 12(kr_{c})\mu_{n}^{2} x^{(-6)}_{n^{\prime}n^{\prime}n^{\prime}n} - 2\mu_{n}^{2} \sum_{j=0}^{+\infty} b_{n^{\prime}n^{\prime}j} b_{j^{\prime}n^{\prime}n}\nonumber\\
    &\hspace{35 pt} - 6 \bigg\{\dfrac{1}{\pi} \int_{-\pi}^{+\pi} d\varphi\hspace{5 pt} (\partial_{\varphi}^{2} A) \varepsilon^{-8}(\partial_{\varphi} \psi_{n})^{4}\bigg\} - \sum_{j=0}^{+\infty} \mu_{j}^{2}b_{n^{\prime}n^{\prime}j}^{2} \label{eq:mu6nnnnQUARTIC1}
\end{align}

\underline{\bf Integral 2:} Next consider the following trivial integral:
\begin{align}
    \dfrac{1}{\pi} \int_{-\pi}^{+\pi} d\varphi\hspace{5 pt} \partial_{\varphi}\bigg\{ \bigg[ \dfrac{3}{2} (\partial_{\varphi} A) \varepsilon^{-2} (\partial_{\varphi}\psi_{n}) - \mu_{n}^{2}\psi_{n} \bigg] \varepsilon^{-6}(\partial_{\varphi}\psi_{n})^{3} \bigg\} = 0
\end{align}
Evaluating this derivative in the same way as we did with the first integral, we find
\begin{align}
    0 &= 12 \bigg\{\dfrac{1}{\pi} \int_{-\pi}^{+\pi} d\varphi\hspace{5 pt} (\partial_{\varphi} A)^{2} \varepsilon^{-8}(\partial_{\varphi} \psi_{n})^{4}\bigg\} - 12(kr_{c})\mu_{n}^{2} x^{(-6)}_{n^{\prime}n^{\prime}n^{\prime}n} - \mu_{n}^{2} c_{n^{\prime}n^{\prime}n^{\prime}n^{\prime}} + 3\mu_{n}^{4} b_{n^{\prime}n^{\prime}nn}  \nonumber\\
    &\hspace{35 pt} + \dfrac{3}{2} \bigg\{\dfrac{1}{\pi} \int_{-\pi}^{+\pi} d\varphi\hspace{5 pt} (\partial_{\varphi}^{2} A) \varepsilon^{-8}(\partial_{\varphi} \psi_{n})^{4}\bigg\}  \label{eq:mu6nnnnQUARTIC2}
\end{align}

\underline{\bf Combining:} Summing Eqs. \eqref{eq:mu6nnnnQUARTIC1} \& \eqref{eq:mu6nnnnQUARTIC2} and then solving for $\sum_{j=0}^{+\infty} \mu_{j}^{2} b_{n^{\prime}n^{\prime}j}^{2}$ immediately yields:
\begin{align}
    \sum_{j=0}^{+\infty} \mu_{j}^{2} b_{n^{\prime}n^{\prime}j}^{2} &= - \mu_{n}^{2} c_{n^{\prime}n^{\prime}n^{\prime}n^{\prime}} + 3\mu_{n}^{4} b_{n^{\prime}n^{\prime}nn} - 2\mu_{n}^{2} \sum_{j=0}^{+\infty} b_{n^{\prime}n^{\prime}j} b_{j^{\prime}n^{\prime}n}\nonumber\\
    &\hspace{35 pt}- \dfrac{9}{2} \bigg\{\dfrac{1}{\pi} \int_{-\pi}^{+\pi} d\varphi\hspace{5 pt} (\partial_{\varphi}^{2} A) \varepsilon^{-8}(\partial_{\varphi} \psi_{n})^{4}\bigg\} \label{eq:mu2bnnj2}
\end{align}
We already know how to rewrite $b_{n^{\prime}n^{\prime}nn}$ and $\sum_{j} b_{n^{\prime}n^{\prime}j}b_{j^{\prime}n^{\prime}n}$ in terms of $c_{n^{\prime}n^{\prime}n^{\prime}n^{\prime}}$ and $a_{nnnn}$; namely, Eqs. \eqref{eq:bnnnnBtoA} \& \eqref{eq:bnnjbjnn}. Using these, Eq. \eqref{eq:mu2bnnj2} becomes,
\begin{align}
    \sum_{j=0}^{+\infty} \mu_{j}^{2} b_{n^{\prime}n^{\prime}j}^{2} = \mu_{n}^{2} c_{n^{\prime}n^{\prime}n^{\prime}n^{\prime}} + \dfrac{1}{3} \mu_{n}^{6} a_{nnnn} - \dfrac{9}{2}\bigg\{\dfrac{1}{\pi} \int_{-\pi}^{+\pi} d\varphi\hspace{5 pt} (\partial_{\varphi}^{2} A) \varepsilon^{-8}(\partial_{\varphi} \psi_{n})^{4}\bigg\}
\end{align}
Finally, applying this result to Eq. \eqref{eq:muj6annj2intermediate} gives us an expression for the desired ``$\mu_{j}^{6}$" sum:
\begin{align}
    \sum_{j=0}^{+\infty} \mu_{j}^{6} a_{nnj}^{2} = 20 \mu_{n}^{2} c_{n^{\prime}n^{\prime}n^{\prime}n^{\prime}} + \dfrac{4}{3} \mu_{n}^{6} a_{nnnn} - 18\bigg\{\dfrac{1}{\pi} \int_{-\pi}^{+\pi} d\varphi\hspace{5 pt} (\partial_{\varphi}^{2} A) \varepsilon^{-8}(\partial_{\varphi} \psi_{n})^{4}\bigg\}~. \label{eq:muj6annj2DERIVED}
\end{align}

Using the ``$\mu^4_j$" sum in Eq. \eqref{eq:mu4annj2} we obtain the result quoted in the text
\begin{align}
    \sum_{j=0}^{+\infty} \bigg[5\mu_{n}^{2}-\mu_{j}^{2}\bigg]\mu_{j}^{4} a^{2}_{nnj} &= \dfrac{16}{3}\mu_{n}^{6}a_{nnnn} + 18\bigg\{\dfrac{1}{\pi} \int_{-\pi}^{+\pi} d\varphi\hspace{5 pt} (\partial_{\varphi}^{2} A) \varepsilon^{-8}(\partial_{\varphi} \psi_{n})^{4}\bigg\}
    ~.\tag{\ref{eq:sr-intermediate}}
\end{align}
This is all that is required for the elastic case discussed in the main text. The inelastic case covered in the next subsubsection is logically similar, but involves longer expressions and significantly more algebra.

\subsubsection{Inelastic}
Before beginning the derivation of the inelastic $\mu_{j}^{6}$ sum rule, it is advantageous to define a well-organized polynomial basis with which we can write our results succinctly. In particular, because the sums we wish to simplify ($\sum_{j} \mu^{2p}\,a_{klj}\,a_{mnj}$) and relevant quartic degrees of freedom all have (at least) the symmetries of $B_{(kl)(mn)}$, it is useful to define a symmetrization operation that forms quantities with symmetries identical to $B_{(kl)(mn)}$:
\begin{align}
    \langle f_{klmn} \rangle &\equiv \bigg\{ \bigg[ f_{klmn} + (k\leftrightarrow l) \bigg] + (m\leftrightarrow n) \bigg\} + (kl\leftrightarrow mn)\\
    &\hspace{-35 pt} = f_{klmn} + f_{lkmn} + f_{klnm} + f_{lknm} + f_{mnkl} + f_{mnlk} + f_{nmkl} + f_{nmlk}
\end{align}
This allows us to quickly construct a finite basis for polynomials of $\mu^{2}\in \{\mu_{k}^{2},\mu_{l}^{2},\mu_{m}^{2},\mu_{n}^{2}\}$ having the aforementioned symmetry structures. For a single power of $\mu^{2}$, there is only one basis element:
\begin{align}
    \alpha_{(kl)(mn)}^{(1,1)} \equiv \langle \mu_{k}^{2} \rangle = 2 \vec{\mu}^{\,2} \equiv 2(\mu_{k}^{2} + \mu_{l}^{2} + \mu_{m}^{2} + \mu_{n}^{2})
\end{align}
For two powers of $\mu^{2}$, there are three:
\begin{align}
    \alpha_{(kl)(mn)}^{(2,1)} \equiv \langle \mu_{k}^{4} \rangle \hspace{35 pt} \alpha_{(kl)(mn)}^{(2,2)} \equiv \langle \mu_{l}^{2}\mu_{k}^{2} \rangle \hspace{35 pt} \alpha_{(kl)(mn)}^{(2,3)} \equiv \langle \mu_{m}^{2}\mu_{k}^{2} \rangle
\end{align}
and for three powers of $\mu^{2}$, there are four:
\begin{align}
    \alpha_{(kl)(mn)}^{(3,1)} \equiv \langle \mu_{k}^{6} \rangle \hspace{18 pt}&\hspace{18 pt} \alpha_{(kl)(mn)}^{(3,2)} \equiv \langle \mu_{l}^{2}\mu_{k}^{4} \rangle\nonumber\\
    \alpha_{(kl)(mn)}^{(3,3)} \equiv \langle \mu_{m}^{2}\mu_{k}^{4} \rangle \hspace{18 pt}&\hspace{18 pt} \alpha_{(kl)(mn)}^{(3,4)} \equiv \langle \mu_{m}^{2}\mu_{l}^{2}\mu_{k}^{2} \rangle
\end{align}
With these, we can generically construct any polynomial of the squared masses (up to cubic degree) having the aforementioned symmetry properties:
\begin{align}
    M_{(kl)(mn)}^{(1)}(c_{1}) &= 2 c_{1}\, \vec{\mu}^{\,2}\\
    M_{(kl)(mn)}^{(2)}(c_{1},c_{2},c_{3}) &= \sum_{i=1}^{3} c_{i}\, \alpha_{(kl)(mn)}^{(2,i)}\\
    M_{(kl)(mn)}^{(3)}(c_{1},c_{2},c_{3},c_{4}) &= \sum_{i=1}^{4} c_{i}\, \alpha_{(kl)(mn)}^{(3,i)} \label{sup-M3Def}
\end{align}
Note that these symbols are intentionally linear in their $c_{i}$ arguments. In this language, Eq. \eqref{sup-muj4sumruleX} may be rewritten as
\begin{align}
    \boxed{\sum_{j} \mu_{j}^{4} \, a_{klj}\, a_{mnj} = 4\, c_{k^{\prime}l^{\prime}m^{\prime}n^{\prime}} -2\, \vec{\mu}^{\, 2} \, B_{(kl)(mn)} + \dfrac{1}{6}\, M^{(2)}_{(kl)(mn)}(1,1,-1) \, a_{klmn}}
\end{align}
We now proceed to the $\sum_{j} \mu_{j}^{6} \, a_{klj} \, a_{mnj}$ rule.

As in the previous subsection, we begin our derivation by applying the B-to-A formulas to a sum of cubic B-type couplings, and then apply existing sum rules:
\begin{align}
    \sum_{j} \mu_{j}^{2}\, b_{k^{\prime}l^{\prime}j}\,b_{m^{\prime}n^{\prime}j} &= \dfrac{1}{4}\sum_{j}\mu_{j}^{6}\, a_{klj}\,a_{mnj} -\vec{\mu}^{\,2}\, c_{k^{\prime}l^{\prime}m^{\prime}n^{\prime}}\nonumber\\
    &\hspace{-70 pt} -\dfrac{1}{12}\,\vec{\mu}^{\,2}\,(\mu_{k}^{2}+\mu_{l}^{2} -\mu_{m}^{2} -\mu_{n}^{2})^{2}\, a_{klmn} + \dfrac{1}{2}\, \bigg[(\vec{\mu}^{\,2})^{2} - (\mu_{k}^{2}+\mu_{l}^{2})\,(\mu_{m}^{2}+\mu_{n}^{2})\bigg] B_{(kl)(mn)} \label{sup-mj2bjklbjmnRHS}
\end{align}
On the RHS, only the desired sum $\sum_{j} \mu_{j}^{6} a_{klj} a_{mnj}$ remains undetermined. However, unlike the previous subsection, we do not yet have a simplification of the LHS of this expression. To find such a simplification, we concoct a vanishing combination of two integrals (namely, $(\text{I1})_{(kl)j}$ and $(\text{I2})_{k(lmn)}$ of Eqs. \eqref{sup-I1klj} and \eqref{sup-I2klmn}), each of which vanishes independently because their integrands are total derivatives.

\underline{\bf Integral 1:} The first integral yields a vanishing combination of cubic quantities, and is defined as
\begin{align}
    (\text{I1})_{(kl)j} &\equiv \dfrac{1}{\pi} \int_{-\pi}^{+\pi} d\varphi\hspace{5 pt} \partial_{\varphi} \bigg\{ \Big[\tfrac{1}{2}(\partial_{\varphi}\psi_{j})-3(\partial_{\varphi}A)\psi_{j}\Big] \varepsilon^{-6}(\partial_{\varphi}\psi_{k})(\partial_{\varphi}\psi_{l})\bigg\}\label{sup-I1klj}\\
    &= \dfrac{1}{\pi} \int_{-\pi}^{+\pi} d\varphi\hspace{5 pt} \partial_{\varphi}\bigg\{\tfrac{1}{2}\varepsilon^{+6} \mathcal{D}_{j}\mathcal{D}_{k}\mathcal{D}_{l} - 3 A^{\prime}\, \varepsilon^{+2}\, \psi_{j} \mathcal{D}_{k} \mathcal{D}_{l} \bigg\}
\end{align}
where $\mathcal{D}_{x} \equiv \mathcal{D}\psi_{x} \equiv \varepsilon^{-4}(\partial_{\varphi}\psi_{x})$. By explicitly applying the differentiation and using the wavefunction of the spin-2 modes ($\partial_{\varphi}\mathcal{D}_{x} = -\mu_{x}^{2} \varepsilon^{-2} \psi_{x}$), we attain
\begin{align}
    (\text{I1})_{(kl)j} &\equiv \dfrac{1}{\pi} \int_{-\pi}^{+\pi} d\varphi\hspace{5 pt}\bigg\{ \Big[-3 A^{\prime\prime} - 6(A^{\prime})^{2}\Big] \varepsilon^{+2}\, \mathcal{D}_{k} \mathcal{D}_{l}\, \psi_{j} - \tfrac{1}{2}\varepsilon^{+4} \Big[\mu_{k}^{2} \psi_{k} \mathcal{D}_{l} + \mu_{l}^{2} \mathcal{D}_{k} \psi_{l}\Big]\mathcal{D}_{j}\nonumber\\
    &\hspace{35 pt} +3 A^{\prime} \Big[\mu_{k}^{2} \psi_{k} \mathcal{D}_{l} + \mu_{l}^{2} \mathcal{D}_{k} \psi_{l} \Big]\psi_{j} - \tfrac{1}{2}\varepsilon^{+4}\, \mathcal{D}_{k}\mathcal{D}_{l}\, \mu_{j}^{2} \psi_{j} \bigg\}
\end{align}
Next, to attain the desired index structure, we multiply by $b_{m^{\prime}n^{\prime}j}$ and sum over all $j$:
\begin{align}
    \sum_{j=0}^{+\infty} (\text{I1})_{(kl)j} b_{m^{\prime}n^{\prime}j} &= 
    \dfrac{1}{\pi} \int_{-\pi}^{+\pi} d\varphi\hspace{5 pt}\bigg\{ \Big[-3 A^{\prime\prime} - 6 (A^{\prime})^{2} \Big] \varepsilon^{+8} \mathcal{D}_{k} \mathcal{D}_{l}\mathcal{D}_{m}\mathcal{D}_{n}\bigg\}\nonumber\\
    &\hspace{35 pt}- \dfrac{1}{2}\sum_{j=0}^{+\infty} \mu_{j}^{2} b_{k^{\prime}l^{\prime}j}b_{m^{\prime}n^{\prime}j} -\dfrac{1}{2}\sum_{j=0}^{+\infty} \Big[\mu_{k}^{2} b_{kl^{\prime}j^{\prime}} b_{m^{\prime}n^{\prime}j} + \mu_{l}^{2} b_{k^{\prime}lj^{\prime}} b_{m^{\prime}n^{\prime}j}\Big]\nonumber\\
    &\hspace{35 pt}+\dfrac{3}{\pi} \int_{-\pi}^{+\pi} d\varphi\hspace{5 pt}\bigg\{ A^{\prime} \, \varepsilon^{+6} \Big[\mu_{k}^{2} \psi_{k} \mathcal{D}_{l} + \mu_{l}^{2} \mathcal{D}_{k} \psi_{l} \Big] \mathcal{D}_{m} \mathcal{D}_{n}\bigg\}
\end{align}
Because $(\text{I1})_{(kl)j}$ vanishes, this sum vanishes too, as does the following combination:
\begin{align}
    (\text{I1}) &\equiv \sum_{j=0}^{+\infty} (\text{I1})_{(kl)j}b_{m^{\prime}n^{\prime}j} + b_{k^{\prime}l^{\prime}j}(\text{I1})_{(mn)j}\\
    &= 
    \dfrac{1}{\pi} \int_{-\pi}^{+\pi} d\varphi\hspace{5 pt}\bigg\{ \Big[-6A^{\prime\prime} -12 (A^{\prime})^{2} \Big] \varepsilon^{+8} \mathcal{D}_{k} \mathcal{D}_{l}\mathcal{D}_{m}\mathcal{D}_{n}\bigg\} - \sum_{j=0}^{+\infty} \mu_{j}^{2} b_{k^{\prime}l^{\prime}j}b_{m^{\prime}n^{\prime}j}\nonumber\\
    &\hspace{35 pt}-\dfrac{1}{2}\sum_{j=0}^{+\infty} \Big[\mu_{k}^{2} b_{kl^{\prime}j^{\prime}} b_{m^{\prime}n^{\prime}j} + \mu_{l}^{2} b_{k^{\prime}lj^{\prime}} b_{m^{\prime}n^{\prime}j} + \mu_{m}^{2} b_{k^{\prime}l^{\prime}j} b_{mn^{\prime}j^{\prime}} + \mu_{n}^{2} b_{k^{\prime}l^{\prime}j} b_{m^{\prime}nj^{\prime}}\Big]\nonumber\\
    &\hspace{35 pt} +\dfrac{3}{\pi} \int_{-\pi}^{+\pi} d\varphi\hspace{5 pt} \bigg\{A^{\prime}\, \varepsilon^{+6} \Big[(\mu_{k}^{2} \psi_{k}) \mathcal{D}_{l} \mathcal{D}_{m} \mathcal{D}_{n} + \mathcal{D}_{k} (\mu_{l}^{2} \psi_{l}) \mathcal{D}_{m} \mathcal{D}_{n}\nonumber\\
    &\hspace{70 pt} + \mathcal{D}_{k} \mathcal{D}_{l} (\mu_{m}^{2} \psi_{m}) \mathcal{D}_{n} + \mathcal{D}_{k} \mathcal{D}_{l} \mathcal{D}_{m} (\mu_{n}^{2} \psi_{n}) \Big]\bigg\} \label{sup-I1}
\end{align}
This completes our manipulations of the first integral quantity.

\underline{\bf Integral 2:} The second integral directly yields a vanishing combination of quartic quantities, and is defined as
\begin{align}
    (\text{I2})_{k(lmn)} &= \dfrac{1}{\pi} \int_{-\pi}^{+\pi} d\varphi\hspace{5 pt} \partial_{\varphi}\bigg\{ \Big[\tfrac{3}{8} (\partial_{\varphi}A) \varepsilon^{-2} (\partial_{\varphi}\psi_{k}) - \mu_{k}^{2} \psi_{k} \Big] \varepsilon^{-6}(\partial_{\varphi}\psi_{l})(\partial_{\varphi}\psi_{m})(\partial_{\varphi}\psi_{n})\bigg\} \label{sup-I2klmn}\\
    &= \dfrac{1}{\pi} \int_{-\pi}^{+\pi} d\varphi\hspace{5 pt} \partial_{\varphi}\bigg\{\tfrac{3}{8} A^{\prime}\, \varepsilon^{+8}\, \mathcal{D}_{k}\mathcal{D}_{l}\mathcal{D}_{m}\mathcal{D}_{n} - \mu_{k}^{2}\, \varepsilon^{+6}\, \psi_{k} \mathcal{D}_{l} \mathcal{D}_{m} \mathcal{D}_{n} \bigg\}
\end{align}
Like the previous integral, we next carry out the differentiation while making use of the spin-2 mode wavefunction ($\partial_{\varphi}\mathcal{D}_{x} = -\mu_{x}^{2} \varepsilon^{-2} \psi_{x}$) as we go, and thereby attain
\begin{align}
    (\text{I2})_{k(lmn)} &= \dfrac{1}{\pi} \int_{-\pi}^{+\pi} d\varphi\hspace{5 pt} \bigg\{ \tfrac{1}{4} \Big[\tfrac{3}{2} A^{\prime\prime} + 12 (A^{\prime})^{2} - \mu_{k}^{2} \varepsilon^{+2}\Big] \varepsilon^{+8} \, \mathcal{D}_{k}\mathcal{D}_{l}\mathcal{D}_{m}\mathcal{D}_{n}\nonumber\\
    &\hspace{35 pt} + \tfrac{1}{4}\varepsilon^{+4} (\mu_{k}^{2} \psi_{k}) \Big[(\mu_{l}^{2}\psi_{l})\mathcal{D}_{m}\mathcal{D}_{n} + \mathcal{D}_{l}(\mu_{m}^{2}\psi_{m})\mathcal{D}_{n} + \mathcal{D}_{l}\mathcal{D}_{m}(\mu_{n}^{2}\psi_{n}) \Big]\nonumber\\
    &\hspace{35 pt} -\tfrac{3}{8}A^{\prime}\, \varepsilon^{+6} \Big[5(\mu_{k}^{2} \psi_{k}) \mathcal{D}_{l} \mathcal{D}_{m} \mathcal{D}_{n} + \mathcal{D}_{k} (\mu_{l}^{2} \psi_{l}) \mathcal{D}_{m} \mathcal{D}_{n}\nonumber\\
    &\hspace{70 pt} + \mathcal{D}_{k} \mathcal{D}_{l} (\mu_{m}^{2} \psi_{m}) \mathcal{D}_{n} + \mathcal{D}_{k} \mathcal{D}_{l} \mathcal{D}_{m} (\mu_{n}^{2} \psi_{n}) \Big]\bigg\}
\end{align}
Next, we symmetrize over the indices, as to attain all unique combinations of indices:
\begin{align}
    (\text{I2}) &\equiv (\text{I2})_{k(lmn)} + (\text{I2})_{l(kmn)} + (\text{I2})_{m(kln)} + (\text{I2})_{n(klm)}\\
    &= \dfrac{1}{\pi} \int_{-\pi}^{+\pi} d\varphi\hspace{5 pt} \bigg\{ \Big[\tfrac{3}{2} A^{\prime\prime} + 12 (A^{\prime})^{2}\Big] \varepsilon^{+8} \, \mathcal{D}_{k}\mathcal{D}_{l}\mathcal{D}_{m}\mathcal{D}_{n} \bigg\} - \dfrac{1}{4}\vec{\mu}^{\,2} c_{k^{\prime}l^{\prime}m^{\prime}n^{\prime}} \nonumber\\
    &\hspace{35 pt} + \dfrac{1}{2}\Big[\mu_{k}^{2}\mu_{l}^{2} b_{klm^{\prime}n^{\prime}} + \mu_{k}^{2}\mu_{m}^{2} b_{kl^{\prime}mn^{\prime}} + \mu_{k}^{2}\mu_{n}^{2} b_{kl^{\prime}m^{\prime}n}\nonumber\\
    &\hspace{70 pt} + \mu_{l}^{2}\mu_{m}^{2} b_{k^{\prime}lmn^{\prime}} + \mu_{l}^{2}\mu_{n}^{2} b_{k^{\prime}lm^{\prime}n} + \mu_{m}^{2}\mu_{n}^{2} b_{k^{\prime}l^{\prime}mn}\Big]\nonumber\\
    &\hspace{35 pt} -\dfrac{3}{\pi} \int_{-\pi}^{+\pi} d\varphi\hspace{5 pt} \bigg\{A^{\prime}\, \varepsilon^{+6} \Big[(\mu_{k}^{2} \psi_{k}) \mathcal{D}_{l} \mathcal{D}_{m} \mathcal{D}_{n} + \mathcal{D}_{k} (\mu_{l}^{2} \psi_{l}) \mathcal{D}_{m} \mathcal{D}_{n}\nonumber\\
    &\hspace{70 pt} + \mathcal{D}_{k} \mathcal{D}_{l} (\mu_{m}^{2} \psi_{m}) \mathcal{D}_{n} + \mathcal{D}_{k} \mathcal{D}_{l} \mathcal{D}_{m} (\mu_{n}^{2} \psi_{n}) \Big]\bigg\} \label{sup-I2}
\end{align}
Because $(\text{I2})_{k(lmn)}$ vanishes, $(\text{I2})$ vanishes as well. This completes our manipulations of the second integral.

\underline{\bf Combining:} We finally add $(\text{I1})$ from Eq. \eqref{sup-I1} and $(\text{I2})$ from Eq. \eqref{sup-I2} to attain a new quantity that, of course, also equals zero. Doing so, we attain
\begin{align}
    0 &= (\text{I1}) + (\text{I2})\nonumber\\
    &=- \dfrac{9}{2 \pi} \int_{-\pi}^{+\pi} d\varphi\hspace{5 pt} \bigg\{ A^{\prime\prime} \varepsilon^{+8} \, \mathcal{D}_{k}\mathcal{D}_{l}\mathcal{D}_{m}\mathcal{D}_{n} \bigg\}  - \dfrac{1}{4}\vec{\mu}^{\,2} c_{k^{\prime}l^{\prime}m^{\prime}n^{\prime}} - \sum_{j=0}^{+\infty} \mu_{j}^{2} b_{k^{\prime}l^{\prime}j}b_{m^{\prime}n^{\prime}j}  \nonumber\\
    &\hspace{35 pt} -\dfrac{1}{2}\sum_{j=0}^{+\infty} \Big[\mu_{k}^{2} b_{kl^{\prime}j^{\prime}} b_{m^{\prime}n^{\prime}j} + \mu_{l}^{2} b_{k^{\prime}lj^{\prime}} b_{m^{\prime}n^{\prime}j} + \mu_{m}^{2} b_{k^{\prime}l^{\prime}j} b_{mn^{\prime}j^{\prime}} + \mu_{n}^{2} b_{k^{\prime}l^{\prime}j} b_{m^{\prime}nj^{\prime}}\Big]\nonumber\\
    &\hspace{35 pt} + \dfrac{1}{2}\Big[\mu_{k}^{2}\mu_{l}^{2} b_{klm^{\prime}n^{\prime}} + \mu_{k}^{2}\mu_{m}^{2} b_{kl^{\prime}mn^{\prime}} + \mu_{k}^{2}\mu_{n}^{2} b_{kl^{\prime}m^{\prime}n}\nonumber\\
    &\hspace{70 pt} + \mu_{l}^{2}\mu_{m}^{2} b_{k^{\prime}lmn^{\prime}} + \mu_{l}^{2}\mu_{n}^{2} b_{k^{\prime}lm^{\prime}n} + \mu_{m}^{2}\mu_{n}^{2} b_{k^{\prime}l^{\prime}mn}\Big]\label{sup-I1plusI2}
\end{align}
The two integrals were intentionally weighted as to ensure all terms containing $A^{\prime}$ exactly cancel between the two expressions. The resulting expression possesses a couple of important features.

First, Eq. \eqref{sup-I1plusI2} contains the desired sum $\sum_{j=0}^{+\infty} \mu_{j}^{2}b_{k^{\prime}l^{\prime}j}b_{m^{\prime}n^{\prime}j}$, which we have already demonstrated generates our ultimate target $\sum_{j=0}^{+\infty} \mu_{j}^{6} a_{klj}a_{mnj}$ via the B-to-A formulas, as made explicit in Eq. \eqref{sup-mj2bjklbjmnRHS}. 

Second, nearly all other terms in Eq. \eqref{sup-I1plusI2} can be expressed in terms of $a_{klmn}$, the $B_{(kl)(mn)}$ and $c_{k^{\prime}l^{\prime}m^{\prime}n^{\prime}}$ using our existing relations. The only exceptional term is
\begin{align}
    - \dfrac{9}{2 \pi} \int_{-\pi}^{+\pi} d\varphi\hspace{5 pt} A^{\prime\prime} \varepsilon^{+8} \, \mathcal{D}_{k}\mathcal{D}_{l}\mathcal{D}_{m}\mathcal{D}_{n} \label{sup-AppDkDlDmDn}
\end{align}
which (as we describe now) contains important information about the radion and the tower of Goldberger-Wise scalars. In the {\it unstabilized} model, $A^{\prime\prime}$ is a sum of brane-localized Dirac deltas (specifically, $A^{\prime\prime} = (kr_{c}|\varphi|)^{\prime\prime} = 2kr_{c}[\delta(\varphi) - \delta(\varphi-\pi)]$) and thus -- because $\mathcal{D}_{x} = \varepsilon^{-4}\partial_{\varphi}\psi_{x}$ vanishes at the branes for each spin-2 state -- Eq. \eqref{sup-AppDkDlDmDn} vanishes in the absence of a stabilization mechanism. This contrasts the Goldberger-Wise-stabilized Randall-Sundrum I model, where the equations of motion demand
\begin{align}
    (\partial_{\varphi}^{2}A) &= \dfrac{1}{12} (\partial_{\varphi}\phi_{0})^{2} + \dfrac{1}{3} \, V_{1} \, r_{c} \, \delta(\varphi) + \dfrac{1}{3} \, V_{2} \, r_{c} \, \delta(\varphi - \pi)\label{sup-SRS1xEOMDAA}
\end{align}
such that Eq. \eqref{sup-AppDkDlDmDn} yields a nonzero contribution directly originating from the nonconstant profile of $\phi_{0}$ through the bulk. That is, a nonzero contribution from Eq. \eqref{sup-AppDkDlDmDn} to our calculations directly reflects the stabilization of the radion.

We may use the equation of motion Eq. \eqref{sup-SRS1xEOMDAA} to simplify Eq. \eqref{sup-AppDkDlDmDn}. In particular,
\begin{align}
    - \dfrac{9}{2 \pi} \int_{-\pi}^{+\pi} d\varphi\hspace{5 pt} A^{\prime\prime} \varepsilon^{+8} \, \mathcal{D}_{k}\mathcal{D}_{l}\mathcal{D}_{m}\mathcal{D}_{n} &= - \dfrac{3}{8\pi} \int_{-\pi}^{+\pi} d\varphi\hspace{5 pt} (\phi_{0}^{\prime})^{2} \varepsilon^{+8} \, \mathcal{D}_{k}\mathcal{D}_{l}\mathcal{D}_{m}\mathcal{D}_{n} \equiv -\dfrac{3}{8} x^{(-8)}_{\phi^{\prime}_{0}\phi^{\prime}_{0}k^{\prime}l^{\prime}m^{\prime}n^{\prime}}
\end{align}

In order to finish rewriting Eq. \eqref{sup-I1plusI2}, all that remains now is the application of the B-to-A formulas, the sum relations, Eq. \eqref{sup-mj2bjklbjmnRHS}, and a lot of algebra. The next section presents the results of this process and summarizes the other inelastic sum relations we have derived. The section thereafter reduces those results to the equivalent elastic rules. Finally, the last section of these supplementary materials reviews additional unproven rules necessary to cancel any ``bad" high-energy behavior from the tree-level helicity-zero $(k,l)\rightarrow(m,n)$ matrix element in the stabilized Randall-Sundrum model.

\subsection{Summary of Proven Sum Rules (Inelastic)}
All B-type couplings $\{b_{l^{\prime}m^{\prime}n},b_{k^{\prime}l^{\prime}mn}\}$ can be eliminated in favor of A-type couplings $\{a_{lmn},a_{klmn}\}$ and new $B_{(kl)(mn)}$ objects via the B-to-A formulas
\begin{align}
    b_{l^{\prime} m^{\prime} n} &= \dfrac{1}{2}\left[\mu_{l}^{2} + \mu_{m}^{2} - \mu_{n}^{2} \right] a_{lmn} \label{sup-eq4152}\\ b_{k^{\prime}l^{\prime}mn} &= B_{(kl)(mn)}+\dfrac{1}{6}\bigg[2(\mu_{k}^{2}+\mu_{l}^{2})-(\mu_{m}^{2}+\mu_{n}^{2})\bigg] a_{klmn}\nonumber
\end{align}
where the $B_{(kl)(mn)}$ are constrained such that $B_{(km)(ln)} + B_{(kn)(lm)} +  B_{(kl)(mn)} = 0$, and are symmetric in each individual pair $(k,l)$ and $(m,n)$ as well as with respect to the pair swap replacement $(k,l)\leftrightarrow (m,n)$. These sums are
\begin{align}
    \sum_{j=0} a_{klj}\, a_{mnj} &= a_{klmn}\\
    \sum_{j=0} \mu_{j}^{2} \, a_{klj}\, a_{mnj} &= -2\, B_{(kl)(mn)} + \dfrac{1}{3}\,\vec{\mu}^{\, 2}\, a_{klmn}\\
    \sum_{j=0} \mu_{j}^{4} \, a_{klj}\, a_{mnj} &= 4\, c_{k^{\prime}l^{\prime}m^{\prime}n^{\prime}} -2\, \vec{\mu}^{\, 2} \, B_{(kl)(mn)} + \dfrac{1}{6}\, M^{(2)}_{(kl)(mn)}(1,1,-1) \, a_{klmn}\\
    \sum_{j=0} \mu_{j}^{6} \, a_{klj}\, a_{mnj}  &= 5\,\vec{\mu}^{\,2} c_{k^{\prime}l^{\prime}m^{\prime}n^{\prime}} +  M^{(2)}_{(km)(ln)}(1,1,1)\,B_{(km)(ln)}  +  M^{(2)}_{(kn)(lm)}(1,1,1)\,B_{(kn)(lm)} \nonumber\\
    &\hspace{-35 pt}-  M^{(2)}_{(kl)(mn)}(0,1,0)\, B_{(kl)(mn)} +\dfrac{1}{6}\,M^{(3)}_{(kl)(mn)}(1,4,-4,0)\, a_{klmn}  - \dfrac{3}{2}\,x^{(-8)}_{\phi^{\prime}_{0}\phi^{\prime}_{0}k^{\prime}l^{\prime}m^{\prime}n^{\prime}}
\end{align}
where $\vec{\mu}^{\,2}\equiv \mu_{k}^{2}+\mu_{l}^{2}+\mu_{m}^{2}+\mu_{n}^{2}$, and
\begin{align}
    c_{k^{\prime}l^{\prime}m^{\prime}n^{\prime}} \equiv \dfrac{1}{\pi} \int d\varphi\hspace{5 pt}\vep^{-6}(\partial_{\varphi}\psi_{k})(\partial_{\varphi}\psi_{l})(\partial_{\varphi}\psi_{m})(\partial_{\varphi}\psi_{n})
\end{align}
\begin{align}
    x^{(-8)}_{\phi^{\prime}_{0}\phi^{\prime}_{0}k^{\prime}l^{\prime}m^{\prime}n^{\prime}} \equiv \dfrac{1}{\pi} \int d\varphi\hspace{5 pt} \vep^{-8}\, (\partial_{\varphi}\phi_{0})^{2}\, (\partial_{\varphi}\psi_{k})(\partial_{\varphi}\psi_{l}) (\partial_{\varphi}\psi_{m})(\partial_{\varphi}\psi_{n}) \label{sup-DEFxm8phi02}
\end{align}
The last two sum rules can be combined as to cancel all factors of $c_{k^{\prime}l^{\prime}m^{\prime}n^{\prime}}$, and thereby yields
\begin{align}
    \sum_{j=0} \mu_{j}^{4} \, \bigg(\mu_{j}^{2} - \dfrac{5}{4}\, \vec{\mu}^{\,2} \bigg) \, a_{klj}\, a_{mnj} &= \dfrac{1}{4}\, M^{(2)}_{(kl)(mn)}(5,1,10) \, B_{(kl)(mn)} \nonumber\\
    &\hspace{-105 pt} + M^{(2)}_{(km)(ln)}(1,1,1)\, B_{(km)(ln)} + M^{(2)}_{(kn)(lm)}(1,1,1) \, B_{(kn)(lm)} \nonumber\\
    &\hspace{-70 pt} - \frac{1}{24}\, M^{(3)}_{(kl)(mn)}(1,-1,16,0) \, a_{klmn}  - \dfrac{3}{2}\,x^{(-8)}_{\phi^{\prime}_{0}\phi^{\prime}_{0}k^{\prime}l^{\prime}m^{\prime}n^{\prime}}
\end{align}
These equations extend and generalize the sum rules derived in \cite{Chivukula:2020hvi}.
 
\subsection{Summary of Proven Sum Rules (Elastic)}\label{sup-SummaryOfSumRulesElastic}Oftentimes, we are particularly interested in the {\it elastic} massive spin-2 KK mode scattering process, wherein $k=l=m=n\,(\neq0)$ and the relations of the previous subsections simplify. Consider, for example, the $B$'s constraint in this context:
\begin{align}
    B_{(km)(ln)} + B_{(kn)(lm)} +  B_{(kl)(mn)} = 0 \hspace{15 pt}\stackrel{\text{elastic}}{\longrightarrow}\hspace{15 pt} B_{(nn)(nn)} = 0
\end{align}
such that all of the $B$'s become identical and vanish. The relevant B-to-A formulas become
\begin{align}
    b_{n^{\prime}n^{\prime}j} &= \dfrac{1}{2} \left[ 2\mu_{n}^{2} - \mu_{j}^{2} \right] a_{nnj}\hspace{35 pt}b_{j^{\prime}n^{\prime}n} = \dfrac{1}{2}\, \mu_{j}^{2}\, a_{nnj}\hspace{35 pt}b_{n^{\prime}n^{\prime}nn} = \dfrac{1}{3}\, \mu_{n}^{2}\, a_{nnnn}
\end{align}
whereas the sum rules reduce to
\begin{align}
    \sum_{j} a_{nnj}^{2} &= a_{nnnn} \label{sup-SumRuleO5}\\
    \sum_{j} \mu_{j}^{2}\, a_{nnj}^{2} &= \dfrac{4}{3}\, \mu_{n}^{2}\, a_{nnnn} \label{sup-SumRuleO4}\\
    \sum_{j} \mu_{j}^{4}\, a_{nnj}^{2} &= 4\, c_{n^{\prime}n^{\prime}n^{\prime}n^{\prime}} + \dfrac{4}{3}\, \mu_{n}^{4}\, a_{nnnn} \label{sup-SumRuleO3x}\\
    \sum_{j} \mu_{j}^{6} a_{nnj}^{2} &= 20\, \mu_{n}^{2} \,c_{n^{\prime}n^{\prime}n^{\prime}n^{\prime}} + \dfrac{4}{3} \,\mu_{n}^{6} \,a_{nnnn} - \dfrac{3}{2}\,x^{(-8)}_{\phi^{\prime}_{0}\phi^{\prime}_{0}n^{\prime}n^{\prime}n^{\prime}n^{\prime}} \label{sup-SumRuleO2x}
\end{align}
with the last two expressions combining to yield
\begin{align}
    \sum_{j} \bigg[\mu_{j}^{2} -5 \mu_{n}^{2}\bigg] \mu_{j}^{4}\,a_{nnj}^{2} &= -\dfrac{16}{3}\, \mu_{n}^{6}\, a_{nnnn} - \dfrac{3}{2}\,x^{(-8)}_{\phi^{\prime}_{0}\phi^{\prime}_{0}n^{\prime}n^{\prime}n^{\prime}n^{\prime}}  \label{sup-SumRule02xO3x}
\end{align}

\subsection{Unproven Rules}
The aforementioned rules are insufficient on their own for ensuring cancellations of the $(k,l)\rightarrow(m,n)$ matrix element (which naively contains $\mathcal{O}(s^{5})$ terms) down to $\mathcal{O}(s)$ growth. This is true in both the fully elastic ($k=l=m=n$) and more general cases.

In the fully elastic case, only one additional rule is required:
\begin{align}
    &3 \bigg[ 9 \sum_{i=0}^{+\infty} a_{n^{\prime}n^{\prime}(i)}^{2} - \mu_{n}^{2}\mu_{n}^{2}  a_{nn0}^{2} \bigg] = 15\, c_{n^{\prime}n^{\prime}n^{\prime}n^{\prime}} + 2\mu_{n}^{4} a_{nnnn} \label{sup-elasticradionrule}
\end{align}
The inelastic case provides a generalization of this rule, as well as two additional rules we have yet to prove analytically. These analytic rules have been attained by calculating the full $(k,l)\rightarrow(m,n)$ matrix element (a nontrivial task), asymptotically series expanding that matrix element in $s$ down to $\mathcal{O}(s^{3/2})$ (also nontrivial; note odd powers of $s$ automatically vanish for this particular process), applying the sum rules we previously derived, and demanding coefficients of any $s^{\sigma}$ for $\sigma > 1$ vanish.

Having done so, we find cancellations of ``bad" high-energy behavior additionally require
\begin{align}
    6 B_{(kl)(mn)} &= (\mu_{k}^{2} - \mu_{m}^{2}) (\mu_{l}^{2} - \mu_{n}^{2}) \sum_{j>0} \dfrac{a_{kmj} a_{lnj}}{\mu_{j}^{2}}\nonumber\\
    &\hspace{35 pt} + (\mu_{k}^{2} - \mu_{n}^{2}) (\mu_{l}^{2} - \mu_{m}^{2}) \sum_{j>0} \dfrac{a_{knj} a_{lmj}}{\mu_{j}^{2}}
\end{align}
to cancel $\mathcal{O}(s^{4})$ growth and, noting the KK indices $(k,l,m)$ are cycled through from term to term,
\begin{align}
    0 &= (\mu_{k}^{2} - \mu_{l}^{2}) (\mu_{m}^{2} - \mu_{n}^{2}) \sum_{j>0} \dfrac{a_{klj} a_{mnj}}{\mu_{j}^{2}}\nonumber\\
    &\hspace{35 pt} + (\mu_{l}^{2} - \mu_{m}^{2}) (\mu_{k}^{2} - \mu_{n}^{2}) \sum_{j>0} \dfrac{a_{lmj} a_{knj}}{\mu_{j}^{2}}\nonumber\\
    &\hspace{35 pt} + (\mu_{m}^{2} - \mu_{k}^{2}) (\mu_{l}^{2} - \mu_{n}^{2}) \sum_{j>0} \dfrac{a_{mkj} a_{lnj}}{\mu_{j}^{2}}
\end{align}
to cancel $\mathcal{O}(s^{3})$ growth.

To simplify writing expressions like those above, define
\begin{align}
    L_{kl;mn} &= (\mu_{k}^{2} - \mu_{l}^{2}) (\mu_{m}^{2} - \mu_{n}^{2}) \sum_{j>0} \dfrac{a_{klj} a_{mnj}}{\mu_{j}^{2}}
\end{align}
$L_{kl;mn}$ is antisymmetric under $k\leftrightarrow l$ and $m\leftrightarrow n$, and is symmetric under $kl\leftrightarrow mn$. The previously-listed new sum rules can thus be written succinctly as
\begin{align}
    6 B_{(kl)(mn)} &= L_{km;ln} + L_{kn;lm} \label{NewSumRule1}\\
    0 &= L_{kl;mn} + L_{lm;kn} + L_{mk;ln} \label{NewSumRule2}
\end{align}
This latter sum rule is mathematically distinct from the defining constraint of $B_{(kl)(mn)}$ (i.e., that the sum of all unique $B$ vanishes). Note that $L_{nn;nn} = 0$, thus explaining the absence of these relations when deriving our elastic sum rules.

Lastly, the $\mathcal{O}(s^{3})$ cancellations also necessitates the following generalization of the elastic radion rule (Eq. \eqref{sup-elasticradionrule}):
\begin{align}
    &12 \bigg[ 9 \sum_{i=0}^{+\infty} a_{k^{\prime}l^{\prime}(i)} a_{m^{\prime}n^{\prime}(i)} - \mu_{k}^{2}\mu_{m}^{2}  a_{kl0} a_{mn0} \bigg] = 60 c_{k^{\prime}l^{\prime}m^{\prime}n^{\prime}} + \dfrac{1}{2} M^{(2)}_{(kl)(mn)}(4,-8,5) a_{klmn}\nonumber\\
    &\hspace{35 pt} - 3 (\mu_{k}^{2} - \mu_{l}^{2})^{2} (\mu_{m}^{2} - \mu_{n}^{2})^{2} \sum_{j>0} \dfrac{a_{klj}a_{mnj}}{\mu_{j}^{4}} + 2 \vec{\mu}^{\,2} \bigg[L_{km;ln}+L_{kn;lm}\bigg]\nonumber\\
    &\hspace{35 pt} -3 \bigg[ (\mu_{k}^{2} + \mu_{l}^{2})\, (\mu_{m}^{2} - \mu_{n}^{2})^{2} +  (\mu_{k}^{2} - \mu_{l}^{2})^{2}\, (\mu_{m}^{2} + \mu_{n}^{2}) \bigg] \sum_{j>0}\dfrac{a_{klj}a_{mnj}}{\mu_{j}^{2}} \label{sup-NewSumRule3x}
\end{align}
where $M^{(2)}_{(nn)(nn)}(4,-8,5) = 8\mu_{n}^{4}$. In the unstabilized inelastic calculation, an identical rule is attained, but with the sum $\sum_{i} a_{(i)k^{\prime}l^{\prime}} a_{(i)m^{\prime}n^{\prime}}$ replaced simply by $a_{(0)k^{\prime}l^{\prime}} a_{(0)m^{\prime}n^{\prime}}$. This completes the rules necessary to ensure cancellations down to $\mathcal{O}(s)$ for 2-to-2 spin-2 mode scattering in the stabilized Randall-Sundrum model.
\section{Eigenvalues and Eigenfunctions of Spin-2 and Spin-0 Modes}
\label{app:PertTheory}

The solutions to the Sturm-Liouville (SL) problem for the spin-0 and the spin-2 parts of the stabilized RS model determine the eigenvalues and eigenfunctions. 
Here we outline two related methods of computing the eigen-values and eigen-functions for both the spin-0 and the spin-2 SL problem in perturbation theory.
In the following we introduced the standard Rayleigh-Schr\"odinger perturbation theory in the context of a  SL problem. Here, the perturbed wavefunctions are expressed as an infinite series in unperturbed wavefunctions. On the other hand, being able to have closed form expressions for the perturbed wavefunctions are extremely useful, especially for the numerical part of our analysis. This leads us to solving the perturbed SL problem directly, by solving an inhomogeneous differential equation. In the end the normalized wavefunction derived in either of these methods are identical and calculating the wavefunction using these two methods serves as a cross-check of our results. 

\subsection{Perturbation theory and a general Sturm-Liouville Problem}

\label{app:PerutrbationTheory}

Here we discuss the application of Rayleigh-Schr\"odinger perturbation theory to a general Sturm-Liouville problem, including one in which the weight function is also perturbed. We compute the first-order shifts to the eigenvalues and eigenfunctions, and demonstrate that completeness holds to the appropriate order.
In section~\ref{sec:SLpertRS} we show how the perturbed eigenfunctions can be calculated as a linear combination of unperturbed eigenfunctions. In practice performing an infinite sum of wavefunctions to determine the perturbed wavefunction is not computationally efficient, so in section~\ref{sec:SLpertODE} we outline an equivalent method of determining the perturbed wavefunctions as closed form expressions and show how it is related to Rayleigh-Schr\"odinger perturbation theory. 
\subsubsection{Mass Corrections}
\label{sec:SLpertRS}

Consider a generic Sturm-Liouville (SL) problem for the Kaluza-Klein modes, which is of the form 
\begin{equation}
    \tilde{\mathcal{L}}\psi_{n} = -\lambda_{n}\tilde{\rho}\psi_{n}
\end{equation}
where  $\tilde{\mathcal{L}}$ is the SL operator (given appropriate boundary-conditions) acting on eigenfunctions $\psi_{n}$ with eigenvalues $\tilde{\lambda}_{n}$ and 
and a weight factor $\tilde{\rho}$.
The solutions to the SL problem are orthogonal with respect to the weight factor $\tilde{\rho}$
\begin{equation}
\dfrac{1}{\pi}\int_{-\pi}^{\pi} d\varphi \hspace{5 pt}\tilde{\rho}(\varphi) \, \psi_{k}(\varphi) \, \psi_{l}(\varphi) = \delta_{k,l}~.
\label{eq:orthonormal}
\end{equation}
These solutions then satisfy the completeness relation\footnote{The symmetry of the $\delta$-function implies that the argument of
$\rho$ in the sum could be either $\varphi$ or $\varphi'$.}
\begin{equation}
\sum_{\ell} \tilde{\rho}(\varphi) \, \psi_{\ell}(\varphi) \, \psi_{\ell}(\varphi') = \pi\delta(\varphi-\varphi')~.
\label{eq:completeness}
\end{equation}
Depending on the nature of the SL problem, the boundary conditions can be Dirichlet or Neumann as pointed out in the main body of the paper. For the rest of this appendix, we will drop the argument $\varphi$ in wave functions and weight factors for simplicity.

In perturbation theory, the SL operator and weight function can be expanded as $\tilde{\mathcal{L}}= \mathcal{L} + \delta\mathcal{L}$ and $\tilde{\rho}=\rho+\delta \rho$, while we expand the eigenvalue $\tilde{\lambda}_{n}= \lambda_{n}^{(0)} +  \lambda_{n}^{(1)} +  \lambda_{n}^{(2)} + \cdots  $ . Here both $\mathcal{L}$ and $\delta\mathcal{L}$ are of Sturm-Liouville form
\begin{align}
    \mathcal{L}=\dfrac{d}{d\varphi}\left[p\dfrac{d}{d\varphi}\right]+q~,\label{eq:SLi}\\
    \delta\mathcal{L}=\dfrac{d}{d\varphi}\left[\delta p\dfrac{d}{d\varphi}\right]+\delta q~. \label{eq:SLii}
\end{align}
In our problems, the perturbations $\delta p$, $\delta q$, and $\delta \rho$ come from expanding Eqs.~\eqref{eq:spin2-perturbative} and \eqref{eq:spin0-perturbative} respectively in powers of $\epsilon$.\footnote{Note that the spin-2 system in Eq.~\eqref{eq:spin2-perturbative} yields a perturbation expansion in $\epsilon^2$, whereas that for the spin-0 system in Eq.~\eqref{eq:spin0-perturbative} gives an expansion in powers of $\epsilon$.}
We then expand the eigenfunction in perturbation theory as, 
\begin{equation}
    \psi_{n}= \psi_{n}^{(0)} + \psi_{n}^{(1)} + \psi_{n}^{(2)} + \cdots
\end{equation}
As usual in perturbation theory, we expand the first order perturbed  wave function as a sum of unperturbed wave functions, 
\begin{equation}
    \psi_{n}^{(1)}=\sum_{m=0}^{\infty} C_{nm}\psi_{m}^{(0)}\ .
    \label{eq:RSperturbedWF}
\end{equation}
The coefficients $C_{nm}$  can be determined using perturbation theory, as we will be described later. Here, as usual in Rayleigh-Schr\"odinger perturbation theory, we assume that $\psi_n^{(1)}$ is chosen to be orthogonal to $\psi_n^{(0)}$.

To the lowest order, the SL equation reads
\begin{equation}
    \mathcal{L} \psi_{n}^{(0)} = -\lambda_{n}^{(0)}\rho \psi_{n}^{(0)} 
    \label{eq:SL}
\end{equation}
where $\lambda_{n}^{(0)}$ is the lowest order (unperturbed) eigenvalue. Additionally, $\psi_{n}^{(0)}$ is the lowest order eigenfunction. 

Expanding the SL equation to first order, and using the fact  the lowest order SL problem satisfies Eq. \eqref{eq:SL} , we obtain, to first order, 

\begin{equation}
    \mathcal{L}\psi_{n}^{(1)} + \delta\mathcal{L}\psi_{n}^{(0)}= -  \left( \lambda_{n}^{(0)}\delta\rho\psi_{n}^{(0)} +\lambda_{n}^{(1)}\rho \psi_{n}^{(0)} + \lambda_{n}^{(0)}\rho \psi_{n}^{(1)} \right)  
    \label{eq:SL-perturbed}
\end{equation}
Multiplying the first order perturbed equation by $\psi_{n}^{0}$  and integrating,
\begin{equation}
   \int d\varphi\ \psi_{n}^{(1)} \mathcal{L}\psi_{n}^{(0)} + \int d\varphi\  \delta\mathcal{L}(\psi_{n}^{0})^2= - \int d\varphi \left( \lambda_{n}^{(0)}\delta\rho(\psi_{n}^{(0)})^2 +\lambda_{n}^{(1)}\rho(\psi_{n}^{(0)})^2 + \lambda_{n}^{(0)}\rho\psi_{n}^{(1)} \psi_{n}^{(0)} \right)\ .  
\end{equation}
In the first term in the equation above, we have used the fact that the operator $\mathcal{L}$ is self adjoint. Using Eq.~\eqref{eq:SL} and rearranging, we obtain, 
\begin{equation}
\lambda_{n}^{(1)}= -\dfrac{1}{\pi}\left(\int_{-\pi}^{\pi} d\varphi ~\delta\rho(\psi_{n}^{(0)})^{2} + \int_{-\pi}^{\pi} d\varphi ~\psi_{n}^{(0)}\delta\mathcal{L}\psi_{n}^{(0)} \right)    
\end{equation}
Since $\delta\mathcal{L}= \frac{d}{dx}\left[\delta p \right] +\delta q$, 
we can integrate the above equation by parts to obtain, 
\begin{equation}
   \lambda_{n}^{(1)} = - \dfrac{1}{\pi}\left[-\int_{-\pi}^{\pi} d\varphi~\delta p\left( \frac{d\psi_{n}^{(0)}}{d\varphi} \right)^{2}  + \int_{-\pi}^{\pi} d\varphi~ \delta q\left(\psi_{n}^{(0)}\right)^{2} + \int_{-\pi}^{\pi} d\varphi~\delta\rho\left(\psi_{n}^{(0)}\right)^{2} \right]~.
   \label{eq:SL-mass-pert}
\end{equation}
Now that we have the perturbed eigenvalue, we can proceed to calculate the perturbed eigenfunctions. We describe two methods to do this. The first one involves directly solving the non- homogenous differential equation in Eq.~\eqref{eq:SL-perturbed}. The second one makes use of standard Rayleigh-Schr\"odinger perturbation theory. In the end, both methods lead to the same eigenfunctions and the use of these two methods serves as a cross-check of our results.



\subsubsection{Solving the In-homogeneous Differential Equation using Variation of Parameters}
\label{sec:SLpertODE}
In this first method, one simply solves the non-homgeneous differential equation that is derived by substituting into Eq.~\eqref{eq:SL-perturbed}, the unperturbed eigenvalue and eigenfunction
\begin{equation}
\left[\frac{d}{d\varphi}\left(p\frac{d}{d\varphi}\right) + q - \lambda_n^{(0)}\rho\right]\psi_n^{(1)} = \left[ -\frac{d}{d\varphi}\left(\delta p\frac{d}{d\varphi}\right) -\delta q +
\lambda_{n}^{(0)}\delta \rho + \lambda_n^{(1)} \rho  \right]\psi_n^{(0)}~.
\label{eq:SLperturbedeq}
\end{equation}
In order to solve this equation for the spin-2 KK modes, we have used the method of variation of parameters.
Using the solution $\psi_n^{(1)}$ found using this method, the wave function $\psi_n^{(0)}+\psi_n^{(1)}$ must then be normalized with respect to $\rho+\delta \rho$ as follows
\begin{align}
\tilde{\psi}_n^{(\text{normalized})} = \psi_n^{(0)} + \left[
 \psi_n^{(1)} - \dfrac{\psi_n^{(0)}}{\pi} \int d\varphi'\left( \rho\ \psi_n^{(0)}\  \psi_n^{(1)} +\frac{1}{2} \delta \rho\  (\psi_n^{(0)})^2\right) 
\right]\ .   
\label{eq:WFnormalized}
\end{align}
 In the following we describe how to determine the same wave function using Rayleigh-Schr\"odinger perturbation theory. We reiterate, that the wave functions determined in either method are identical and that the advantage of determining the wave function in this way is that we get closed form solutions for the perturbed wave functions and masses.
\subsubsection{Wave functions in Rayleigh-Schr\"odinger perturbation theory}

The perturbed wave function is determined as a sum over unperturbed wave function as shown in Eq.~\eqref{eq:RSperturbedWF}.
In order to determine the coefficients $C_{nm}$, we multiply the perturbed SL equation ~\eqref{eq:SL-perturbed} by $\psi_{m}^{(0)}$ ($m\neq n$) and integrate, 
\begin{equation}
-\lambda_{m}^{(0)}\pi C_{nm} + \int_{-\pi}^{\pi} d\varphi~\psi_{m}^{(0)}(\delta\mathcal{L}\psi_{m}^{(0)}) = -\left(\lambda_{n}^{(0)}\int_{-\pi}^{\pi} d\varphi~\psi_{m}^{(0)}\delta\rho\psi_{n}^{(0)} + \lambda_{n}^{(0)}\pi C_{nm}\right)    
\end{equation}
leading to 
\begin{equation}
    C_{nm}=- \dfrac{1}{\pi} \frac{\left( \lambda_{n}^{(0)}\int_{-\pi}^{\pi} d\varphi~\psi_{m}^{(0)}\delta\rho\psi_{n}^{(0)} + \int_{-\pi}^{\pi} d\varphi~\psi_{m}^{(0)}(\delta\mathcal{L}\psi_{m}^{(0)})  \right)}{\lambda_{n}^{(0)} - \lambda_{m}^{(0)}}\ .
\end{equation}
Using the definition of $\delta\mathcal{L}$ and integrating by parts, we get, 
\begin{equation}
    C_{nm}= -\dfrac{1}{\pi} \left[\frac{-\int_{-\pi}^{\pi} d\varphi~\delta p\frac{d\psi_{n}^{(0)}}{d\varphi}\frac{d\psi_{m}^{(0)}}{d\varphi} + \int_{-\pi}^{\pi} d\varphi~\delta q\psi_{n}^{(0)}\psi_{m}^{(0)} +\lambda_{n}^{(0)}\int_{-\pi}^{\pi} d\varphi~\delta\rho\psi_{n}^{(0)}\psi_{m}^{(0)} }{ \lambda_{n}^{(0)} - \lambda_{m}^{(0)} }\right]\ .
    \label{eq:Cnm}
\end{equation}
 
 In Rayleigh-Schr\"odinger perturbation theory, we usually assume that the first order perturbed solutions are orthogonal to the lowest order solution so that $C_{nn}=0$. However, in the presence of a perturbation to the weight function, that is no longer the case. Here, to obtain the coefficient $C_{nn}$, we use the normalization condition,
\begin{align}
\int_{-\pi}^{\pi} d\varphi\, \tilde{\rho} \tilde{\psi}_n^2=\int_{-\pi}^{\pi} d\varphi (\rho + \delta \rho) \left(\psi^{(0)}_n + \sum_{m} C_{nm} \psi^{(0)}_m\right)^2 =\pi~,
\end{align}
which, to first order, implies 
\begin{equation}
C_{nn} =-\,\frac{1}{2\pi}  \int_{-\pi}^{\pi} d\varphi\, \delta \rho (\psi^{(0)}_n)^2 ~.
\label{eq:cnn}
\end{equation}
We have checked, numerically, that the wave functions derived using Eq.~\eqref{eq:RSperturbedWF} and Eq.~\eqref{eq:Cnm} are identical to the ones derived using Eq.~\eqref{eq:SLperturbedeq} and ~\eqref{eq:WFnormalized}.

\subsection{Wavefunction and Masses of KK modes in the DFGK model }
\label{app:WFandMasses}

Here we present the wavefunctions and mass corrections for the spin-2 as well as the scalar sector for the DFGK model in the stiff-wall limit. These expressions are derived by solving the differential equations described in section~\ref{app:PertTheory} and specifically using Eq.~\eqref{eq:SL-mass-pert} and Eq.~\eqref{eq:SLperturbedeq}. The expressions presented here are relevant for the large $k r_c$ limit. The general expressions, valid for all $k r_c$, are quite cumbersome and are provided in supplementary Mathematica files on \href{https://github.com/kirtimaan/Stabilized-Extra-Dimenion}{GitHub}.


\subsubsection{Spin-2 mass and wavefunction corrections}
In order to verify sum rules to order $\epsilon^2$ we need the spin-2 wavefunction and masses to order $\epsilon^2$. We start by expanding the spin-2 Sturm-Liouville equation in \eqref{eq:spin2-SLeqn} up to order $\epsilon^2$. We also expand the wavefunction and masses, as described earlier, to order $\epsilon^2$
\begin{align}
\psi_n &= \psi_n^{(0)} + \psi_n^{(2)} + \cdots ~, \\
\mu_n^2 &= (\mu_n^{(0)})^2 + \delta \mu_n^2 + \cdots ~.
\end{align}
We have dropped the $\psi_n^{(1)}$ term since the corrections to the spin-2 Sturm-Liouville problem start at order $\epsilon^2$ as can be seen from its expanded form below 
\begin{align}
    0 &=\bigg[\partial_{\varphi}^{2} - 4\tilde{k}r_{c}\, \partial_{\varphi} + \left(\mu _n^{(0)}\right)^{2}\,e^{2\tilde{k} r_{c} \varphi} \bigg]\psi_{n}^{(0)}\nonumber\\
    &\hspace{15 pt} + \epsilon^{2}\Bigg\{\bigg[ -8\varphi\,\partial_{\varphi} + e^{2\tilde{k}r_{c} \varphi} \bigg(\delta \mu_{n}^{2} + 2\,\varphi^{2} \left(\mu_{n}^{(0)}\right)^{2} \bigg)\bigg] \psi^{(0)}_{n} \nonumber\\
    &\hspace{35 pt} +\bigg[ \partial_{\varphi}^{2} - 4\tilde{k}r_{c}\,\partial_{\varphi} + e^{2\tilde{k}r_{c}\varphi} \left(\mu_{n}^{(0)}\right)^{2}\bigg] \psi^{(2)}_{n}\Bigg\} +\mathcal{O}(\epsilon ^3)
\label{eq:spin2-SL-expanded}
\end{align}
Here we see that the leading term for $\psi_n^{(0)}$is the usual one that we encounter in the unstabilized limit. Solutions to the leading order differential equation are well known and can be found in Ref.~\cite{Chivukula:2020hvi}. We reproduce some of these results here later. After solving for $\psi_n^{(0)}$, 
we then proceed to solve the above differential equation at order $\epsilon^2$. 
\subsubsection{The massless graviton to order \texorpdfstring{$\epsilon^2$}{Lg}}
The massless graviton is the easiest, since it does not acquire a mass, its wavefunction is derived by setting $\mu_0^{(0)} = \delta \mu_0 = 0$. The wavefunction is a constant and any $\epsilon$ dependence comes from the normalization condition \eqref{eq:spin2-mode-norm}. 
The normalized wavefunction for the massless graviton  to order $\epsilon^2$ is of the form
\begin{align}
    \psi_{0} &= \bigg[ \dfrac{1}{\pi}\int d\varphi\ e^{-2 A}\bigg]^{-1/2} \nonumber\\ &=  e^{\pi  \tilde{k}
   r_c} \sqrt{\frac{ \pi \tilde{k} r_c}{e^{2 \pi  \tilde{k}
   r_c}-1}} + \epsilon^2 \frac{\sqrt{\pi }  e^{\pi  \tilde{k} r_c} \Big[-2
   \pi ^2 \tilde{k}^2 r_c^2-2 \pi  \tilde{k} r_c+e^{2 \pi 
   \tilde{k} r_c}-1\Big]}{8 r_c^{3/2} \Big[\tilde{k}
   (e^{2 \pi  \tilde{k}
   r_c}-1)\Big]^{3/2}}\ + \mathcal{O}(\epsilon^3).
   \label{eq:GravitonWF}
\end{align}

\subsubsection{Massive spin-2 modes to order \texorpdfstring{$\epsilon^2$}{Lg}}
Before we present results on the perturbed wavefunctions and masses of the massive spin-2 modes, we remind the reader about the unperturbed wave functions and masses that are identical to the unstabilized case. Since the full expressions can be quite lengthy, we present only results that are valid in the large $\tilde{k} r_c $ limit and provide the full expressions in supplementary Mathematica files on \href{https://github.com/kirtimaan/Stabilized-Extra-Dimenion}{GitHub}.
The wavefunction for the massive spin-2 modes to order $\epsilon^0$ is the same as those derived in the unstabilized RS model and in the large $\tilde{k}r_c$ limit is of the form
\begin{align}
    \psi_n^{(0)}= 
    e^{2 \varphi  \tilde{k} r_c} \sqrt{\frac{ \pi r_c \tilde{k}}{e^{2  \tilde{k} r_c \pi} \, J_2\Big(j_{1,n}\Big)^2 -J_2\Big(e^{-\pi  \tilde{k} r_c} j_{1,n}\Big)^2}}\,\, J_2\Big(e^{\tilde{k} (\varphi r_c-\pi  r_c)} j_{1,n}\Big)\ .
\end{align}
Here $J_i$ are Bessel-J functions. The mass of the spin-2 modes are determined by solving a transcendental equation that is derived from the Neummann boundary condition satisfied by the spin-2 modes:
\begin{align}
    J_1\bigg(\frac{e^{\pi  \tilde{k} r_c} \mu _n}{\tilde{k} r_c}\bigg)\, Y_1\bigg(\frac{\mu _n}{\tilde{k} r_c}\bigg)-J_1\bigg(\frac{\mu _n}{\tilde{k} r_c}\bigg)\, Y_1\bigg(\frac{e^{\pi  \tilde{k} r_c} \mu _n}{\tilde{k} r_c}\bigg)=0\ .
\end{align}
In the large $k r_c$ limit, the solution to the transcendental equation reduces to a simple form
\begin{align}
m_n= \dfrac{\mu_n^{(0)}}{r_c} = \tilde{k}  e^{-\pi  \tilde{k} r_c}\, j_{1,n}\ .
\end{align}
Here $j_{1,n}$ are the roots of $J_1$.

 Substituting the above form of the leading order spin-2 wavefunction and masses into Eq.~\eqref{eq:spin2-SL-expanded}, we end up with a  non-homogeneous differential equation which can be solved using the method of variation of parameters. Alternately, the same differential equation can also be derived from Eq.~\eqref{eq:SLperturbedeq} and in the large $\tilde{k} r_c$ limit is of the form given below.
 \begin{align}
    &\Big[\partial_{\varphi}^{2} - 4\tilde{k}r_{c} \partial_{\varphi} + \left(\mu_{n}^{(0)}\right)^{2} e^{2 \tilde{k} r_c \varphi}\Big] \psi^{(2)}_{n}\\
   &\hspace{35 pt}= 2 c_1 e^{3 \varphi  \tilde{k} r_c} \Bigg[ 8 \mu _n^{(0)} \, \varphi \, J_1\bigg(\frac{e^{ \tilde{k} r_c \varphi} \mu _n^{(0)}}{\tilde{k} r_c}\bigg)- e^{\tilde{k} r_c \varphi} \Big(\delta \mu _n^2+2 (\mu _n^{(0)})^2 \varphi ^2\Big) J_2\bigg(\frac{e^{ \tilde{k} r_c \varphi} \mu _n^{(0)}}{\tilde{k} r_c}\bigg) \Bigg]\nonumber
 \end{align}
Here $c_1$ corresponds to the normalization of the leading order wavefunction $\psi_n^{(0)}$. Below we write down the perturbation of the spin-2 wavefunction $\psi_n^{(2)}$.
The resulting expressions at order $\epsilon^2$ are quite lengthy and are provided also provided in supplementary Mathematica files on \href{https://github.com/kirtimaan/Stabilized-Extra-Dimenion}{GitHub}. 
\begin{align}
    \psi_n^{(2)}=& \frac{c_1 Y_2(z)}{384\mu_n^4} \epsilon ^2 \Bigg\{\mu _n^2 \pi  z^6 \hspace{5 pt}\,
   _3F_4\left(\tfrac{3}{2},2,2;1,3,3,4;-z^2\right) \Big[1-4 \log(\beta z)\Big]
\nonumber \\   
   &+2 \pi z^6 \hspace{5 pt}\,
   _2F_3\left(\tfrac{3}{2},2;1,3,4;-z^2\right) \Big[2 \log
   (\beta z) -1\Big]
\nonumber \\   
   &+384 \pi 
   z^2 \bigg[\bigg(\, _1F_2\left(\tfrac{1}{2};1,1;-z^2\right)-2 \hspace{5 pt}\,
   _1F_2\left(\tfrac{1}{2};1,2;-z^2\right)\bigg) \log (\beta z)+\log (\beta)\bigg]
\nonumber \\   
   &+\pi  z^6 \hspace{5 pt}\,
   _4F_5(\tfrac{3}{2},2,2,2;1,3,3,3,4;-z^2)
\nonumber \\   
   &+192 \sqrt{\pi
   } z^2 \left[G_{2,4}^{2,1}\Bigg(z^2\Bigg|
\begin{array}{c}
 \frac{1}{2},1 \\
 0,0,0,0 \\
\end{array}
\Bigg)-2\,G_{2,4}^{2,1}\Bigg(z^2\Bigg|
\begin{array}{c}
 \frac{1}{2},1 \\
 0,0,-1,0 \\
\end{array}
\Bigg)\right]
\nonumber \\
&-48\pi  \mu_{n}^{2} z^4 J_2(z){}^2 \bigg[4
   \beta^{2} \delta \mu _n^2+ \bigg(2 \Big[\log
   (\beta z)-1\Big] \log
   (\beta z) +1\bigg)\bigg]
\nonumber\\
   &+48\pi \beta^{2} z^4 J_1(z) J_3(z) \bigg[4 \beta^{2}
   \delta \mu_{n}^{2} + \bigg(2 \Big[\log (\beta z)-1\Big] \log (\beta z) +1\bigg)\bigg]
   \Bigg\}
\nonumber\\
  & +\frac{z^2 J_2(z)}{8 \mu
   _{n}^{2}} 
\Bigg\{2 \sqrt{\pi }\,
   G_{3,5}^{2,2}\Bigg(z,\tfrac{1}{2}\Bigg|
\begin{array}{c}
 1,\frac{3}{2},\frac{1}{2} \\
 1,3,-1,0,\frac{1}{2} \\
\end{array}
\Bigg) \Big[2 \beta^{2} \delta \mu _{n}^{2}+ \log
   ^2(\beta z)\Big] \nonumber \\
   &+
   2\mu_{n}^{2}\sqrt{\pi} \left[4\,
   G_{3,5}^{3,1}\Bigg(z,\tfrac{1}{2}\Bigg|
\begin{array}{c}
 \frac{1}{2},-\frac{1}{2},1 \\
 0,0,2,-1,-\frac{1}{2} \\
\end{array}
\Bigg)-G_{4,6}^{2,3}\Bigg(z,\tfrac{1}{2}\Bigg|
\begin{array}{c}
 1,1,\frac{3}{2},\frac{1}{2} \\
 1,3,-1,0,0,\frac{1}{2} \\
\end{array}
\Bigg)\right] \log (\beta z)
\nonumber \\
&+\mu_{n}^{2}\sqrt{\pi} \left[4\,
   G_{4,6}^{4,1}\Bigg(z,\tfrac{1}{2}\Bigg|
\begin{array}{c}
 \frac{1}{2},-\frac{1}{2},1,1 \\
 0,0,0,2,-1,-\frac{1}{2} \\
\end{array}
\Bigg)+G_{5,7}^{2,4}\Bigg(z,\tfrac{1}{2}\Bigg|
\begin{array}{c}
 1,1,1,\frac{3}{2},\tfrac{1}{2} \\
 1,3,-1,0,0,0,\frac{1}{2} \\
\end{array}
\Bigg)\right]
\nonumber \\
&+8 \mu _n^2 \log (z) \Big[2 \log (\beta)+\log (z)\Big]
   \Bigg\}   + \beta^{2} z^{2} \Big[ c_{3}  J_2(z)+c_{4} Y_2(z)\Big]~.
\end{align}
Here $\beta \equiv \tilde{k}r_{c}/\mu_{n}$, $z\equiv(\mu_n/\tilde{k} r_c) e^{\tilde{k} r_c \varphi }$, $G_{i,j}^{k,l}$ are Meijer-G functions~\cite{gradshteyn2014table} , $_i F_j$ are Hypergeometric functions. $c_1$ is a normalization constant given in Eq.~\ref{eq:cnn} and $c_3$ and $c_4$ are constants that are determined from the Neummann boundary condition that the wavefunction must satisfy. Additionally, we provide these expressions as well as those valid also for arbitrary values of $\tilde{k}r_c$ in supplementary Mathematica files on \href{https://github.com/kirtimaan/Stabilized-Extra-Dimenion}{GitHub}.

Correction to the mass to order $\epsilon^2$ are calculated using Eq.~\eqref{eq:SL-mass-pert}. In the large $k r_c$ limit is given in terms of Bessel functions and hypergeometric functions as follows:
\vspace{-0.4 cm}
\begin{align}
    \delta \mu_n^2 =&  -\frac{c_1^2 \epsilon^{2}}{48 \tilde{k}} \Bigg\{\left(j_{1,n}\right)^2 \Bigg[16\,
   J_0\left(j_{1,n}\right)^2 \Big(6 \pi  \tilde{k}
   r_c-1\Big)-48\, J_2\left(j_{1,n}\right)^2 \Big(2 \pi 
   \tilde{k} r_c (\pi  \tilde{k} r_c-1)+1\Big)\nonumber \\ &
   +96\hspace{5 pt}
   \,
   _3F_4\Big(1,1,\tfrac{3}{2};2,2,2,2;-\left(j_{1,n}\right)^
   2\Big)-96 \hspace{5 pt}\,
   _3F_4\Big(1,1,\tfrac{3}{2};2,2,2,3;-\left(j_{1,n}\right)^
   2\Big) \nonumber \\ 
   &+\left(j_{1,n}\right)^2 \hspace{5 pt}\,
   _4F_5\Big(\tfrac{3}{2},2,2,2;1,3,3,3,4;-\left(j_{1,n}\right
   )^2\Big)\Bigg]-32 \Big(3 \,
   J_0\left(j_{1,n}\right)^2-3\Big)\Bigg\} \ .
   \label{eq:delta-mu-squared}
\end{align}
\vspace{-0.75 cm}
\subsection{Spin-0 mass and wave functions in the DFGK model}
In order to verify the sum rules to order $\epsilon^2$ we need the radion mass squared and wavefunction corrections to order $\epsilon^2$. On the other hand, due to the normalization condition in Eq.~\eqref{eqScalar:Norm1} in the stiff-wall limit, the GW scalar wavefunctions do not have a $\epsilon^0$ piece, but instead start at order $\epsilon$. Therefore it is only necessary to calculate their masses and wavefunctions to leading order in $\epsilon$. In order to determine the wavefunctions to the required order in $\epsilon$, 
we start with the Sturm-Liouville problem for the scalar modes defined in Eq.~\eqref{eq:spin0-SLeqn}, and perform and expansion of the same up to order $\epsilon^2$ as follows
\vspace{-0.5 cm}
 \begin{align}
& \bigg[\partial_{\varphi}^{2} + 2\tilde{k}r_{c} \partial_{\varphi} + \left(\mu _{(n)}^{(0)}\right)^{2}\,e^{2\tilde{k}r_{c}\varphi}\bigg] \gamma_{n}^{(0)} + \epsilon \bigg[\dfrac{4\sqrt{6}}{\phi_{1}}\partial_{\varphi}\bigg] \gamma_{n}^{(0)}\nonumber \\
  &+ \epsilon^{2} \Bigg\{\bigg[4\varphi\, \partial_{\varphi} - 4 + e^{2\tilde{k}r_{c}\varphi} \bigg(\delta\mu^{2}_{(n)} + 2 \varphi^{2} \left(\mu^{(0)}_{(n)}\right)^{2}\bigg) \bigg]\gamma^{(0)}_{n} +  \bigg[\partial_{\varphi}^{2} + 2\tilde{k} r_{c}\, \partial_{\varphi} + e^{2\tilde{k}r_{c}\varphi} \left(\mu^{(0)}_{(n)}\right)^{2}\bigg]\gamma^{(2)}_{n}\Bigg\}  = 0
   \label{eq:radion-SL-expanded}
\end{align}


\subsubsection{Radion wavefunction and mass to order \texorpdfstring{$\epsilon^2$}{Lg}}
We expand the wavefunction and mass in perturbation theory as described earlier. For the massless radion we start the with the ansatz
\begin{align}
    \gamma_n &=  \gamma_n^{(0)} + \gamma_n^{(2)} + \cdots \nonumber \\
    \mu_{(n)}^2 &= \left(\mu_{(n)}^{(0)}\right)^2 + \delta \mu_{(n)}^2 + \cdots
\end{align}
Note the absence of order $\epsilon$ term in the expansion above although there is an explicit order $\epsilon$ term in the differential equation. It is easy to see that up to order $\epsilon$, the radion wavefunction is constant and only acquires non-trivial dependence at order $\epsilon^2$. We can substitute the above expansion into Eq.~\eqref{eq:radion-SL-expanded} and solving for $\gamma_n^{(0)}$,  $\gamma_n^{(2)}$, $(\mu_{(n)}^{(0)})^2$ and $\delta \mu_{(n)}^2$ order by order. This amounts to determining the unperturbed wavefunction and using Eq.~\eqref{eq:SLperturbedeq} to determine the perturbed wavefunction by solving the resulting non-homogeneous differential equation
\begin{align}
    \bigg[\partial_{\varphi}^{2} + 2\tilde{k}r_{c} \partial_{\varphi}  +\Big(\delta\mu _{(0)}^2\,e^{2 \tilde{k} r_c \varphi} -4\Big) \bigg] \gamma_{0}^{(0)} = 0
\end{align}
We find the normalized radion mass and wavefunction to order $\epsilon^2$ to be
\begin{align}
\mu_{(0)} &= 2\,\epsilon\,\sqrt{\frac{2}{1+ e^{2 \pi k r_c}}}~, \\
\gamma_{0} = 
&
   \sqrt{\frac{\pi\tilde{k} r_c}{e^{2 \pi  \tilde{k}
   r_c}-1}}  \nonumber \\ 
&+\epsilon ^2\frac{
    \sqrt{\pi \left(e^{2 \pi  \tilde{k}
   r_c}-1\right)}}{6 (\tilde{k} r_c)^{3/2} \left(e^{4 \pi 
   \tilde{k} r_c}-1\right){}^2}
   \Bigg[
   -5 +6 e^{2 \varphi 
   \tilde{k} r_c}+3 e^{2 \pi  \tilde{k} r_c}\nonumber \\ 
&-6 e^{
   (2\varphi +4\pi ) \tilde{k} r_c}-6 \tilde{k} r_c \Big(e^{2 \pi  \tilde{k} r_c}+1\Big)^2
    \Big(e^{2 \pi  \tilde{k} r_c}
   (\pi ^2 \tilde{k} r_c-2 \varphi +\pi )+2 \varphi
   \Big)\nonumber \\ 
&+5 e^{6 \pi \tilde{k} r_c}-6
   e^{2 (\pi - \varphi)  \tilde{k} r_c}+6 e^{2( 3\pi - \varphi ) \tilde{k}
   r_c} -3 e^{4 \pi 
   \tilde{k} r_c}
   \Bigg]
   +\mathcal{O}(\epsilon ^3) \ .
   \label{eq:RadionWF}
\end{align}
Note that that the expression is valid for arbitrary values of $\tilde k r_c$. It is possible to simplify this expression further in the large $\tilde k r_c$ limit to 
\begin{align}
    \gamma_0\bigg|_{\tilde k r_c \gg 1} = \gamma_{0}^{(0)}\left\{
    1 
    +\frac{\epsilon ^2 }{4 \tilde{k}^2 r_c^2} \left(- \tilde{k}
   r_c \left(\pi ^2 \tilde{k} r_c-2 \varphi +\pi \right)+
   e^{-2 \varphi  \tilde{k} r_c}- e^{2 (\varphi -\pi )
   \tilde{k} r_c}+\frac{5}{6}\right)\right\} +\mathcal{O}(\epsilon ^3)
\end{align}
In the large $\tilde k r_c$ limit, the wavefunction for the unstabilized radion is $\gamma_{0}^{(0)} = e^{-\pi \tilde k r_c}\sqrt{\pi \tilde k r_c}$.
\subsubsection{GW-scalar mass and wavefunction to leading order}
As remarked upon earlier, in order to verify the sumrules to order $\epsilon^2$, we need the GW-scalar wavefunction to order $\epsilon$, which, due to the normalization condition in Eq.~\eqref{eqScalar:Norm1}, is in fact the leading order for the massive GW-scalars. Hence we only need to solve the leading term in Eq.~\eqref{eq:radion-SL-expanded}.
The normalized GW-scalar wavefunction to order $\epsilon$ is 
\begin{align}
 \gamma_i=\dfrac{2 \epsilon}{\mu_{(i)}}\, e^{ - \tilde{k}
   r_c \varphi} \sqrt{\frac{ \tilde{k} r_{c}\pi}{e^{2 \pi  \tilde{k} r_c}\,
   J_1\Big(\frac{e^{\pi  \tilde{k} r_c} \mu _{(i)}}{\tilde{k}
   r_c}\Big)^2-J_1\Big(\frac{\mu _{(i)}}{\tilde{k}
   r_c}\Big)^2}}\hspace{5 pt} J_1\bigg(\frac{e^{\tilde{k} r_c \varphi}
   \mu _{(i)}}{\tilde{k} r_c}\bigg)
   \label{eq:GWscalarWF}
\end{align}
Looking at the normalization condition of Eq.~\eqref{eqScalar:Norm1}, we see that there is no $\epsilon^0$ piece. Here we have used the large $k r_c$ limit and omitted Bessel-Y functions.
The mass of GW-scalars are determined by solutions of the following transcendental equation:
\begin{align}
J_2\bigg(\frac{e^{  \tilde{k} r_c \pi} \mu _{(i)}}{\tilde{k} r_c}\bigg)\, Y_2\bigg(\frac{\mu _{(i)}}{\tilde{k} r_c}\bigg)-J_2\bigg(\frac{\mu _{(i)}}{\tilde{k} r_c}\bigg)\, Y_2\bigg(\frac{e^{\pi  \tilde{k} r_c} \mu _{(i)}}{\tilde{k} r_c}\bigg)=0~. \label{eq:GWscalarMassCond}
\end{align}
In the large $k r_c$ limit, this reduces to the simple form
\begin{align}
    m_{(i)}= \dfrac{\mu_{(i)}}{r_c} = \tilde{k}  e^{-\pi  \tilde{k} r_c}\, j_{2,i}\ ,
    \label{eq:GWScalarMass}
\end{align}
where $j_{2,i}$ are roots of $J_2$.

\bibliographystyle{utphys}
\bibliography{references}

\end{document}